\pdfoutput=1

\documentclass[journal]{IEEEtran}
\IEEEoverridecommandlockouts

\usepackage{cite}
\usepackage{amsmath,amssymb,amsfonts}
\usepackage{algorithmic}
\usepackage{graphicx}
\usepackage{textcomp}
\usepackage{xcolor}
\usepackage{todonotes}
\usepackage{epsfig}
\PassOptionsToPackage{hyphens}{url}
\usepackage{hyperref}
\usepackage{array}
\usepackage{hhline}
\usepackage{caption}
\usepackage{subcaption}
\usepackage{epstopdf}
\usepackage{multirow}
\usepackage{fixltx2e}
\usepackage{balance}
\usepackage{booktabs}

\newcommand{\etal}{\textit{et al}. }

\begin{document}

\title{Kinetic Song Comprehension: Deciphering Personal Listening Habits via Phone Vibrations}

\author{\IEEEauthorblockN{Richard Matovu, Isaac Griswold-Steiner, Abdul Serwadda} \\
	\IEEEauthorblockA{Texas Tech University, Lubbock, TX 79409 \\
		Email: \{richard.matovu, isaac.griswold-steiner, abdul.serwadda\}@ttu.edu }
}

\maketitle

	\begin{abstract}
	Music is an expression of our identity, showing a significant correlation with other personal traits, beliefs, and habits. If accessed by a malicious entity, an individual's music listening habits could be used to make critical inferences about the user. In this paper, we showcase an attack in which the vibrations propagated through a user's phone while playing music via its speakers can be used to detect and classify songs.
	
	Our attack shows that known songs can be detected with an accuracy of just under 80\%, while a corpus of 100 songs can be classified with an accuracy greater than 80\%. We investigate such questions under a wide variety of experimental scenarios involving three surfaces and five phone speaker volumes. Although users can mitigate some of the risk by using a phone cover to dampen the vibrations, we show that a sophisticated attacker could adapt the attack to still classify songs with a decent accuracy.
	
	This paper demonstrates a new way in which motion sensor data can be leveraged to intrude on user music preferences without their express permission. Whether this information is leveraged for financial gain or political purposes, our research makes a case for why more rigorous methods of protecting user data should be utilized by companies, and if necessary, individuals.
	\end{abstract}


	\section{Introduction}
	\label{intro1}
	Recent research has shown that the music we listen to is strongly correlated with several core attributes of our social life, including our personality traits, moral and political beliefs, criminal behavior,  and  interpersonal relationships, to mention but a few \cite{Adrian10}. For example, in a recent study conducted about the music tastes and political affiliations of 1,007 Americans\cite{Akshay2018}, it was found that  republicans were twice as likely as democrats and independents to listen to country music. 
In other studies that have explored a wide range of dynamics of how music taste relates to social traits \cite{Rentfrow2003, North2010, Adrian10}, a number of interesting patterns have been reported, including that, lovers of classical music tend to be creative and introverted while 
hip-hop fans tend to be extroverted and have high self-esteem; fans of the latest chart-topping songs are likely to be low in age; fans of hip-hop/rap were more likely to support the UK being part of the EU, while fans of country music were not; fans of opera, country and classical music were more likely to own a home while fans of hip-hop were not; and, the range of hard drugs tried by fans of hip-hop/rap was likely to be much wider than that of rock and classical music fans.  

These music-behavior correlations, coupled with the fact that people are increasingly dedicating sizable chunks of their time to listening to music on their mobile devices, are the driving force behind why users' music preferences have recently emerged as one of the key markers that drive online advertisement engines. For example, music platforms such as Pandora and Spotify now heavily rely on user's listening habits to determine which kinds of adverts to push out to the user \cite{Amadou2018}.

These music-behavior correlations could however also enable privacy abuse --- e.g., if the user installs an arbitrary app on their device that somehow learns their music selections and exploits them to make inferences on highly personal information about the user. Imagine for instance if an insurance company's app accessed the user's music choices and made determinations about whether they might be very likely to take hard drugs. Even worse, imagine if an app owned by a state actor made inferences on the possible political leanings of the end-user (e.g., democrat vs republican) and then made targeted advertisements aimed to influence voter opinions and potentially skew an election. Depending on how well the individual user's music habits predict their social behavior, such attacks could have significant impacts. 

In this paper, we argue that the accelerometer sensors inbuilt in a smartphone provide a highly reliable channel for a malicious app to mine the user's music preferences, which might then be usable to launch the kind of attacks described in the previous paragraph. In particular, we show that, depending on factors such as the volume at which one plays music, the surface on which the phone rests, the learning paradigm and kind of data used for training, the vibrations of the phone while playing music can be used to identify the song being played with an accuracy of over 80\%. 
Because some phones (namely, Android devices) do not require explicit user permissions to access motion sensor data, this attack would potentially happen without any sort of suspicion from the user.

The general problem of the inference of audio signals based on a phone's vibrations (or motion sensors) has been studied in several previous works (e.g., see \cite{Michalevsky:2014:, zhang2015accelword, AnandS18, anand2019spearphone}). However, all these works focused on speech signals, either in the form of individual spoken words (e.g., a number such as {\it{one}}, {\it{two}} or {\it{three}} --- see \cite{Michalevsky:2014:}) or short phrases (e.g., {\it{OK Google}} --- see \cite{zhang2015accelword}). Compared to a spoken word (or short sequence of spoken words), music poses a much different form of pattern recognition problem given its more complex acoustic and structural properties. For example, while speech operates in a narrow frequency range due to constraints imposed by the physical limits of the human vocal cords, music is a multidimensional superposition of different voices and instruments, which implies a diverse range of frequencies and time-varying patterns in properties such as beats, pitch and tone, to mention but a few. The word-specific traits driving a typical word classification engine would thus hold very little sway in a music classification problem. The search space for music could be narrowed by identifying the individual singer's articulation of certain words. However, a robust classification engine would have to go beyond this and capture song-specific acoustic and structural attributes that are way beyond speech.

In our attack, this is complicated by the nature of the data being consumed by the attacking system. The typical smartphone samples accelerometer and gyroscope data at a rate of 100-200Hz, which is of much lower magnitude than the sampling rate of the music itself (typically 44.1 kHz). The very low sampling rate combined with the way in which different surfaces may emphasize or disguise high level characteristics of a song raises questions as to whether identifying properties of a song may be captured at all by these imprecise sensors. Our work is the first to provide answers to these questions.

Beyond our focus on the previously unexplored music signal, we provide insights into several other new problems in this space, including the extensive evaluation of a possible defense mechanism against the attack, attacker counter measures against the defense, and the detection of novel classes, among others. Consistent with our assumption of an adversary who leverages an app located on the victim's phone to learn and exploit the earlier described music-behavior correlations, we perform all these evaluations under the threat scenario of a malicious app which is located on the phone itself (i.e., no secondary device speakers involved), which is another key variation of our work from the majority of works that evaluated speech signals under different threat models (a more detailed description of how we significantly differ from the state-of-the-art is provided in Section \ref{rela1}).

The paper makes the following contributions:
\begin{enumerate}
	\item {\bfseries{Music identification via motion sensor side-channel attack}}: We design a motion-sensor-based side-channel attack against mobile phones which allows an attacker to fingerprint music played on the mobile phone. The attack takes advantage of the way in which underlying patterns in the audio manifest as vibrations of the phone when the device plays the audio through its speakers. Using a corpus of 100 songs on the Billboard top 100 during the month of June 2018, we show that, given a song that is part of the training set, the attack is able to identify the song with an average F score of over 80\% for certain attack configurations. Because one can only build a training set with a limited number of songs, we also build an anomaly detector that returns a binary decision of whether a given test song is part of the corpus or not. We show the anomaly detection mode to attain F scores of up to 80\%. These results point to motion sensors as a powerful side-channel for leakage of information on the music which a smartphone user listens to. 
	
	\item {\bfseries{Evaluating defensive technique and attacker counter-measures}}: Because the attack is centered on vibrations of the phone caused by the music, it is instructive to evaluate whether the damping effect of smartphone covers might mitigate the attack. Using one of the most popular phone covers (i.e., the Otterbox \cite{CaseSurvey}), we reevaluated the attack and found that its impact can be somewhat curtailed if the attacker does not take into consideration the usage of the phone cover during training. For attackers who incorporate phone covers into their training process, we show that the defense is much less effective however. 
	
	\item \label{attack-contrib} {\bfseries{Sensitivity analysis of the attack for various determinant factors}}: To understand the limits of the attack, we rigorously studied its behavior under a wide range of conditions, including, (1) {\it{phone placement surfaces}} --- we studied the three commonest surfaces on which people place their phones when listening to music, namely the top of a wooden table, on a bed and on a leather couch, (2) {\it{variations in music volume}} --- we rerun the attack for each of the volumes 7, 9, 11, 13 and 15 on our Samsung Galaxy S6 phones, (3) {\it{learning paradigms}} --- we performed a comparative analysis of deep learning and traditional machine learning in order to get an understanding of how machine-generated features compare with learned features (3) {\it{training data compositions}} --- we had a wide range of training data configurations, including those where data collected from all volumes and surfaces were combined to build a composite training dataset, those where individual volumes and surfaces were each used to build a training model, and those were subsets of volumes and surfaces were used to drive the training. In practice the attacker could select from a wide range of configurations, hence the wide array of training configurations gives us insights into how well various kinds of attackers might perform.
	
\end{enumerate}

	\section{Related Work}
\label{rela1}
Below, we describe the two streams of past research which relate to our research, namely: (1) work that studied how a smartphone's motion and orientation sensors can be leveraged for the inference of words spoken by humans on the phone, or other audio signals in the vicinity of the phone, and (2) the broader body of research that studied smartphone motion sensors as a side-channel threat on non-audio information. 

\subsection{Smartphone motion and orientation sensors as a side-channel for audio information leakage}
One of the earliest works investigating the inference of audio information through mobile device motion and orientation sensors was that by Michalevsky \etal \cite{Michalevsky:2014:}. In that work, a subwoofer and two tweeters played the audio of 11 previously recorded spoken words while the gyroscope sensor of a nearby phone captured gyroscope sensor patterns caused by the audio signal. The spoken words in question were the ten numbers {\it{zero}} through {\it{nine}}, as well as the exclamation ``{\it{oh}}''. Using the captured gyroscope data and a series of machine learning algorithms that included a Support Vector Machine (SVM), a Gaussian Mixture Model (GMM) and Dynamic Time Warping (DTW), the authors were able to recognize the 11 words with accuracies of 6-23\% for the speaker-independent scenario and 5-65\% for the speaker-dependent scenario.  

In a study \cite{zhang2015accelword} closely related to that by Michalevsky \textit{et al}., the authors instead focused on the smartphone accelerometer sensor and the hot-words ``{\it{OK Google}}'' and ``{\it{Hi Galaxy}}''. These two hot-words are respectively used by the Google Assistant and Samsung Galaxy devices to initiate voice command-based interaction. The aim of the study was to investigate whether the acceleromter might provide a more energy efficient way to detect these words than the conventionally used (always-on) microphone. The study thus involved an investigation of both the hot-word detection accuracy, as well as the associated energy consumption relative to when a microphone is used. Using data collected from 10 volunteers and a classification engine based on the Decision Tree Classifier, the authors were able to separate these two hot-words from a corpus of other random words with accuracies of 80-85\%, depending on whether the speaker was moving or stationary. Compared to traditional microphone-based hot-word detection mechanisms, this approach was shown to be twice as energy-efficient. 

More recently, Anand \textit{et al}., in their conference paper \cite{AnandS18}, and a follow-up paper that is currently posted on arXiv \cite{anand2019spearphone}, revisited the research question posed in \cite{Michalevsky:2014:} and \cite{zhang2015accelword}, and carried out a set of experiments that studied several scenarios not covered in \cite{Michalevsky:2014:} and \cite{zhang2015accelword}. For example, they, (a) modified the experiment in \cite{Michalevsky:2014:} to explore scenarios of less powerful speakers (e.g., laptop speakers) and scenarios where the phone's on-board speakers produce the audio signals causing the vibrations, (b) evaluated cases of live human speech (i.e., where the hot-words are not played from a speaker), (c) studied cases where the phone playing the speech signals was located on the same surface as the phone on which motion sensor patterns are being monitored, and (d) studied how vibrations caused by speech could be leveraged for both speaker identification and gender classification.  

\begin{figure*}[h]
	\centering
	\includegraphics[width=.98\linewidth]{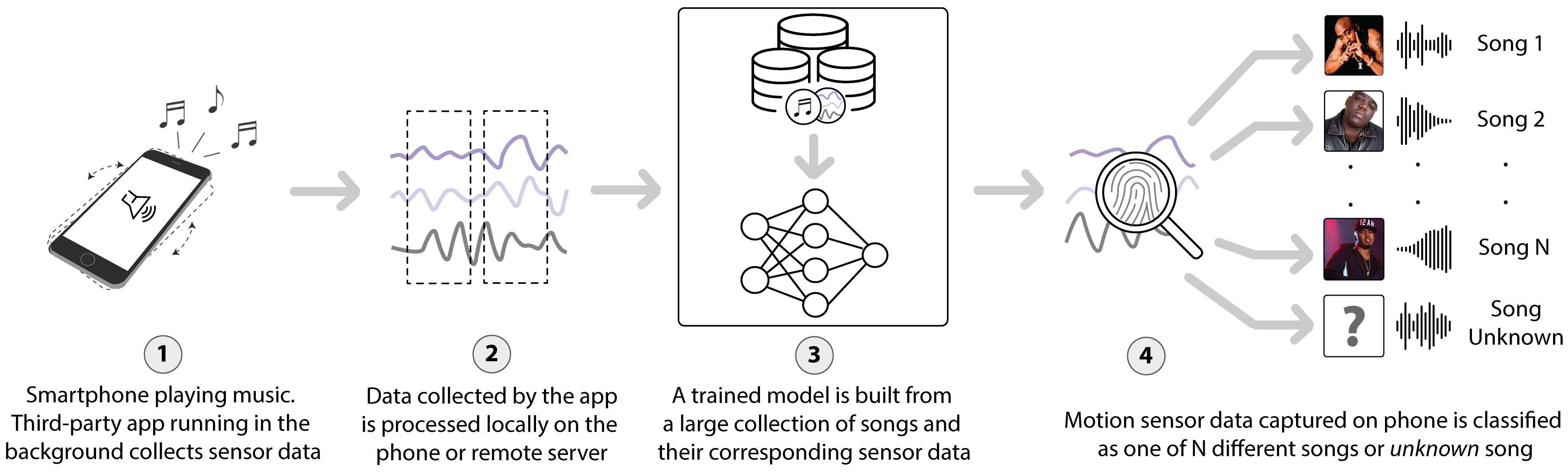}
	\caption{Attack threat model.}
	\label{fig:threat_model}
\end{figure*} 

By virtue of studying the question of audio signal classification based on a smartphone's accelerometer (and) or gyroscope sensors, the above cited four works have some commonality in objective with our work. That said, these works have several significant differences from our work, which include: 

(1) {\it{Properties of Audio Signals Under Investigation}}: The above four papers studied human speech while our work is focused on music. As earlier described in Section \ref{intro1}, the wide range of acoustic and structural differences between music and spoken language signals (see \cite{Muller5709966}) introduce intricacies that make ours a different pattern recognition problem from that studied in these previous works. 

Take the embedded frequency components for instance. Due to the inherent human constraints on the size and structure of the vocal tract, the fundamental frequencies of female and male speech are on average about 210 Hz and 120 Hz respectively  \cite{traunmuller1995frequency}. This limitation in frequencies creates a well-posed problem in which the required discriminative traits are located within a narrow, predictable range of frequency bands. 
Music on the other hand is a superposition of multiple instruments and voices which often overlap in time. A single song can depict high variability in embedded frequencies depending on the instruments and singers involved, and can exhibit complex dynamics in properties such as pitch, tone, beats, etc, depending on factors such as the genre of the song (See \cite{Muller5709966} for detailed descriptions of these properties and how they separate music from speech).  
While studies such as \cite{Michalevsky:2014:}, \cite{AnandS18},  \cite{anand2019spearphone} and \cite{zhang2015accelword} have shown that spoken words recorded during a typical human conversation could be inferred from smart phone motion sensor patterns, {\it{it is not known whether, or how well}}, these frequency-limited sensors\footnote{Mobile device Operating Systems limit them to a maximum frequency of about 200Hz \cite{Michalevsky:2014:} in order to save battery power}, subject to noise from variations in phone placement surfaces, might be able to capture the wealth of information required to uniquely identify a song from a pool of songs that potentially have a wide range of patterns in common with it. {\it{This paper is to our knowledge the first to address this question}}. 

(2) {\it{Evaluation of Defence Mechanism}}:
None of the above four papers studied defenses to the attacks showcased therein. Our paper fronts the vibration damping effect of phone covers as a possible defense mechanism that might be deployed by users, and dedicates a significant amount of experiments to studying the performance of this defence under different assumptions about the capabilities of the attacker. Given that phone covers are, for other purposes, already widely used by smartphone users, insights into their defensive credentials are very critical to realistically understanding the threat posed by these audio recognition attacks in practice.  

(3) {\it{Wide Variety of Experiments that Simulate Different Attacker Choices}}: To rigorously understand the behavior of different flavors of our attacks, we experiment with a wide range of design choices, such as, (a) surfaces of phone placement (i.e., table, bed and couch), (b) data sources driving the training module (e.g., surface-specific vs mixed surface training), and, (c) mis-matches between training and testing surfaces. Additionally, we go beyond the primary music identification attack and study two complimentary attacks (i.e., the novelty detection attack, and the phone cover detection attack) that sophisticated adversaries might in practice use to augment the primary music identification. Because the works in \cite{Michalevsky:2014:, zhang2015accelword, AnandS18} and \cite{anand2019spearphone}, tackle a significantly different instance of the audio inference problem, they do not provide any of this kind of analysis.

(4) {\it{Comparing of Feature-Learning and Feature-Engineering-based Classification Paradigms}}:
Given its ability to learn highly powerful features that humans were previously unable to formulate using traditional approaches, feature (or deep) learning has recently had a revolutionary impact on the performance of machine learning-centric systems. All four above works entirely used traditional machine learning-based schemes for their attack design. By additionally employing deep learning in this paper, we not only provide a view of the attack from the perspective of a sophisticated attacker who is aware of the latest developments in machine learning,  but also provide a comparison with an attacker who might employ the more traditional feature-engineering based approaches.

Three of the above cited four papers (i.e., \cite{Michalevsky:2014:, zhang2015accelword, AnandS18}) have an additional fundamental difference from our work, namely,

(5) {\it{Threat Scenario Studied}}: 
Our work is focused on the scenario of a malicious entity such as an advertising company that has a rogue app which seeks to make inferences on the kind of multimedia content consumed by the owners of the phones who install the app. Our experiments thus involve a smartphone playing music while a rogue app running on the same phone records the sensor data emanating from the vibrations caused by the music. 
On the other hand, all experiments in the above cited 3 studies involved vibrations caused by audio generated from a secondary speaker that was not integral to the phone being attacked. In \cite{Michalevsky:2014:}, a subwoofer and two speakers were used while in \cite{zhang2015accelword} a second phone was used to generate the audio, which was then sensed, across the air medium, by the sensors in the target phone. In \cite{AnandS18}, a conventional loudspeaker with
subwoofers, laptop speaker, and smartphone speaker placed on the same surface and different surface with the target phone were used, as well as people speaking in the neighborhood of the target phone. These three publications hence provide little or no insights in to the threat studied in our paper. The forth paper \cite{anand2019spearphone}, by virtue of studying vibrations produced from the victim phone's on-board speakers, explores a scenario similar to our work. However, our work varies significantly in almost every other aspect. This paper demonstrates multiple attacks against a different type of audio with a significantly larger set of experimental scenarios and configurations. We also rigorously explore a likely defensive measure against our own attack and how an attacker might react to such a response. These differences (detailed in points (1) to (4) above), set our work significantly apart from their research.

\subsection{Smartphone motion and orientation sensors as a side-channel for non-audio information leakage}
A more distantly related line of works to our research are those that studied other forms of motion sensor side-channel attacks that target non-audio information. Among the earliest of these was the work by 
Cai \etal \cite{cai2011touchlogger}, in which a tool called TouchLogger, was shown to infer keystrokes on a smartphone using smartphone orientation sensor data. Other flavors of this text inference attack have since been studied --- e.g., Xu \etal  \cite{xu2012taplogger} used both the accelerometer and orientation sensors for keystroke inference, Marquardt \etal  \cite{marquardt2011sp} focused on inference of text typed on nearby keyboards, and more recently Tang \etal \cite{tang2018niffler}  and Hodges \etal \cite{hodges2018reconstructing} respectively focused on validating the inference of PINs in a much larger user-independent setting and the inference of bigrams within text using both accelerometer and gyroscope data. Beyond the inference of keystrokes, other works have used motion and orientation sensors for a wide range of attacks, including, tracking metro paths \cite{hua2016we}, inferring objects printed on nearby 3D printers \cite{song2016my}, fingerprinting smart phone identities \cite{das2018every},  and prying into private spaces \cite{zakery2019prying}, to mention but a few.  

These works share our motivation of showcasing the threats posed by motion and orientation sensors on mobile devices. However, because they focus on a target variable that is completely different from ours (i.e., non-audio information), we do not discuss them in details due to space limitations.

\begin{figure*}[h]
	\centering
	\begin{subfigure}[t]{0.98\textwidth}
		\begin{subfigure}[t]{0.45\textwidth}
			\centering
			\includegraphics[width=1\linewidth]{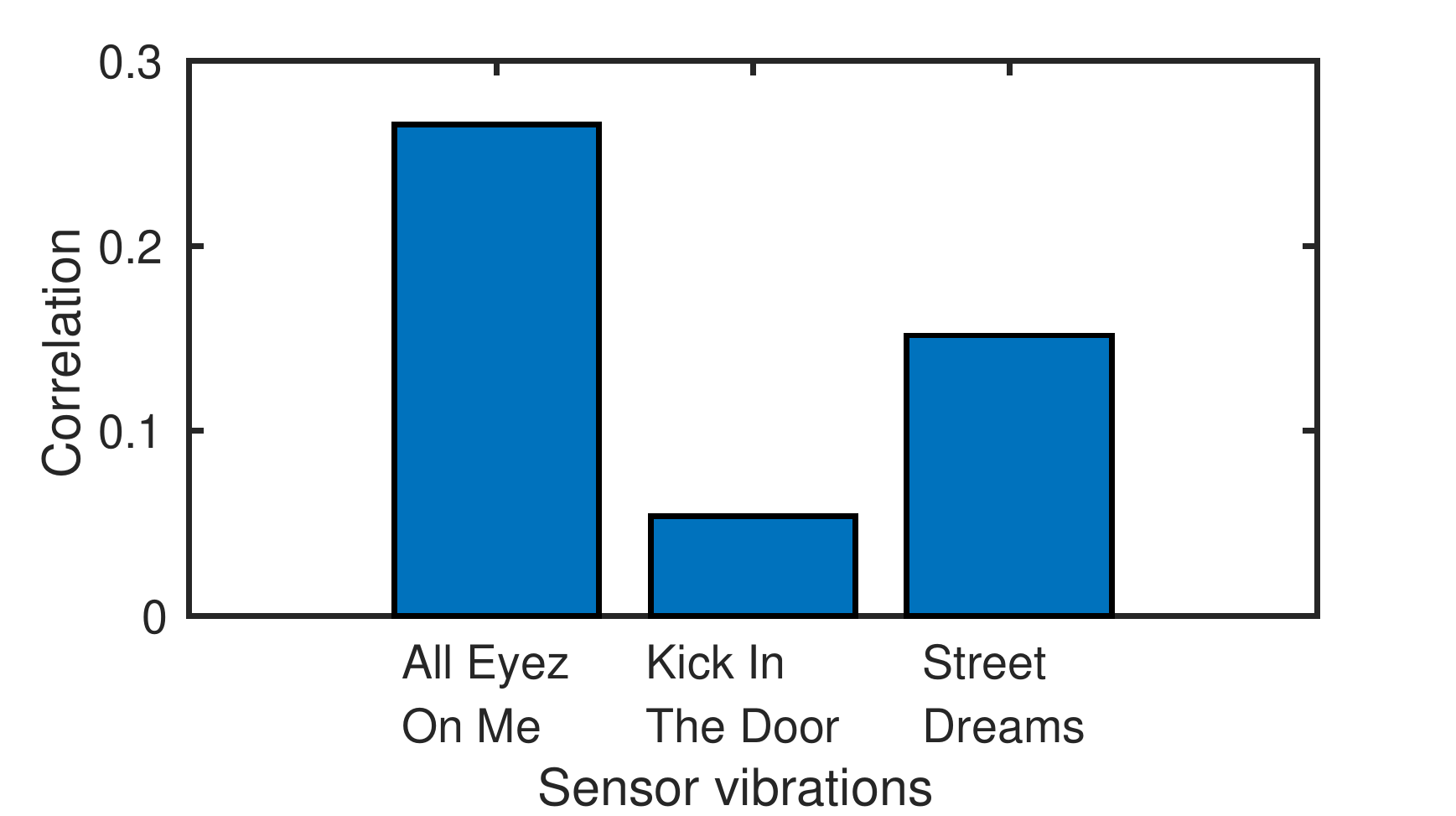}
			\caption{Correlation between audio stream and phone vibrations. }
			\label{fig:audio_sensor_correlation}
		\end{subfigure} \hfill%
		~
		\begin{subfigure}[t]{0.45\textwidth}
			\centering
			\includegraphics[width=1\linewidth]{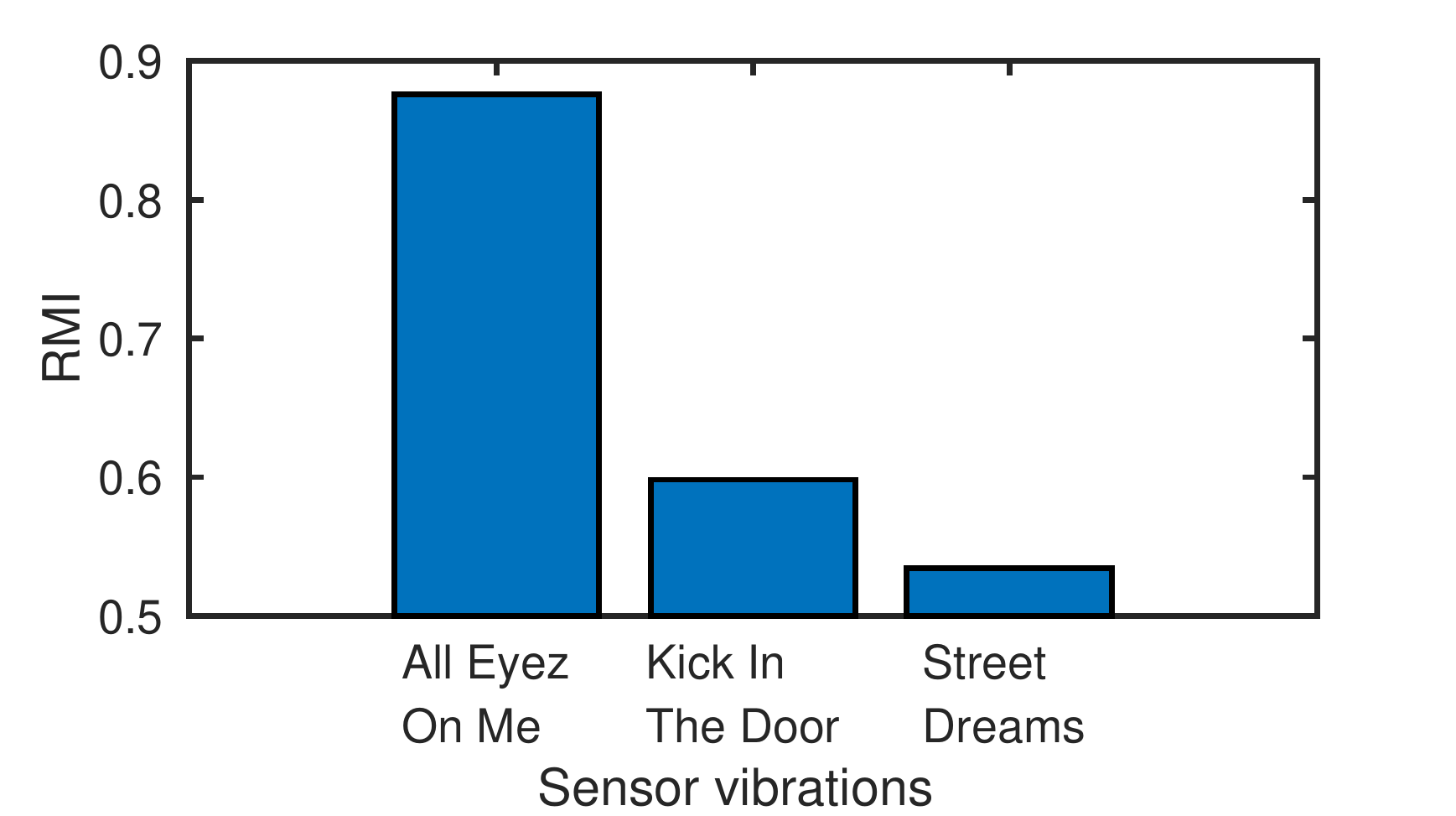}
			\caption{Relative mutual information between audio stream and phone vibrations.
			}
			\label{fig:audio_sensor_rmi}
		\end{subfigure}
	\end{subfigure}%
	\caption{Comparing the audio data stream of the song, \textit{All Eyez on Me}, with the sensor vibration data stream produced when a Samsung Galaxy S6 phone resting on a table-top respectively plays the songs, \textit{All Eyez on Me} (Tupac), {\it{Kick In the Door}} (Notorious BIG) and {\it{Stream Dreams}} (Nas). All three songs are in the same genre while the first and third songs have exactly the same beat. 
	}
	\label{fig:audio_sensor_comparison}
\end{figure*}

\section{Threat Model}
\label{sec:threat_model}
The attack studied in this work assumes a user who listens to music on their smartphone (e.g., via a web browser, or via an app built for a platform such as YouTube, Spotify or Pandora). The music is played from the phone's speakers (as opposed to earbuds or headphones) since our interest are the vibrations of the phone due to the music. As the user listens to the music, some third party app running on the phone in the background stealthily captures motion sensor data being produced by the phone to perform analytics on it and identify the song being played (see Figure \ref{fig:threat_model} for high level overview of process). Note that while apps such as YouTube or Pandora also perform analytics on the user's music selections, these apps do this analysis with the user's permission (as specified in the privacy agreements that the user consents to while installing the apps). The third party app specified here seeks to get access to a similar kind of information without the user's knowledge or consent. 

During the attack itself, the rogue third party app might: (1) perform the song identification analytics locally on the phone and send the result to the attacker's server, or, (2) send the raw motion sensor data to some remote server where the analytics would be performed, or, (3) employ some hybrid of the above two approaches in order to optimize the balance between resource consumption on the phone and bandwidth usage when sending data over the Internet. Note that the analytics referred to above are all steps entailed in the classification stage of the machine learning process. The much more resource-intensive training stage would likely be implemented on a large music database at the attacker's server. On identifying the music played over a given length of time (say, several weeks or months), the attacker will proceed to profile the user without their knowledge and potentially execute one or more threat vectors, such as those highlighted in Section \ref{intro1} (e.g., sending customized ads to the victim, setting insurance premiums based on the user profile, etc.). 

Given a tech-savvy victim who is knowledgeable about the side-channel threat posed by motion sensors on mobile devices, it is possible that they might house their phone in a cover that has a damping effect on the vibrations. Our threat model assumes that certain attackers would be aware of such potential precautionary measures and might hence configure the attack to attempt to work around them. In the performance evaluation section, we study both the attacker who considers the possibility of a damping-based defence (i.e., includes such data in the training process) and the attacker who performs a plain attack that assumes no such defences. 

It is noteworthy that while our study is focused on music, this kind of attack could in practice generalize to other multi-media content that produces audio on the phone (e.g., cable news, sports events, podcasts, etc.). All that the attacker would have to do is to train their classifiers on the specific multimedia content and then later compare the victim's motion sensor data with the trained model.

\section{Experimental Design}
\subsection{Preliminary exploration of dependence of phone vibrations on music being played from its speakers}
As discussed in Section \ref{rela1}, an underlying question that is critical to the feasibility of our study is that of whether the limited sampling rates of a phone's motion sensors could capture the structural and acoustic complexities of music well enough to uniquely identify a song. Before committing to the fully-fledged data collection experiments, we sought to gain insights into this question through a series of mini experiments designed to study the associations between the vibrations exhibited by a phone's motion sensors and the music being played from the phone's speakers. In these experiments, we took small subsets of carefully selected songs and studied the association between the songs and the vibrations (i.e., using correlation and mutual information), and also made simple visual illustrations of how well simple hand-crafted features separated the songs in low dimensional spaces.

\begin{figure}[h]
	\centering
	\includegraphics[width=.9\linewidth]{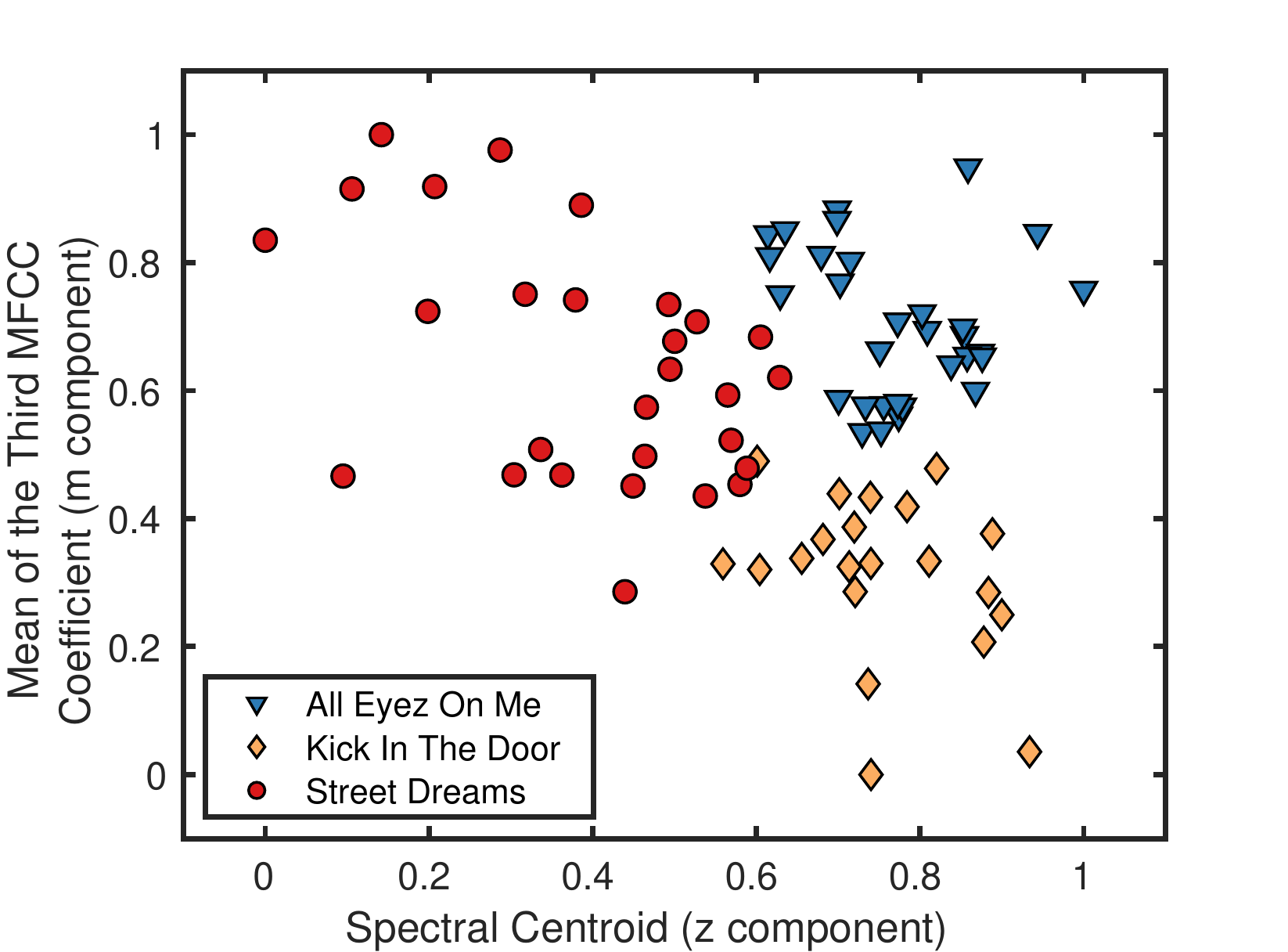}
	\caption{Visual illustration of a two-feature representation of the sensor vibration data stream produced when a Samsung Galaxy S6 phone resting on a table-top respectively plays the songs, \textit{All Eyez on Me} (Tupac), {\it{Kick In the Door}} (Notorious BIG) and {\it{Street Dreams}} (Nas). In this very low dimensional space, the three songs separate into 3 distinct clusters.}
	\label{fig:song_features}
\end{figure}

For a subset of the songs studied in these preliminary experiments, Figure \ref{fig:audio_sensor_comparison} shows the correlation and mutual information analysis, while Figure \ref{fig:song_features} shows the simple features.  To do the mutual information and correlation computations represented in Figure \ref{fig:audio_sensor_comparison}, we first pre-process the raw audio times series and vibration data in order to bring them down to a similar number of samples (recall that the raw audio has a sample rate of 44.1kHz, while the sensor vibrations have a sample rate of 100Hz). The pre-processing is done as follows. First, we use the Librosa library \cite{LibrosaLibrary} to compute the number of beats per minute (bpm) and locate the positions of the beats in the audio time series of each song. Let $t_i$ and $A_i$ respectively represent the timestamp at which the $i_{th}$ beat is detected, and the corresponding value of the audio time series at that time. For a window of 0.5 seconds centered at $t_i$, we compute, $P_i$, the mean of the magnitude of the phone's acceleration (i.e., accelerometer data) generated by the song in question. Over the length of a song, the vector containing the $A_i$ values and the vector containing the corresponding $P_i$ values respectively represent the song audio signal and vibration signal, and have the same number of samples. The correlation and mutual information between the song audio and song vibrations (see Figure \ref{fig:audio_sensor_comparison}) are computed based on these two vectors. The simple hand-crafted features (Figure \ref{fig:song_features}) are computed on the raw acceleration data (over windows of 0.5 seconds).

The two figures are based on 3 songs, namely, {\it{All Eyez on Me}} \cite{AllEyesonMeYoutube} by Tupac Shakur, {\it{Kick In the Door}} \cite{KickInTheDoorYoutube} by the Notorious B.I.G and {\it{Street Dreams}} \cite{StreetDreamsYoutube} by Nas. The songs are all in the hip-hop/rap genre and thus provide an interesting example of whether the statistical measures highlighted in the previous paragraph could depict class separability for songs in the same genre. Further, the songs, {\it{All Eyez on Me}} and {\it{Street Dreams}}, have exactly the same instrumental (or beat), providing a good case study of whether phone vibrations can capture their separating subtleties beyond the beat.  

Figure \ref{fig:audio_sensor_correlation} shows results from the correlation analysis. The first bar in this figure represents the Pearson correlation coefficient computed between the audio signal of the song, {\it{All Eyez on Me}}, and the phone vibrations caused by the same song on a Samsung galaxy S6 phone. For the other two bars, the correlation coefficients are computed in such a way that the audio signal is still that of the song, {\it{All Eyez on Me}}, while the vibrations are respectively obtained when songs, {\it{Kick In the Door}} and {\it{Street Dreams}} are played on the Samsung Galaxy S6 phone. {\it{The plot thus shows the correlation coefficients between a song and its own vibrations (shown in the first bar), and between the same song and the vibrations caused by songs other than it (shown in the second and third bars)}}. 

Observe that while all correlations are weakly positive, the correlation coefficient between the song and its own vibrations (i.e., bar \#1) is almost twice as high as that when the song and vibrations are cross-matched (i.e., bars \#2 and \#3). 
This result (depicted in several other song selections not shown here) provides the first evidence of a song's audio signal having a stronger association with its vibrations than it does with the vibrations of other songs that have a significant amount of similarity with it (e.g., in terms of similar beats or genre). 

A deficiency with correlation analysis is that it only captures the linear dependencies between the variables in question \cite{james2013introduction}.  With music having many structural complexities emanating from the sometimes multidimensional integration of various sounds, it is highly likely that any connections between a song and its vibrations might be more than just linear. 

To more rigorously explore the music-vibration associations, we proceeded and studied the mutual information between the songs and the vibrations. Mutual information captures the reduction in uncertainty of one variable (in this case the song) given information about the other variable (in this case the vibrations) and captures both the linear and non-linear relationships between the variables in question. 

Figure \ref{fig:audio_sensor_rmi} shows the mutual information results for the same three songs. 
The first bar in this figure shows results for the mutual information between the song {\it{All Eyez on Me}}, and its vibrations, while the other two bars capture the mutual information between the same song and the vibrations of the two other songs. Note that we use the relative mutual information (RMI), a normalized form of mutual information that is often used for feature comparisons (e.g., see \cite{frank2012touchalytics}). By constraining the mutual information metric to the range $[0,1]$, the RMI metric eases our comparisons and analysis. 

Figure \ref{fig:audio_sensor_rmi} shows that the song, {\it{All Eyez on Me}}, has an RMI with its vibrations that is almost equal to 1, indicating that the phone vibration pattern almost entirely removes all uncertainty about the identity of the song. It s noteworthy however that the vibrations of the other two songs have fairly high RMI with the song, {\it{All Eyez on Me}}. We conjecture that this could be due to the fact that the mutual information might have possibly picked up some other genre-specific patterns common to the three hip-hop songs. This fact notwithstanding however, a song such as {\it{Street Dreams}} that has a beat very similar beat to that of {\it{All Eyez on Me}} is still clearly distinct from the reference song, {\it{All Eyez on Me}} itself, given an RMI of just slightly over 0.5. 

A final step of our preliminary analysis was a visual inspection of feature plots to get some idea of class separability based on simple human-engineered features. Figure \ref{fig:song_features} shows an example plot from this analysis. The features in question are: (1) the third MFCC coefficient of the magnitude of accelerometer measurements, and, (2) the  spectral centroid of the z component of the acceleromter measurements. Observe that these two features separate the three songs into 3 distinct clusters. Several other combinations of songs produced similarly distinct clusters in these low dimensional spaces, a fact that prompted us to hypothesize that with a larger carefully crafted feature-set (i.e., high dimensional space), the motion sensor measurements might indeed be able to discriminate between a large number of songs. With these preliminary results, we proceeded to the fully-fledged data collection experiment which is described next.

\begin{figure*}[h]
	\centering
	\begin{subfigure}[t]{0.98\textwidth}
		\begin{subfigure}[t]{0.3\textwidth}
			\centering
			\includegraphics[width=1\linewidth]{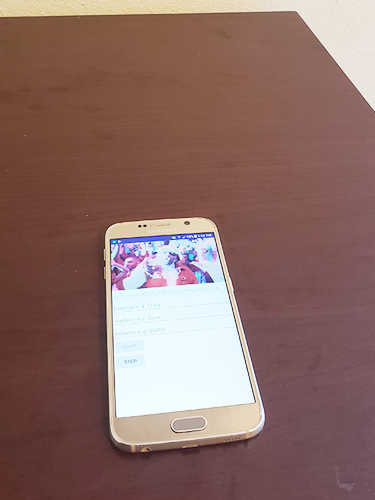}
			\caption{Table. }
			\label{fig:surface_table}
		\end{subfigure} \hfill%
		~
		\begin{subfigure}[t]{0.3\textwidth}
			\centering
			\includegraphics[width=1\linewidth]{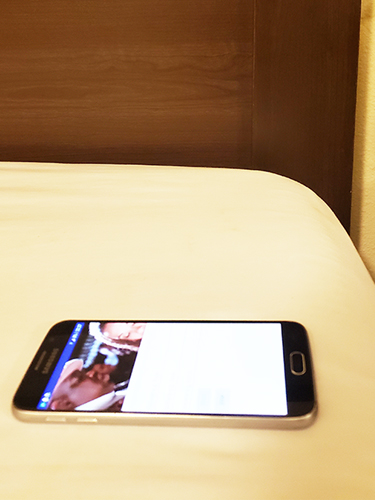}
			\caption{Bed.}
			\label{fig:surface_bed}
		\end{subfigure}
		~
		\begin{subfigure}[t]{0.3\textwidth}
			\centering
			\includegraphics[width=1\linewidth]{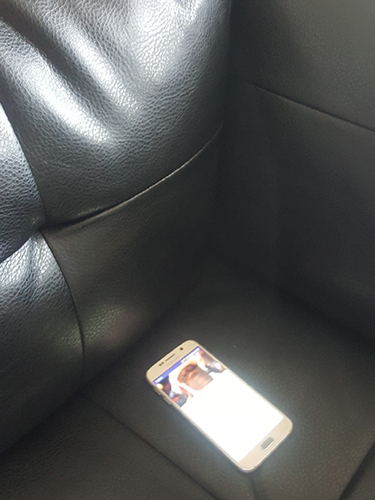}
			\caption{Leather couch.}
			\label{fig:surface_chair}
		\end{subfigure}
	\end{subfigure}%
	\caption{Surfaces on which the phone was placed during our experiments.}
	\label{fig:expt_surfaces}
\end{figure*}

\subsection{Fully-Fledged Data Collection Experiment}
\subsubsection{Data Collection App Implementation}
The basic functionality required for our smartphone data collection app was the ability to: (1) load and play a song, (2) log the song's unique identifier (e.g., song name) in a database, (3) provide cues on the song's start and end points, and, (4) capture motion sensor data generated by the phone and use the cues specified in (3) above to delimit the segment of motion sensor data that precisely maps to the end-points of the song in question.  

To meet these requirements, we leveraged the YouTube and Android Sensor APIs \cite{YouTubeAndroidPlayerAPI, MotionSensorsAPI} and built an app that consists of a YouTube player that addressed requirements (1) through (3) above, and a background sensor service that collects motion sensor data in the background. The application directly plays the songs from the internet. The songs are played one after another until the end of the playlist. Each song is played in its entirety before the next song in the playlist is loaded and played. When the playback of a particular song in the playlist starts, the application records the name of the song and starts writing the sensor data into a local file on the phone. When the song ends, the collection of sensor data is paused, before resuming once the next song starts.

\subsubsection{Sensor Data Recorded by our App} 
\label{subsubsec:sensordata}
Our background sensor service collected accelerometer data (i.e., phone accelerations with and without gravity) and the rotation rates from the gyroscope. For each of these, we collected the $x$, $y$ and $z$ streams along with the corresponding time-stamps. During preliminary experiments, we found that the gyroscope stream had very low accuracy on the song identification task and thus discontinued it from our analysis. We found the accelerations with and without gravity to both perform well, however, we found that their combination did not add much to the performance. For all analysis performed in this work therefore, we thus focused on the linear acceleration only (i.e., acceleration without gravity). Because the most commonly used mobile web browsers (e.g., chrome, safari, opera) use a sensor  sampling rate in the range 100Hz - 120Hz \cite{das2018every}, we instrumented our app to capture sensor data at a rate of 100Hz. This way, our analysis captures the attack in the extreme case of an imprecise sensor that, in theory,  represents some form of lower bound of the attack performance on the average user's device. 

\begin{figure*}[h!]
	\centering
	\includegraphics[width=.5\linewidth]{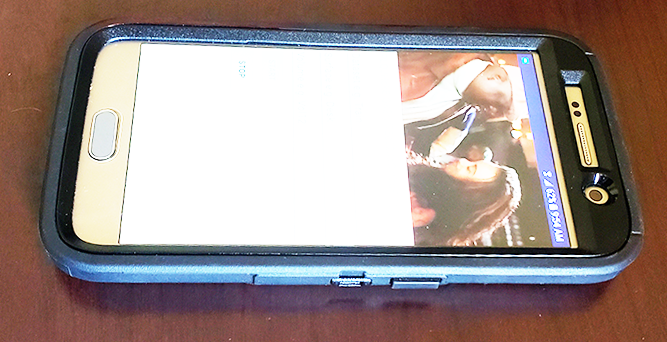}
	\caption{Phone housed in a cover during our attack defense experiments.}
	\label{fig:attack_defense_setup}
\end{figure*}

\subsubsection{Data Collection Process}
\label{subsubsec:data_collection_setup}
Our data collection process basically involved a phone placed on a given surface playing a wide range of songs while sensor data is collected in the background. The two key design decisions in this process were the choice of music to be played in the experiment, and the choice of the surface on which to place the phone as it played the music. Our primary consideration for the selection of music to be played was that it should be highly popular music in order for our study to showcase the attack from the perspective of music that people actually listen to (or frequently listened to during a certain time period).

To meet this requirement, we used the Billboard Hot 100 chart \cite{Billboard} for our study. The Billboard top 100 ranks songs according to the airplay they get on radio, social media and streaming statistics and the number of album sales \cite{BillboardFAQ}. It thus solidly supports our music popularity requirement. Another advantage offered by this chart is that it has a fair mix of some of the most popular music genres \cite{BillboardGenres} (i.e., country, pop, rock, rap, etc.), enabling us to minimize genre bias by studying our attack based on a music selection that exhibits a reasonable amount of diversity in acoustic and structural attributes. 

The specific Billboard chart that we used was that of the week of 18th June, 2018 \cite{Billboard18thJune2018} (see full list of 100 songs in the Appendices, Table \ref{tab:song_list}). There was no particular reason for selecting that week's music chart, other than its being the current chart at the time when we initiated our data collection experiments. Because this top 100 list has song movements happening on a daily basis (both in ranking and on/off the list), we cloned the playlist to a private Youtube account to ensure that we have the same songs in all our experiments.

For the surface on which to place the phone during our experiments, we selected three surfaces on which people often place their phones in home and (or) office settings. These surfaces are the top of a wooden table, on a bed covered by a comforter, and on a leather couch (see Figure \ref{fig:expt_surfaces} for images of the 3 surfaces). The table-top provides an instance of a hard surface that should, in theory, cause the phone to vibrate more when music plays, while the other two surfaces provide two variants of soft surfaces that might provide different forms of dampened vibrations (i.e., the really soft bedding material on one hand, and the somewhat harder leather material on the other). While peoples' homes in practice have a much wider range of surfaces on which phones might be placed to play music, we conjecture that these three surfaces provide a fairly representative mix whose attack performance should provide some measure of the general impact of the attack that is not so far from what other surfaces might produce (e.g., plastic tables, floor tiles, etc.). 

The data collection was done using two Samsung Galaxy S6 phones which were run concurrently. At any given point in time each phone was placed on a separate surface that was either in a  different room or at the furthest extreme end of the same room so as to minimize cross-vibrations. The 100 songs were played at five different volume levels, namely, Volumes, 15, 13, 11, 9, and 7. Volume 15 is the highest volume supported on our Samsung Galaxy phones, while Volume 7 is the volume 8 steps below the maximum volume (i.e., the volume level after 8 presses of the volume reduction switch on the side of the phone). We left out volumes lower than 7 from the full study since our preliminary experiments showed them to provide very low predictive accuracy (i.e., they caused very weak vibrations). For each surface, each volume level and each phone, we collected two sessions of data on two separate days to capture any variability that might occur while the songs are playing (e.g., variability due to random music-independent vibrations of the surface itself). For the defence experiment we dressed the phone in an Otterbox cover \cite{OtterBox} (see Figure \ref{fig:attack_defense_setup}) and run all the experiments again. In total we conducted 60 different instances of the attack experiment ($=$ 2 sessions $\times$ 2 phones $\times$ 3 surfaces $\times$ 5 volume levels) and 60 different instances of the defence experiment\footnote{A complete instance of the experiment is when the full playlist of 100 songs is played from start to end}. The complete data collection experiment lasted about 780 hours (i.e., slightly over a month) as each instance of the experiment lasted about 6 hours and 30 minutes.

\begin{figure*}[h]
	\centering
	\begin{subfigure}[t]{0.98\textwidth}
		\centering
		\includegraphics[width=.95\linewidth]{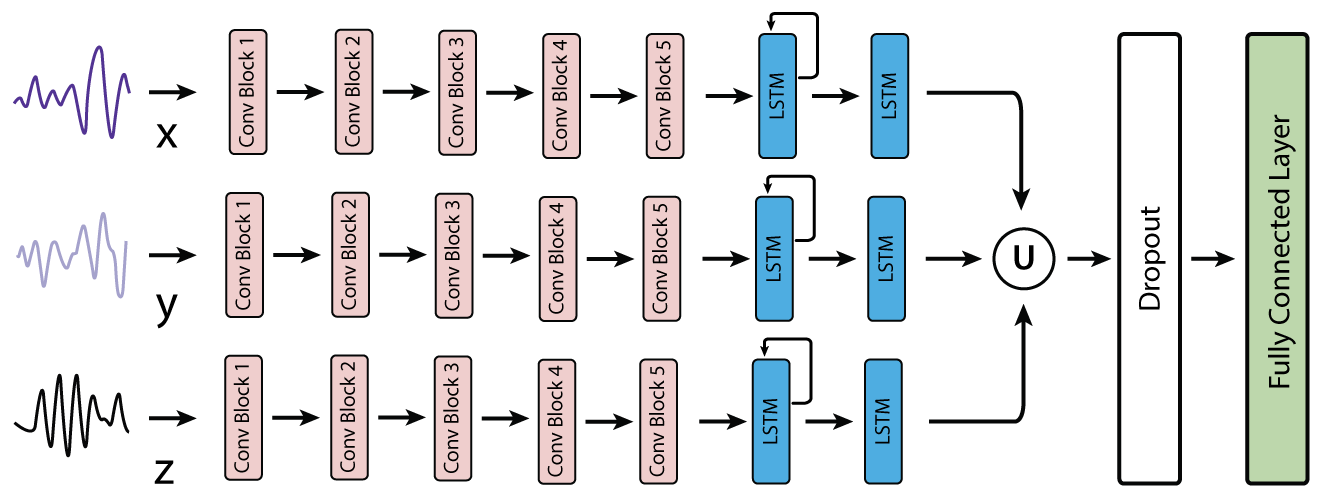}
		\caption{Architecture of our DNN. \textit{Conv Block} represents a convolution block while $U$ represents concatenation. }
		\label{fig:dnn_architecture}
	\end{subfigure} 
	
	\begin{subfigure}[t]{0.75\textwidth}
		\centering
		\includegraphics[width=.40\linewidth]{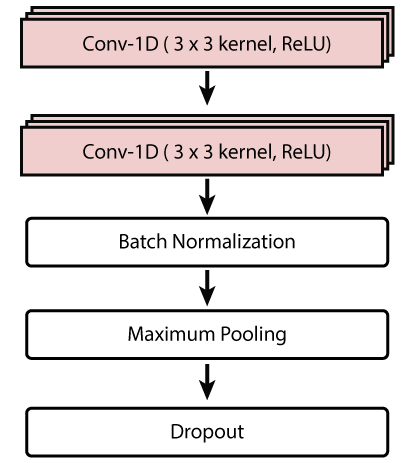}
		\caption{Detailed view of the convolutional block (Conv Block) used in Figure \ref{fig:dnn_architecture}.}
		\label{fig:conv_block}
	\end{subfigure}
	\caption{Overall architecture of the deep neural network (DNN) used in our feature-learning framework. Our DNN consisted of three different CNN-LSTM stacks (with five \textit{Conv Blocks} and two LSTM layers) that were concatenated, then followed by dropout, full connected layer, batch normalization layer and finally a softmax layer for outputting the predicted song label.
	}
	\label{fig:overall_dnn_architecture}
\end{figure*}

\section{Machine Learning Frameworks Driving the Attack}
\subsection{Overview}
In this section (subsections \ref{ml} through \ref{dm}) we describe the machine learning mechanisms used to build the attack, and the structure of our performance evaluation design for both the attack and its defence. As previously highlighted in Section \ref{intro1},  an adversary launching our attack can only have a finite number of songs for training, meaning that every now and then, a test song to be identified will not have a matching sample in the training set. To address this situation, the machine learning framework driving the attack has to have a provision to identify a song whenever possible and to mark a song as unknown if a classification decision is not possible. One could achieve this dual functionality in multiple ways. For example, one could build a single consolidated model in which the unknown class is one of the target classes. Alternatively, one could build two separate systems, one of which determines whether a song is known or unknown (we refer to this as the {\it{novelty detection}} problem), and the other system matching known songs to identities in a training database (we refer to this as the {\it{music identification}} problem).  

In this research we used the latter approach due to its ease of implementation and have thus structured this section in such a way to present a description of how we designed the music identification system, and then a description of how we designed the novelty detection system. Finally, we describe the machine learning elements of the phone cover-based defense mechanism. The design descriptions of these three components generally cover the machine learning algorithms used and their associated configurations, the data pre-processing steps, and the various choices made during the training and testing processes. Having described the design of these three components in subsections \ref{ml} through \ref{dm}, we will later present their performance in Sections \ref{sec:attack_expt_results}, \ref{novelty-12} and \ref{cover-def}. 

\subsection{Implementation and Evaluation of Music Identification Attack}
\label{ml}
To run the music identification attack, we used two different learning paradigms: a feature learning based scheme and feature-engineering based scheme. The former scheme is particularly advantageous in terms of model explainability (i.e., the features map to certain traits that the engineer can understand and map to physical phenomena). The latter scheme on the other hand can potentially learn features that despite not being meaningful to the human, can be very powerful. Because attackers might, depending on expertise and preference, implement either of the two schemes, it is instructive to explore how they would perform for this attack. The next 2 subsections provide our implementation details for these schemes. 

\subsubsection{Feature Learning-Based Framework}
\label{learn1}
Here, we describe our deep neural network architecture, as well as the data pre-processing and augmentation steps performed on the data before feeding it to the network. We finally present details of the training and testing processes.

{\bfseries{Data Pre-processing:}} Recall from Section \ref{subsubsec:sensordata} that our attack design focused on the linear acceleration measurements (i.e., accelerometer readings without gravity) since these performed best during our preliminary analysis. Given the $x$, $y$ and $z$ streams of this linear accelerometer data at each time stamp, $t$, our primary pre-processing step was to normalize the data from each axis to the range [-1, 1] for each song. We also converted all of the data to float16 \cite{float16} to reduce the space required in memory and increase the number of instances that we could fit in memory at the same time during the training process.

To prepare our data for training the deep neural network (DNN) we randomly select windows of data (each window comprised of the $x$, $y$ and $z$ streams) to feed into the input layer of the DNN. During this process we randomly select an equal number of windows for each song instance, with 2048 samples per window. This is done to avoid causing an imbalance in the number of windows per song instance. After extracting a window of data we then applied data augmentation, as described further below.

\begin{table*}[h]
	\centering
	\begin{tabular}{ccccccccc}
		\toprule
		\multicolumn{1}{c}{\multirow{2}{*}{\textbf{\begin{tabular}[t]{@{}c@{}}Training \\ configuration\end{tabular}}}} & \multicolumn{3}{c}{\textbf{\begin{tabular}[t]{@{}c@{}}Surface on which the phone \\ is placed during training\end{tabular}}} & \multicolumn{5}{c}{\textbf{Volume levels used to build training set}} \\
		\multicolumn{1}{c}{}                                                                                       & Table                                    & Bed                                    & Couch                                    & Vol 07     & Vol 09     & Vol 11     & Vol 13     & Vol 15    \\ \midrule
		I                                                                                                          &  \checkmark                                        &     \checkmark                                   &       \checkmark                                   &    \checkmark        &     \checkmark       &     \checkmark       &      \checkmark      &     \checkmark       \\
		II                                                                                                         &     \checkmark                                     &                                        &                                          &     \checkmark       &     \checkmark       &    \checkmark        &     \checkmark       &     \checkmark       \\
		III                                                                                                        &                                          &                \checkmark                        &                                         &     \checkmark       &     \checkmark       &      \checkmark      &    \checkmark        &     \checkmark       \\
		IV                                                                                                         &                                          &                                        &         \checkmark                                 &      \checkmark      &     \checkmark       &     \checkmark       &    \checkmark        &     \checkmark      \\
		\bottomrule
	\end{tabular}
	\caption{Training configurations}
	\label{tab:training_configurations}
\end{table*}

\begin{table*}[h]
	\centering
	\begin{tabular}{ccccccccc}
		\toprule
		\multirow{2}{*}{\textbf{\begin{tabular}[c]{@{}c@{}}Testing\\ Configuration\end{tabular}}} & \multicolumn{3}{c}{\textbf{\begin{tabular}[c]{@{}c@{}}Surface on which the phone\\ is placed during testing\end{tabular}}} & \multicolumn{5}{c}{\textbf{\begin{tabular}[c]{@{}c@{}}Volume level used \\ during testing\end{tabular}}}                                       \\
		& \multicolumn{1}{c}{Table}                & \multicolumn{1}{c}{Bed}               & \multicolumn{1}{c}{Couch}               & \multicolumn{1}{c}{Vol 07} & \multicolumn{1}{c}{Vol 09} & \multicolumn{1}{c}{Vol 11} & \multicolumn{1}{c}{Vol 13} & \multicolumn{1}{c}{Vol 15} \\ \toprule
		I                                                                        & \checkmark                     &                   &                     &        &        &        &        &   \checkmark    \\ 
		II      &                        \checkmark                  &                                       &                                         &                            &                            &                            &               \checkmark             &                            \\
		III     &                      \checkmark                    &                                       &                                         &                            &                            &          \checkmark                  &                            &                            \\
		IV      & \checkmark                     &                   &                     &        &    \checkmark    &       &        &        \\
		V   &               \checkmark                           &                                       &                                         &        \checkmark                    &                            &                            &                            &                            \\ \midrule
		VI                                                                       &                                          &         \checkmark                              &                                         &                            &                            &                            &                            &            \checkmark                \\
		VII     &                                          &          \checkmark                             &                                         &                            &                            &                            &             \checkmark               &                            \\
		VIII    &                                          &             \checkmark                          &                                         &                            &                            &           \checkmark                 &                            &                            \\
		IX      &                                          &            \checkmark                           &                                         &                            &               \checkmark             &                            &                            &                            \\
		X   &                                          &           \checkmark                            &                                         &                      \checkmark      &                            &                            &                            &                            \\ \midrule
		XI                                                                      &                                          &                                       &                    \checkmark                     &                            &                            &                            &                            &             \checkmark               \\
		XII     &                                          &                                       &                       \checkmark                  &                            &                            &                            &          \checkmark                  &                            \\
		XIII    &                                          &                                       &                      \checkmark                   &                            &                            &                  \checkmark          &                            &                            \\
		XIV     &                                          &                                       &                           \checkmark              &                            &           \checkmark                 &                            &                            &                            \\
		XV      &                                          &                                       &                 \checkmark                        &                    \checkmark        &                            &                            &                            &                           \\ \bottomrule
	\end{tabular}
	\caption{Testing configurations}
	\label{tab:testing_configurations}
\end{table*}

{\bfseries{Data Augmentation:}} For deep neural networks to learn weights that generalize well, they often require large amounts of data. It's also important that the data distribution used for training the neural network encompasses factors such as variability in the environment or user behavior that is likely to be encountered in the wild. In our case the broader distribution of data might include scenarios in which the volume is suppressed, the phone is not perfectly flat against the surface on which it rests, or the vibrations of the device are dulled by some external factor, to mention but a few. Applying data augmentation to the training data can introduce variability in the dataset that captures some of the above factors and enables the neural network to generalize more effectively. Although the dataset we collected is large and diverse in its range of environmental conditions, recent research (e.g., see \cite{um2017data, urban2018deep, salamon2017deep}) has shown that data augmentation can reduce overfitting and enable the use of deeper neural networks that can learn features that better differentiate between classes.

For data augmentation in this work we used rotation and scaling \cite{um2017data}. Augmentation was applied to 75\% of the windows extracted from the scaled data, with 25\% for each of rotation and scaling, and 25\% using both. Signals from each axis were rotated by a random amount between $-\pi$ and $\pi$. Scaling was applied by multiplying the sequence of values in a signal by a random number in the interval [0.9, 1.1]. It should be assumed throughout the rest of this paper that the deep neural network (DNN) uses data augmentation during the training process (but not for validation or test data).

{\bfseries{Model Architecture:}} For the deep neural network architecture we used convolutional layers (CNN) to extract spatially dependent features before extracting time variant information from the data using long short-term memory (LSTM) layers. Our architecture is as follows. First, we created a Convolutional Block (see Figure \ref{fig:conv_block}) by combining two CNN layers (with relu activations), batch normalization, 1-dimensional pooling, and then dropout. We then stacked five Convolutional Blocks for each sensor axis. The convolutional layers used increasing kernel sizes [32, 64, 128, 256, 512], a kernel size of 3, a pooling size of 2, and 20\% dropout. After the last convolutional block we stacked two LSTM layers, the first of which returned the full sequence. A final 50\% dropout layer was applied after concatenating the activations for each axis of the accelerometer and before a fully connected layer (using softmax) with 100 neurons.

\begin{table*}[h]
	\begin{center}
		\begin{tabular}[t]{l>{\raggedright\arraybackslash}p{0.62\linewidth}}
			\toprule
			\textbf{Category} & \textbf{Feature} \\
			\toprule\begin{tabular}[c]{@{}l@{}} Statistical metrics computed on MFCC Coefficients \end{tabular} & \begin{tabular}[c]{@{}l@{}} Minimum, Maximum, Mean, Median, Standard Deviation, Skewness, Kurtosis \end{tabular} \\ \midrule
			Spectral Features & \begin{tabular}[c]{@{}l@{}} Spectral Centroid, Spectral RMS, Spectral Skewness, Spectral Kurtosis, Spectral Entropy, Spectral \\ Spread, Spectral Crest, Spectral Energy, Spectral Rolloff, Spectral Flatness \\   \end{tabular} \\
			\bottomrule
		\end{tabular}
	\end{center}
	\caption{A list of features used in the feature engineering-based framework of our attack.}
	\label{tab:feature-set}
\end{table*}

{\bfseries{Training and Testing Details:}} 
We implemented 4 different training configurations in order to capture different flavors of the attack. Table \ref{tab:training_configurations} summarizes these configurations, with the difference between them being the choice of data used for training. Configuration \#1 assumes an attacker who combines data from 3 different surfaces with the hope to build a single generalized model that embodies patterns exhibited by the different surfaces. Configurations \#2 through \#4 on the other hand assume an attacker who uses data collected from a phone placed on a single surface. Such a choice might for instance be made if the attacker aims to train a precise model that targets a certain type of victim who is known to use the surface in question. For all configurations \#1 through \#4, we use data collected while the phone plays music at the five highest volumes. We assume the attacker will in practice also use data from the highest phone volumes for training since the lower volumes produce much weaker vibrations that might not be captured accurately enough by the phone's sensors. 

For each of these four training configurations, we use 15 different testing configurations (see Table \ref{tab:testing_configurations}), each of which might map to a victim who is listening to music on their phone. For example, Testing Configuration \#1 assumes a victim who places the phone on the table and plays music at Volume 7 (which is the 7th phone volume setting after the zero volume point). Again, we ignore the scenarios having very low phone volumes as the vibrations caused at such volumes are largely noise.  In all cases the data collected on the first three days was used for training while the data collected on the fourth day was used for testing.  

For each of the training configurations, we trained the DNN using our augmented training data over 200 epochs and used the validation data (not augmented) to evaluate the quality of the model as it is trained. The final version of the model was selected based on the best validation accuracy and we used early stopping (with a patience of 25) to preemptively stop the training process if the model stopped improving. 

\subsubsection{Feature Engineering-Based Framework}
\label{feat-eng-1}
This framework follows the traditional machine learning process, namely: (1) sensor data pre-processing, (2) feature extraction, (3) classifier training and testing. Below, we describe the details underlying each of these steps.

{\bfseries{Data Pre-processing:}} 
In addition to the $x$, $y$ and $z$ streams used for the feature learning configuration (see Section \ref{learn1}), we computed the magnitude, $m=\sqrt{x^2+y^2+z^2}$, at each timestamp, $t$. We then smoothed the emergent 4-dimensional data stream  using a high-pass Butterworth filter with a 1.5Hz cut-off frequency and order 3. For each of the 4 data streams and each song, we then normalized the smoothed data to a range of [-1 1]. Finally, we split the data into non-overlapping windows of 2048 samples each, that were then used for feature extraction. 

{\bfseries{Feature Extraction and Analysis:}}
For each of the 4 data streams, we extracted 59 features from each data window (which gave us a total of 236 features for the 4 streams combined). A significant proportion of these features were Mel Frequency Cepstral Coefficients (MFCC) features \cite{foote1997content}, which we used due to their well known ability to capture patterns embedded in audio signals (e.g., see \cite{rabiner1993fundamentals}, \cite{tzanetakis2002musical}). Our full feature-set was computed as follows. For each data window, the first 7 MFCC features were computed for each of 7 sub-windows. From a vector comprised of the first MFCC coefficient drawn from each sub-window, 7 statistical metrics were computed (see Table \ref{tab:feature-set} for these metrics). This was repeated for the $2^{nd}$ through the $7^{th}$ MFCC coefficient, creating a total of 49 (=7 $\times$ 7) features. From the same window, the power spectrum was computed and 10 spectral features (also shown in Table \ref{tab:feature-set}) were generated from it. This created a total of 59 features for each of the $x$, $y$, $z$ and $m$ dimensions, and hence the earlier mentioned total of 236 features form the data window.

To reduce the number of weakly discriminative and highly correlated features among our feature-set, we used \textit{ReliefF} \cite{kononenko1997overcoming} to perform a feature analysis and selection step on the 236 features. This feature analysis and selection was done based on a subset of the data reserved for training. We attained the best performance when the top 150 ranked features were selected (see Appendices, Figure \ref{fig:top_150_features} for the list of 150 selected features). 

The majority of these top ranked features came from $z$ component followed by the $m$ component. The outstanding performance of the  $z$ component is particularly unsurprising because music playing on a phones resting on a given surface would likely cause the phone to mainly vibrate on and off the surface (i.e., along the $z$ direction). The $m$ component also likely performs well because it channels some of the information contained in the $z$ component. In terms of feature category, about 80\% of these top ranked features were from the MFCC category, with the statistical features of third and fifth MFCC co-efficient generally ranking the highest. The non-MFCC features which performed well include, the spectral centroid of the $z$ and $m$ components, as well as the spread, roll-off and root mean square of the power spectrum. 

{\bfseries{Classifier Details and Configurations:}} 
For classification, we explored several classification algorithms in the Python's Scikit-Learn framework. We attained the best prediction accuracies using k-Nearest Neighbors (kNN) \cite{cover1967nearest} and Extra Trees (ET) \cite{geurts2006extremely} classifiers. The kNN classifier was configured with $n\_neighbors = 15$, $weights = "distance"$, $p = 1$ and other default parameters while the ET classifier was configured with $n\_estimators = 1000$ and the other default parameters. To further improve the fingerprinting accuracy, we then combined these two best performing classifiers by fusing the classifier scores using the weighted sum-rule \cite{ross2003information}. The ET classifier was assigned a weight of 0.6 while the kNN classifier was assigned a weight of 0.4. We derived these classifier weights through a grid search of different weight combinations. The combined classifier ($ET + kNN$) was then used to make inference prediction. Combining these two best performing classifiers improved our prediction accuracy by about 5\% to 7\%. In the results section (i.e., Section \ref{sec:combined_results}), we only report results from this combined classifier ($ET + kNN$) for the attack performance based on the feature engineering-based framework.

{\bfseries{Training and Testing Details:}} The data configurations used for training and testing are exactly the same as those previously described in Section \ref{learn1} and represented in Tables \ref{tab:training_configurations} and \ref{tab:testing_configurations}. Again, all vibration data collected on the first 3 days is used for training, while data collected on the fourth day is used for testing.

\subsection{Implementation and Evaluation of Novelty Detection Attack}
\label{nd}
Given sensor data captured from the victim's phone, the novelty detection framework (or attack) renders a decision of whether the sensor data is caused by a song which is part of the corpus of songs used in training, or by an unknown song (a song not represented in the training set). For this attack we only present a feature-engineering-based approach (i.e., traditional machine learning engine) as DNN-based approaches always performed much worse than it during our preliminary experiments. Below, we describe elements of this novelty detection framework. 

{\bfseries{Data Pre-processing and features used:}} The pre-processing of raw sensor data was done in exactly the same way as was previously described under the feature-engineering-based song identification framework (recall Section \ref{feat-eng-1}). Following pre-processing, the same set of features described in Section \ref{feat-eng-1} were extracted and reused for the novelty detection attack. We hence do not describe these two components again. 

{\bfseries{Classifier Details and Configurations:}} We cast the novelty detection problem as a one-class classification problem. In particular, we labeled a given song as a single class, and then built a classifier that is trained to learn a model that represents this class. A song seen during testing would then be classified as belonging to this single class or being unknown.  We explored several one-class classification algorithms in the Scikit learn firework and attained the best results using the one-class SVM \cite{scholkopf2001estimating}. The one-class SVM classifier was configured with $nu = 0.05$, $kernel="rbf"$, $gamma="scale"$ and other default parameters. Details of the training and testing process follow:

{\bfseries{Training and Testing Details:}} 
For each song $i$, four different instances of a one-class SVM classifier $C_i$ were trained based on data collected for this song. The difference between these instances was the surfaces and volume levels used for training (see Table \ref{tab:training_configurations} for all four training configurations). During testing, data from the song in question was drawn from the test dataset and presented to the classifier as the {\textit{known class}}, while randomly selected data from the remaining 99 songs in the testing dataset was presented to the classifier as the \textit{unknown class}. The number of testing samples from both the \textit{known class} and \textit{unknown class} varied from 14 to 30 depending on the duration of the song that was used as the \textit{known class}.

\subsection{Implementation and Evaluation of Defense Mechanism}
\label{dm}
\label{novelty_classification}
The final piece of our analysis is on the defense mechanism against the attack. This piece has two sub-components, namely: (1) defense impact --- here we study the effect that vibration damping due to a phone cover has on the attack, and, (2) Detection of defense deployment --- this sub-component is meant to evaluate whether the attacker can detect the usage of a phone cover by the victim. We provide more details on each of these sub-components in the ensuing subsections. 

\subsubsection{Defense impact}
This sub-component is studied under two different sets of assumptions. In one set of assumptions, we assume that the attacker is aware that the intended victim is implementing the defence (i.e., has the phone housed in a cover). In this case the attacker therefore trains the learning models with sensor data collected from phones which have covers on them. In the second set of assumptions, the attacker is unaware that the victim uses a phone cover and thus trains the learning algorithms using data from phones having no covers. Beyond the variations in the kinds of data used, the system under both sets of assumptions is evaluated in exactly the same way (i.e.,  all the data processing, surfaces and volumes, feature engineering/learning and training and testing configurations mirror those already described in Sections \ref{ml} and \ref{nd} before the defense implementation).

\subsubsection{Detection of defense deployment}
Given a savvy attacker who seeks to maximize the performance of the attack, a natural first step in the attack would be to train learning algorithms to detect whether the intended victim is using the defense mechanism (i.e., whether the victim uses a phone cover or not). The basic idea behind this step is that the phone vibrates in a different way if a phone cover is being used (relative to when it is not), and that this cover-induced vibration pattern might be distinct enough to learned by classifiers from motion sensor data. Depending on whether the victim uses a phone cover or not, the attacker might then accordingly tune the attack's training module to attain maximum success against the user.  To study this sub-component, we again use the data processing configurations reported in Sections  \ref{ml} and \ref{nd} (i.e., the same data processing, surfaces and volumes, feature engineering/learning and training and testing configurations). The only variation from the descriptions given in Sections  \ref{ml} and \ref{nd} is that the target variable (i.e., class label) is the binary variable of whether the  phone has a cover or not.

\section{Experimental Results --- Music Identification Attack}
\label{sec:attack_expt_results}
In this section we discuss the results from the music identification attack experiments. 
We use the term {\it\underline{{baseline}} }to refer to the scenario where data from all surfaces and all volumes is combined into a single training set (recall training Configuration \#1 in Table \ref{tab:training_configurations}, Page \pageref{tab:training_configurations}). Section \ref{sec:combined_results} is dedicated to showcasing results for different testing configurations when training is done using the baseline configuration. In Section \ref{surface-specific-1}, we change the training configuration to a surface-specific configuration (recall training Configurations II through IV in Table \ref{tab:training_configurations}, Page \pageref{tab:training_configurations}) and again showcase results for different testing configurations of the music identification attack.

\begin{figure*}[h]
	\centering
	\begin{subfigure}[t]{0.95\textwidth}
		\begin{subfigure}[t]{.5\textwidth}
			\centering
			\includegraphics[width=1\linewidth]{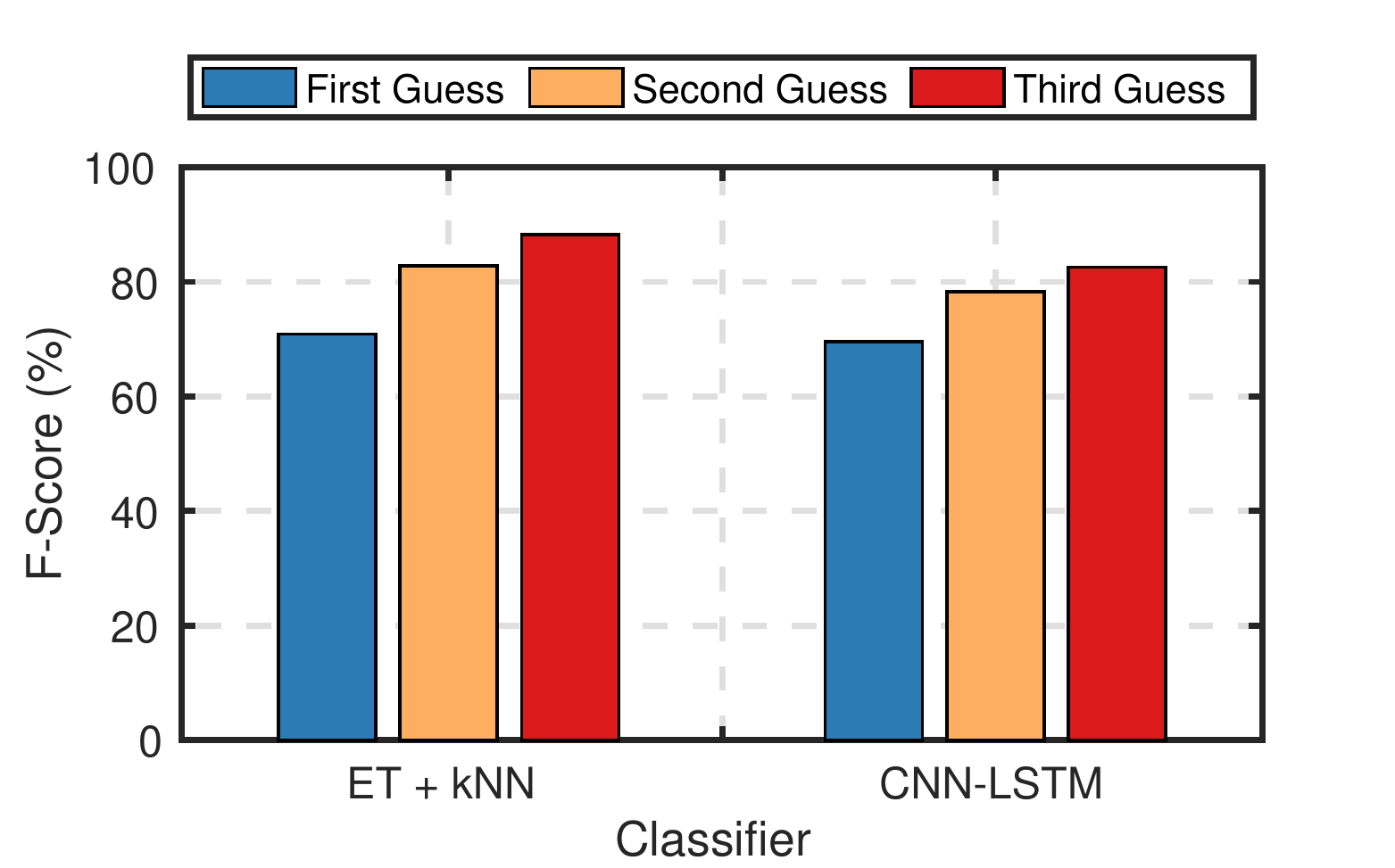}
			\caption{Attack performance attained over three guessing attempts.}
			\label{fig:guess_baseline_attack_performance}
		\end{subfigure}%
		~ 
		\begin{subfigure}[t]{0.5\textwidth}
			\centering
			\includegraphics[width=1\linewidth]{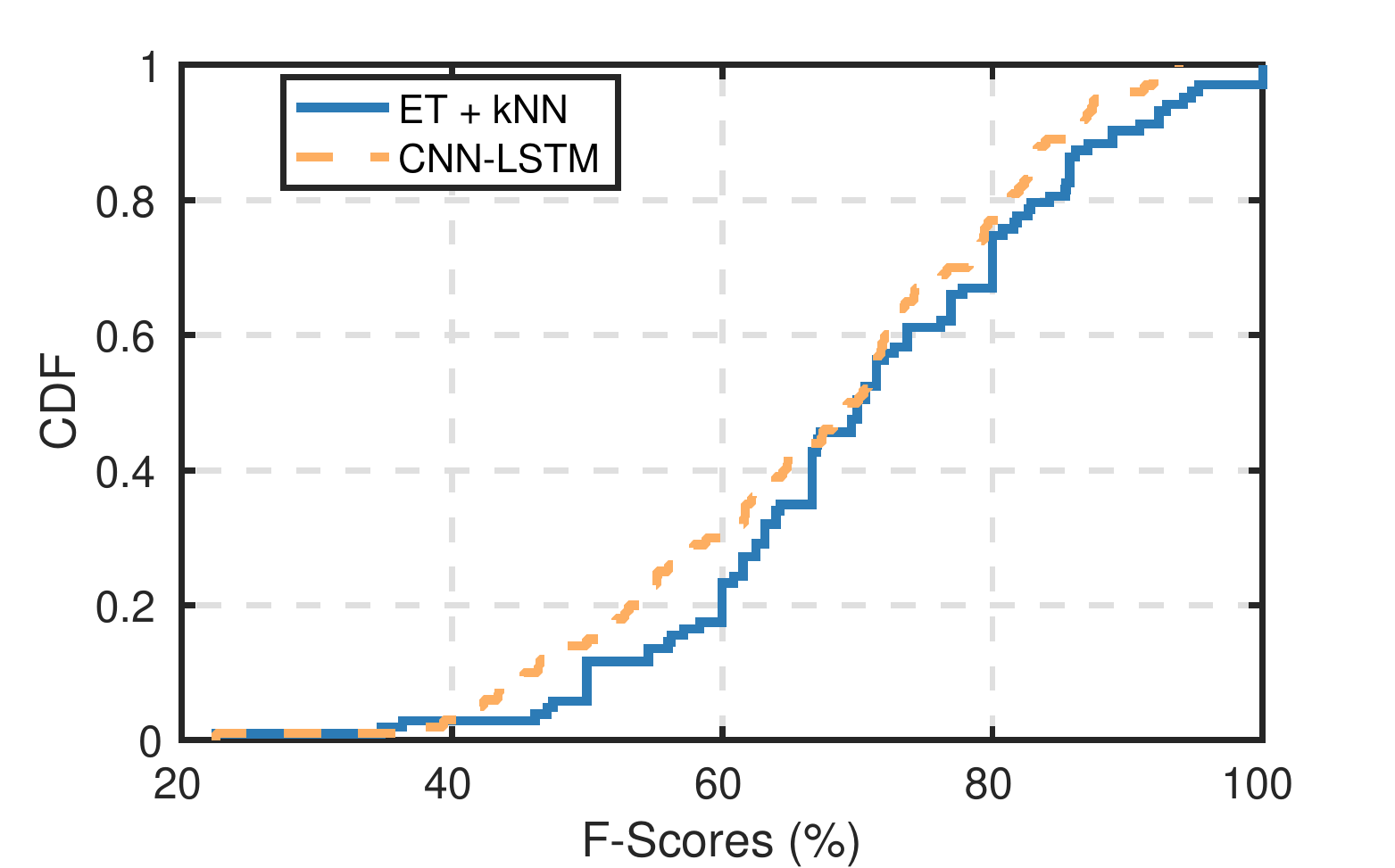}
			\caption{CDF of the song identification accuracies obtained in the first guessing attempt.}
			\label{fig:cdf_baseline_attack_performance}
		\end{subfigure}
	\end{subfigure}
	
	\caption{Performance of baseline training scenario when testing is done using a phone placed on the table surface while playing music at the highest volume.}
	\label{fig:baseline_attack_performance}	
\end{figure*}	

\subsection{Performance of Music Identification Attack Under Baseline Training Scenario (i.e., Training Configuration I, Table \ref{tab:training_configurations})}
Under the baseline training scenario, the following subsections give the performance for each of our testing configurations. 	

\label{sec:combined_results}
\subsubsection{Phone placed on table surface and music playing at maximum volume during testing}
Figure \ref{fig:baseline_attack_performance} shows the baseline performance of the attack when the victim placed the phone on the table and played music at the highest volume (i.e., testing configuration V in Table \ref{tab:testing_configurations}, Page \pageref{tab:testing_configurations}). Figure \ref{fig:guess_baseline_attack_performance} shows the F-scores of the ET+kNN and CNN-LSTM classifiers for guessing attempts \#1, 2 and 3 (i.e., the 3 songs returned by the classifier as being most likely given the vibration pattern). We report our results based on the F-score (as opposed to the classification accuracy) in order to compensate for cases where our test dataset is not perfectly balanced. 

Figure \ref{fig:guess_baseline_attack_performance} shows that the ET+kNN classifier attains about a 70\% F-score if the attacker has to make only one guess on the song being played, and up to 80\% and 90\% respectively when the attacker seeks to have the correct song in a pool of 2 or 3 guesses made. The CNN-LSTM classifier depicts a similar trend, albeit with slightly lower F-scores at each of the 3 guesses. This being a 100-class problem, a random guess attack would achieve only about 1\% F-score. The results in Figure \ref{fig:guess_baseline_attack_performance} hence show that this attack is highly effective when the user plays music at maximum volume with a phone placed on a table (or a hard surface in general).

Figure \ref{fig:cdf_baseline_attack_performance} digs deeper into the consolidated F-scores reported in Figure \ref{fig:guess_baseline_attack_performance}. In particular, the figure shows a CDF of the per-song F-scores obtained across the full corpus of 100 songs for each of the classification approaches. The figure is focused on the first guessing attempt since the other two attempts revealed no new patterns. The figure reveals that about 50\% of the songs had an F-score of over 75\% while only about 5\% of the songs had an F-score of less than 30\%. This indicates that the consolidated F-score number (of about 70\% for the first guessing attempt) reported in Figure \ref{fig:guess_baseline_attack_performance} is actually representative of the individual prediction performances seen with a sizable number of individual songs (i.e., we do not have a small number of songs disproportionately skewing the mean upwards). Differently stated, this pattern indicates that few songs see a very low song identification accuracy. This points to the potency of the attack. The other pattern depicted in Figure \ref{fig:cdf_baseline_attack_performance} is that the CDFs for the CNN-LSTM and ET+kNN classifiers mostly overlap, further explaining why these two classification approaches did not differ much in terms of the global performance reported in Figure \ref{fig:cdf_baseline_attack_performance}.

\begin{figure*}[h]
	\centering
	\begin{subfigure}[t]{1\textwidth}
		\begin{subfigure}[t]{0.5\textwidth}
			\centering
			\includegraphics[width=1\linewidth]{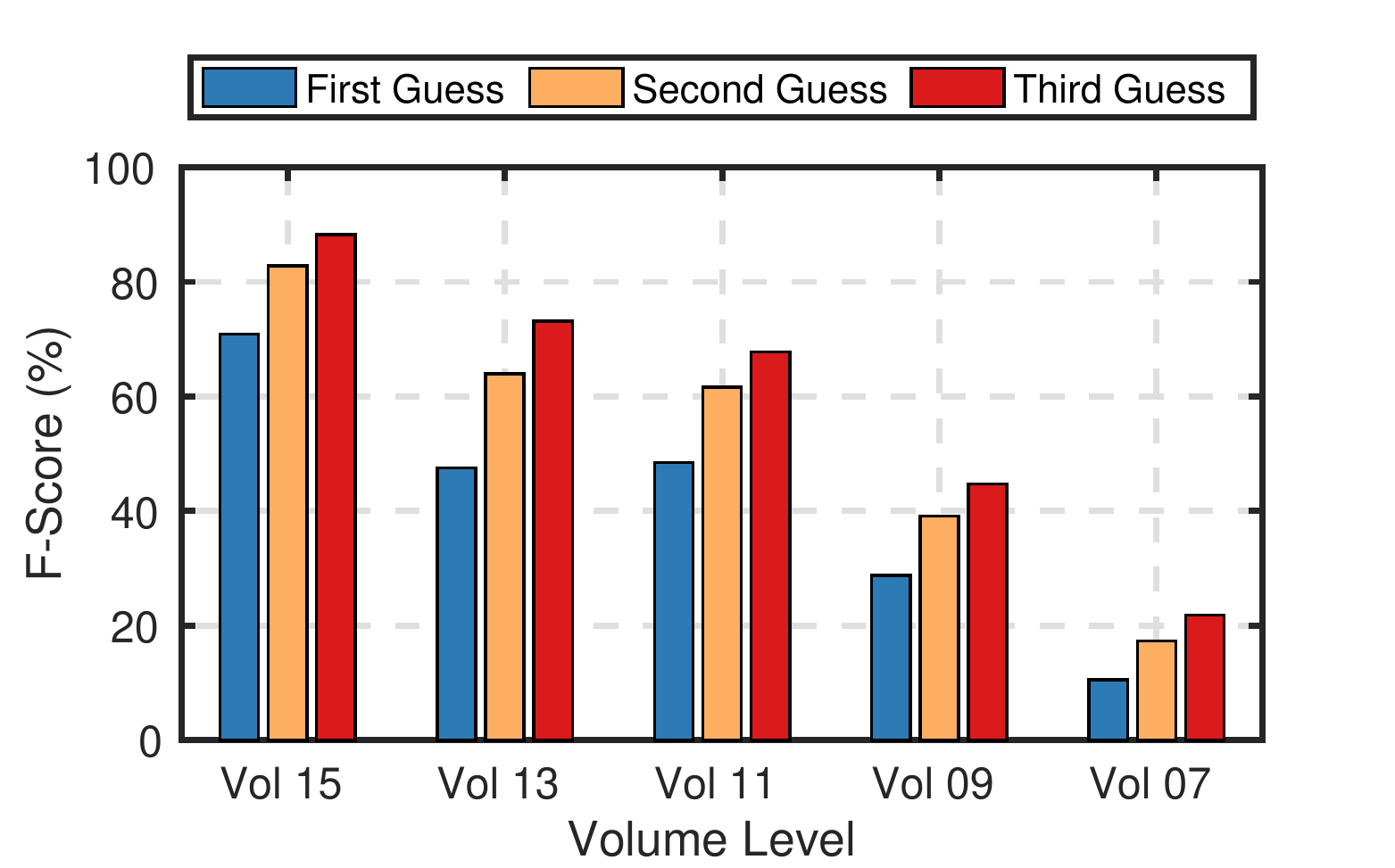}
			\caption{$ET + kNN$ classifier performance at various test volumes}
			\label{fig:volume_et_knn_results}
		\end{subfigure}%
		~ 
		\begin{subfigure}[t]{0.5\textwidth}
			\centering
			\includegraphics[width=1\linewidth]{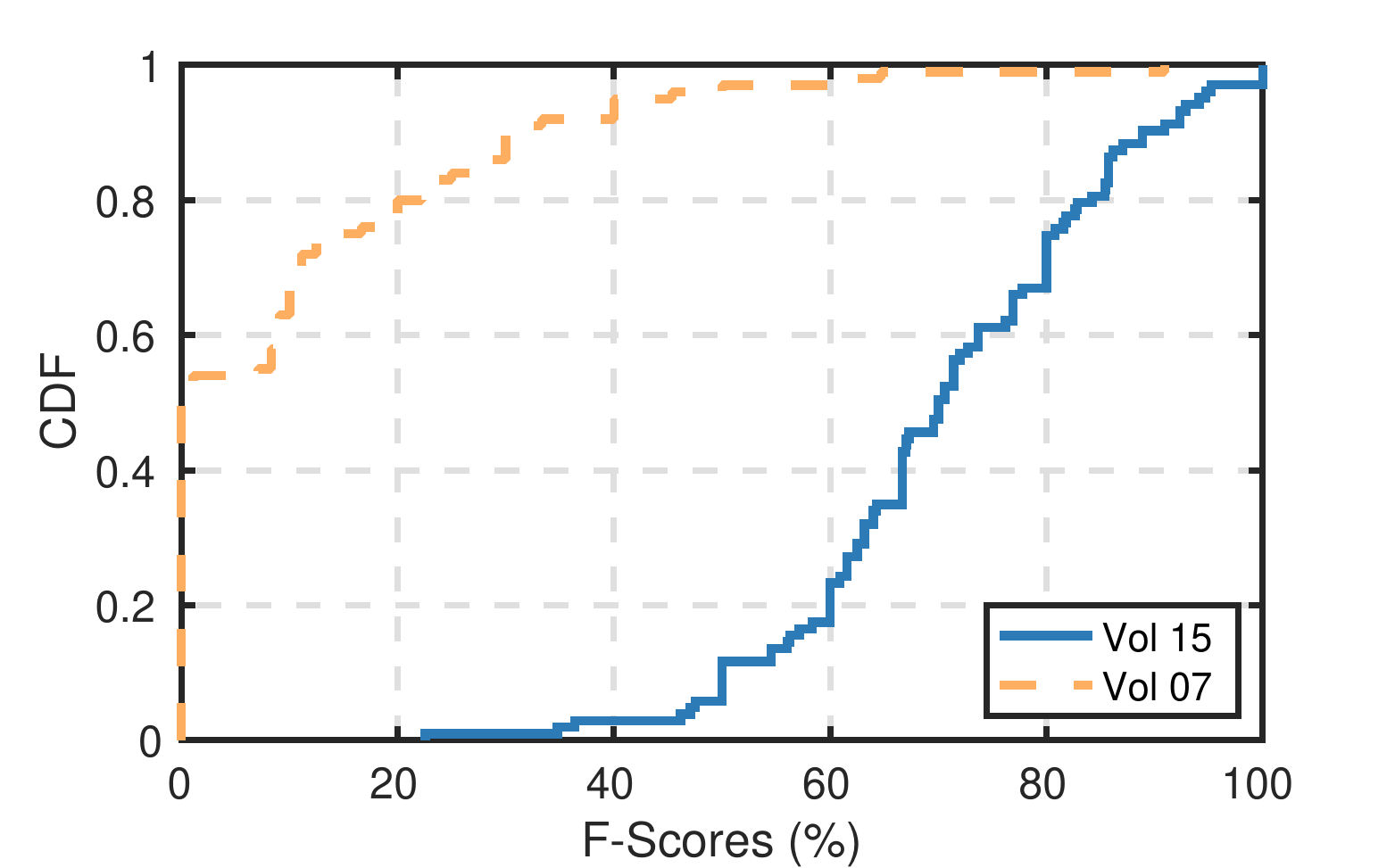}
			\caption{CDFs of F-Scores returned by $ET + kNN$ classsifier for the highest and lowest test volumes}
			\label{fig:cdf_volume_et_knn_results}
		\end{subfigure}
	\end{subfigure}%
	
	\begin{subfigure}[t]{1\textwidth}
		\begin{subfigure}[t]{.5\textwidth}
			\centering
			\includegraphics[width=1\linewidth]{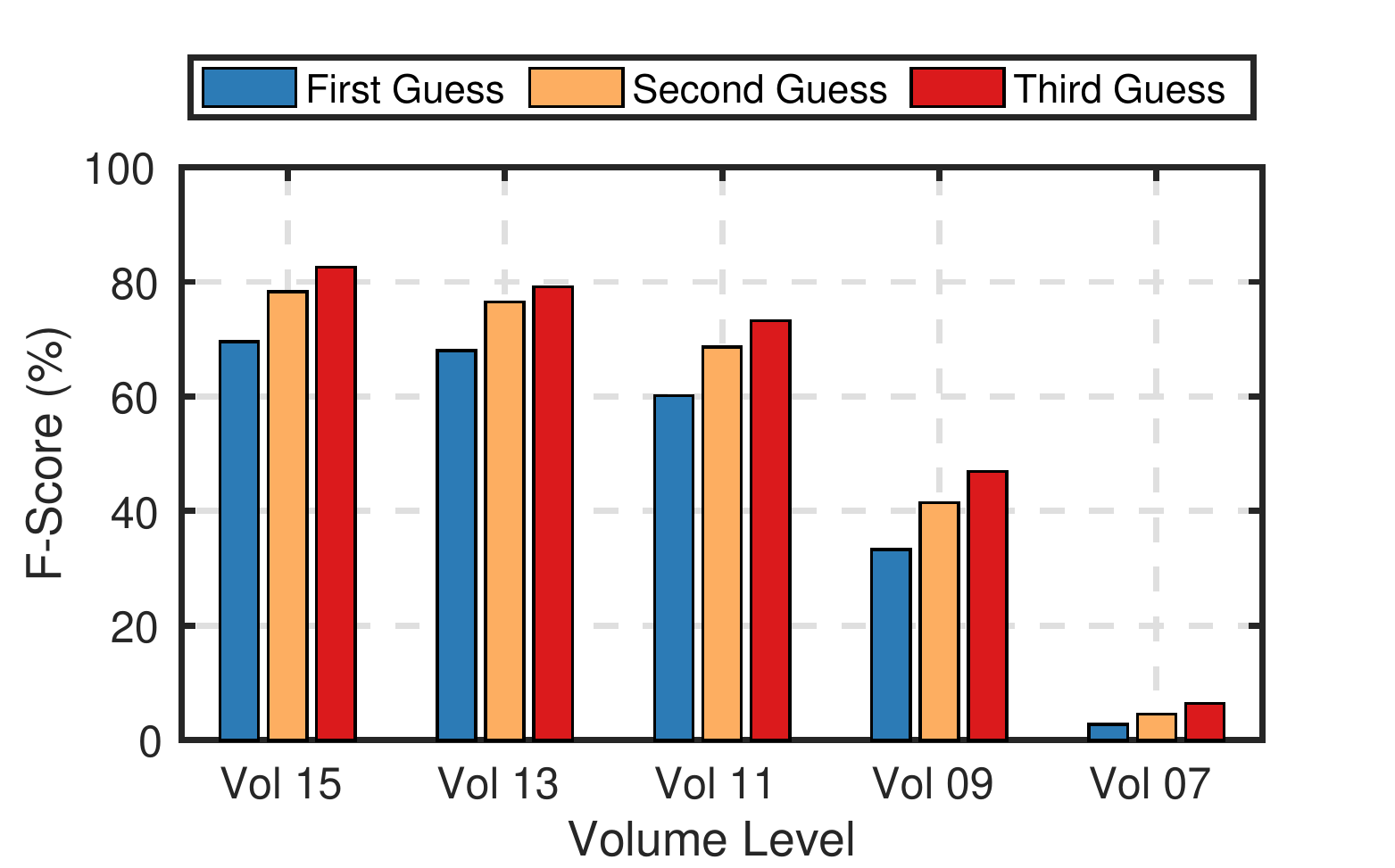}
			\caption{$CNN - LSTM$ classifier performance at various test volumes}
			\label{fig:volume_cnn_lstm_results}
		\end{subfigure}%
		~ 	
		\begin{subfigure}[t]{0.5\textwidth}
			\centering
			\includegraphics[width=1\linewidth]{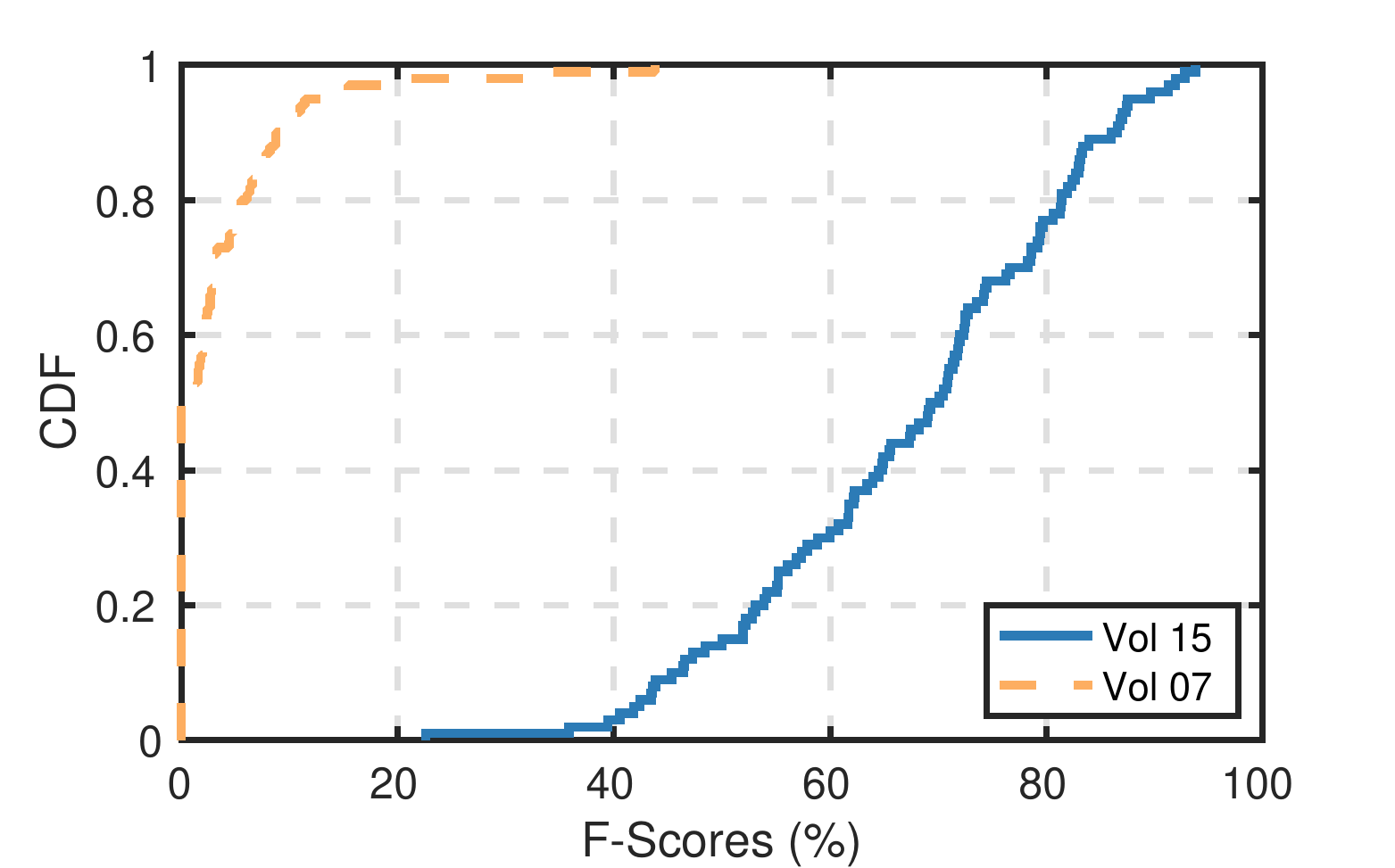}
			\caption{CDFs of F-Scores returned by $CNN - LSTM$ classsifier for the highest and lowest test volumes}
			\label{fig:cdf_volume_cnn_lstm_results}
		\end{subfigure}
	\end{subfigure}%
	
	\caption{Impact of changes in test volumes under the baseline training scenario. Testing is done using data collected when the phone plays music at various volumes while resting on the table.}
	\label{fig:volume_attack_performance}		
\end{figure*}

\subsubsection{Impact of changes in volume during testing (i.e., phone still placed on table surface but  music playing at different volumes)} 
To understand the impact of playback volume on the music identification accuracy, we kept the phone placement surface constant (i.e., table) but then varied the volume of music used to test the baseline classifier. Figure \ref{fig:volume_attack_performance} summarizes the results from this scenario. As done previously, we again show 3 guessing attempts side-by-side. 

Figures \ref{fig:volume_et_knn_results} and \ref{fig:volume_cnn_lstm_results} reveal a systematic reduction in the music identification F-score on average for both classifiers as the volume at which music is played gets reduced. This F-score reduction is unsurprising since reductions in volume imply less powerful phone vibrations which in turn have a reduced likelihood to be picked by the phone's sensors in enough detail to identify the subtleties intrinsic to a song.  It is worthy noting though that at our lowest volume (i.e., Vol 07 --- 8 steps below the highest volume supported on our Samsung phones), the attack still attains F-scores in the range 10-20\%, a whole order of magnitude above the random guessing F-score of about 1\%. This suggests that the attacker would still derive some value from this attack even for a victim who plays music at a relatively low volume. 

To zoom in into the individual song patterns that drive the overall pattern seen in Figures \ref{fig:volume_et_knn_results} and \ref{fig:volume_cnn_lstm_results}, we plotted Figures \ref{fig:cdf_volume_et_knn_results} and \ref{fig:cdf_volume_cnn_lstm_results} to show CDFs of the individual song F-scores for the highest and lowest song volumes. Relative to Volume 15, Volume 7 has a CDF which is skewed towards the top left corner of the plot. This trait tells us that relative to Volume 15, Volume 7 has a higher number of songs with low F-scores. This general trait is again unsurprising. However, what is noteworthy is that at the lowest volume, 40-50\% of the songs still attain an individual F-score that exceeds what a random guess would attain. This further supports our earlier argument about the attack being appealing even for victims who play music at volumes several notches lower than the peak volume supported on a phone. 

\begin{figure*}[h]
	\centering
	\begin{subfigure}[t]{1\textwidth}
		\begin{subfigure}[t]{0.42\textwidth}
			\centering
			\includegraphics[width=1\linewidth]{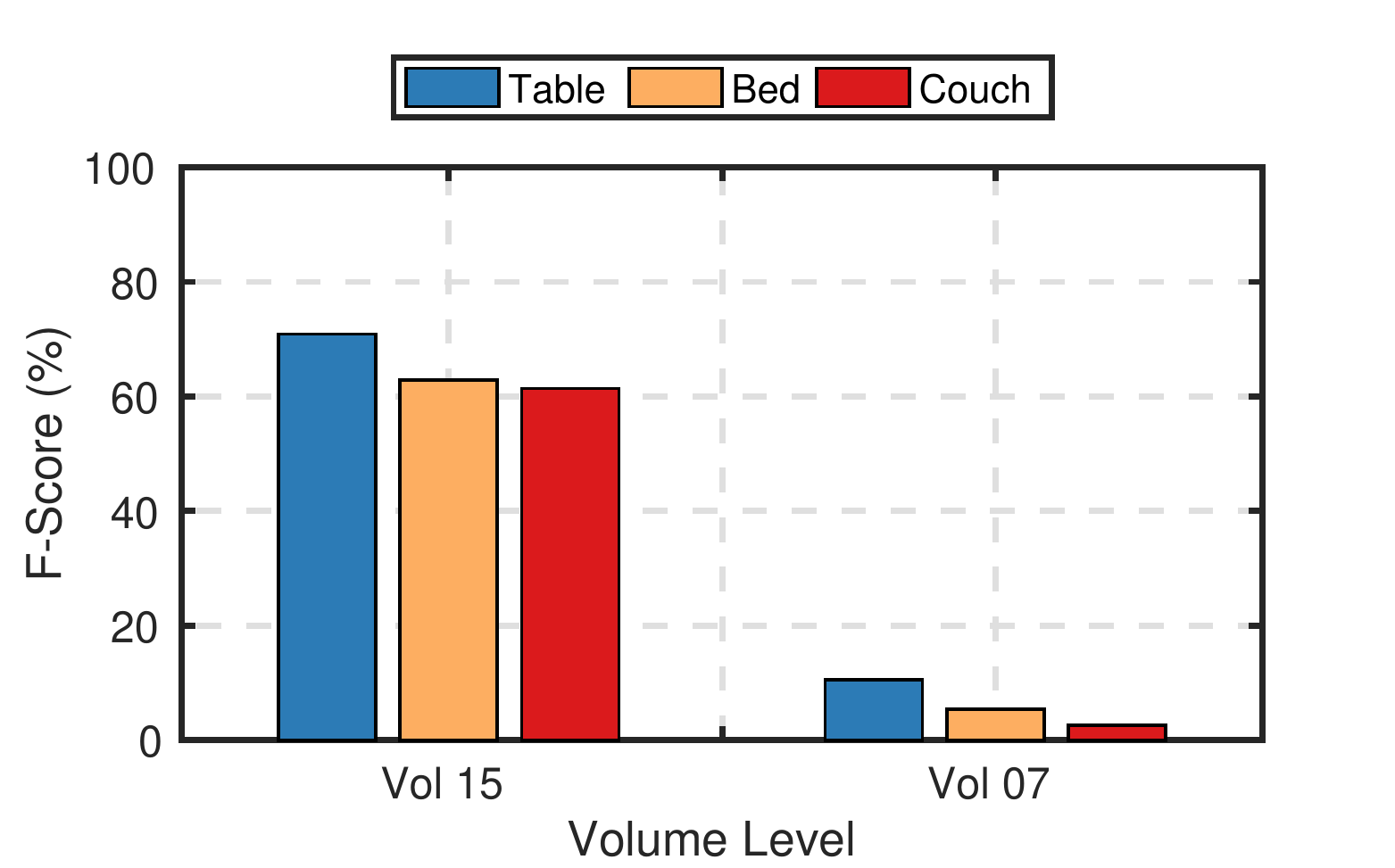}
			\caption{$ET + kNN$ classifier F-Scores for various\\ test surfaces at max and min volumes}
			\label{fig:attack_surface_etkNN_results}
		\end{subfigure}%
		~ 
		\begin{subfigure}[t]{0.52\textwidth}
			\centering
			\includegraphics[width=1\linewidth]{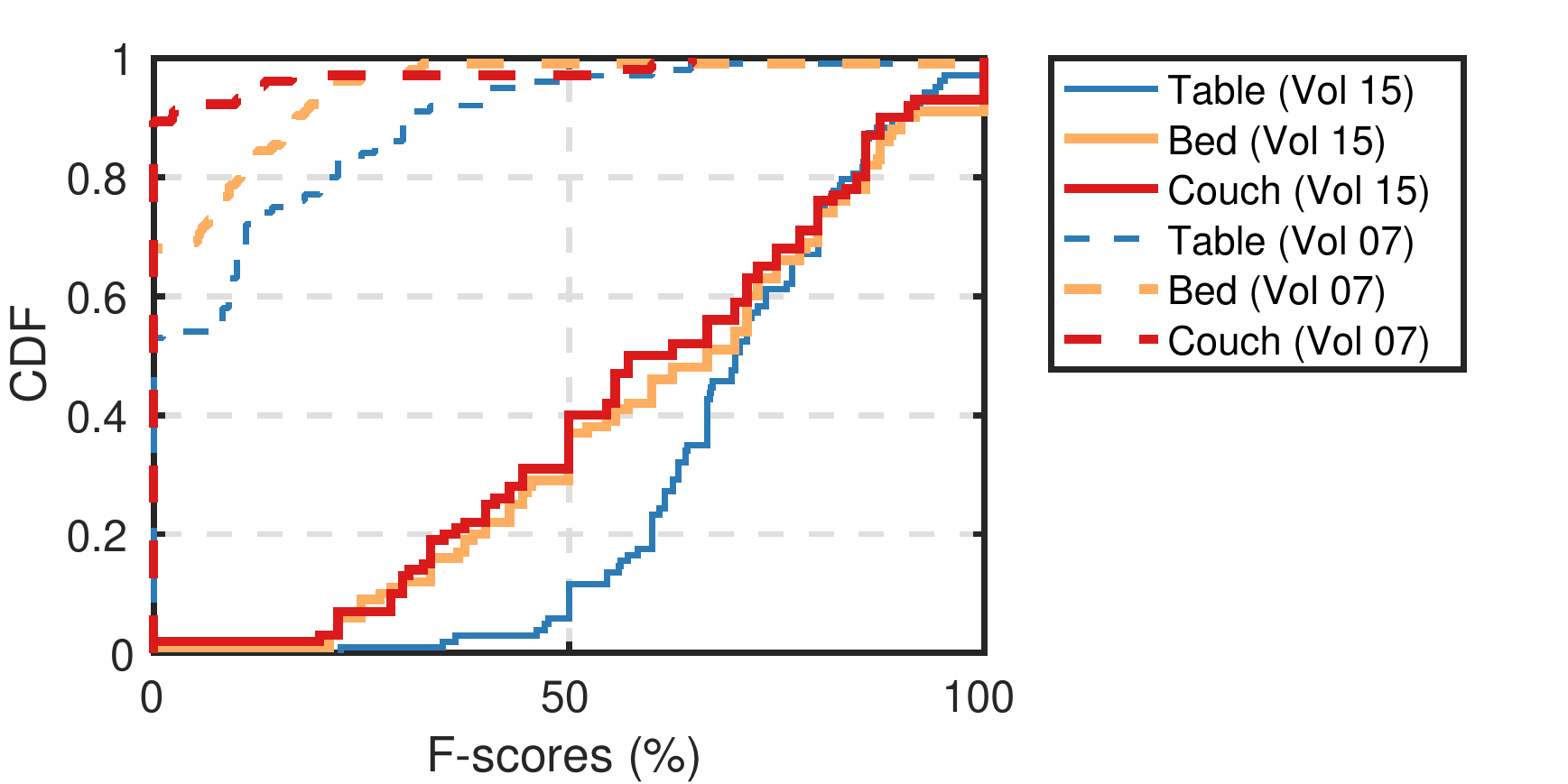}
			\caption{CDF of $ET + kNN$ classifier F-Scores for various test surfaces at max and min test volumes}
			\label{fig:cdf_surface_etkNN_results}
		\end{subfigure}
	\end{subfigure}%
	
	\begin{subfigure}[t]{1\textwidth}
		\begin{subfigure}[t]{.42\textwidth}
			\centering
			\includegraphics[width=1\linewidth]{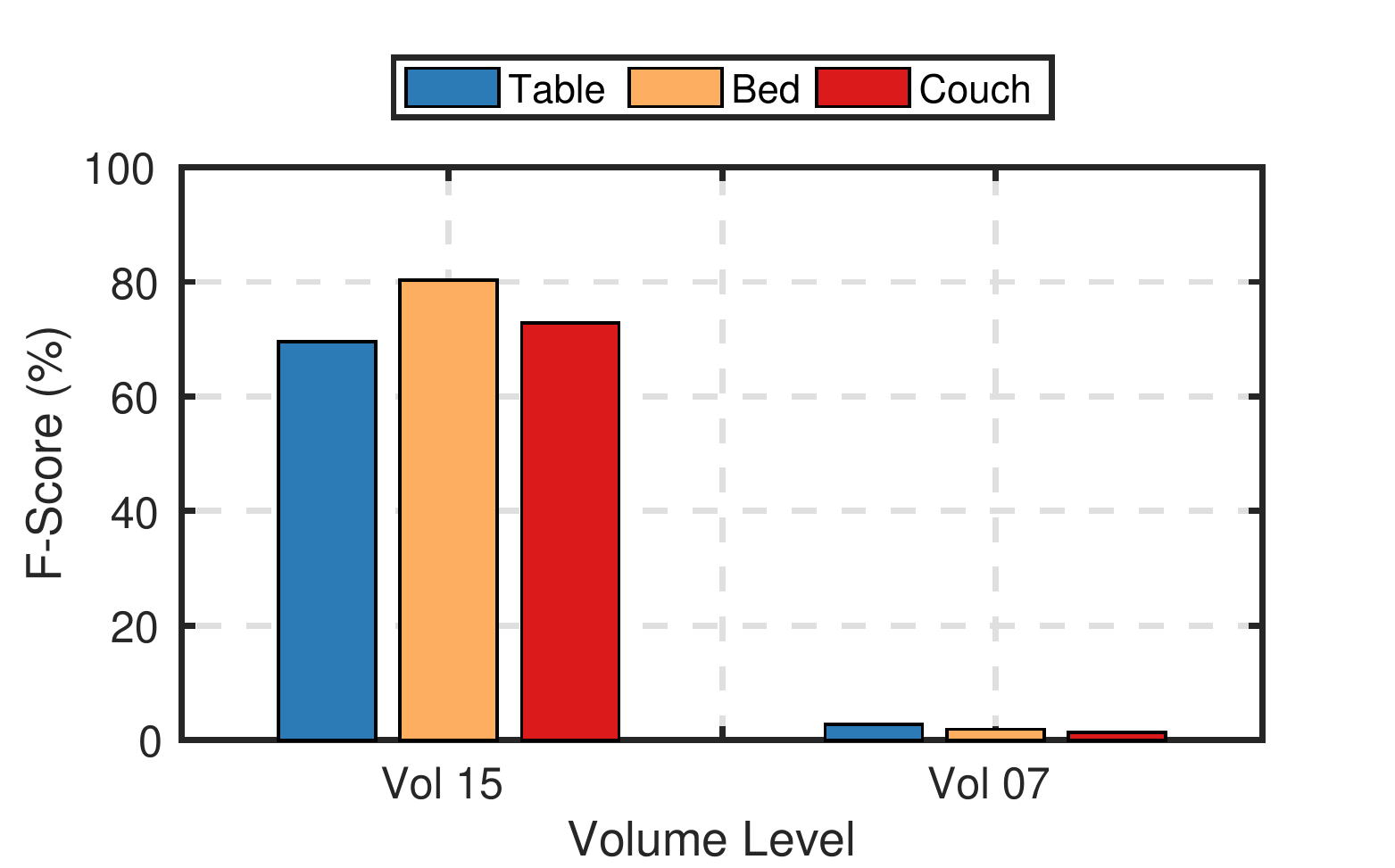}
			\caption{$CNN - LSTM $ classifier F-Scores for various \\test surfaces at max and min volumes}
			\label{fig:attack_surface_cnnLSTM_results}
		\end{subfigure}%
		~ 	
		\begin{subfigure}[t]{0.52\textwidth}
			\centering
			\includegraphics[width=1\linewidth]{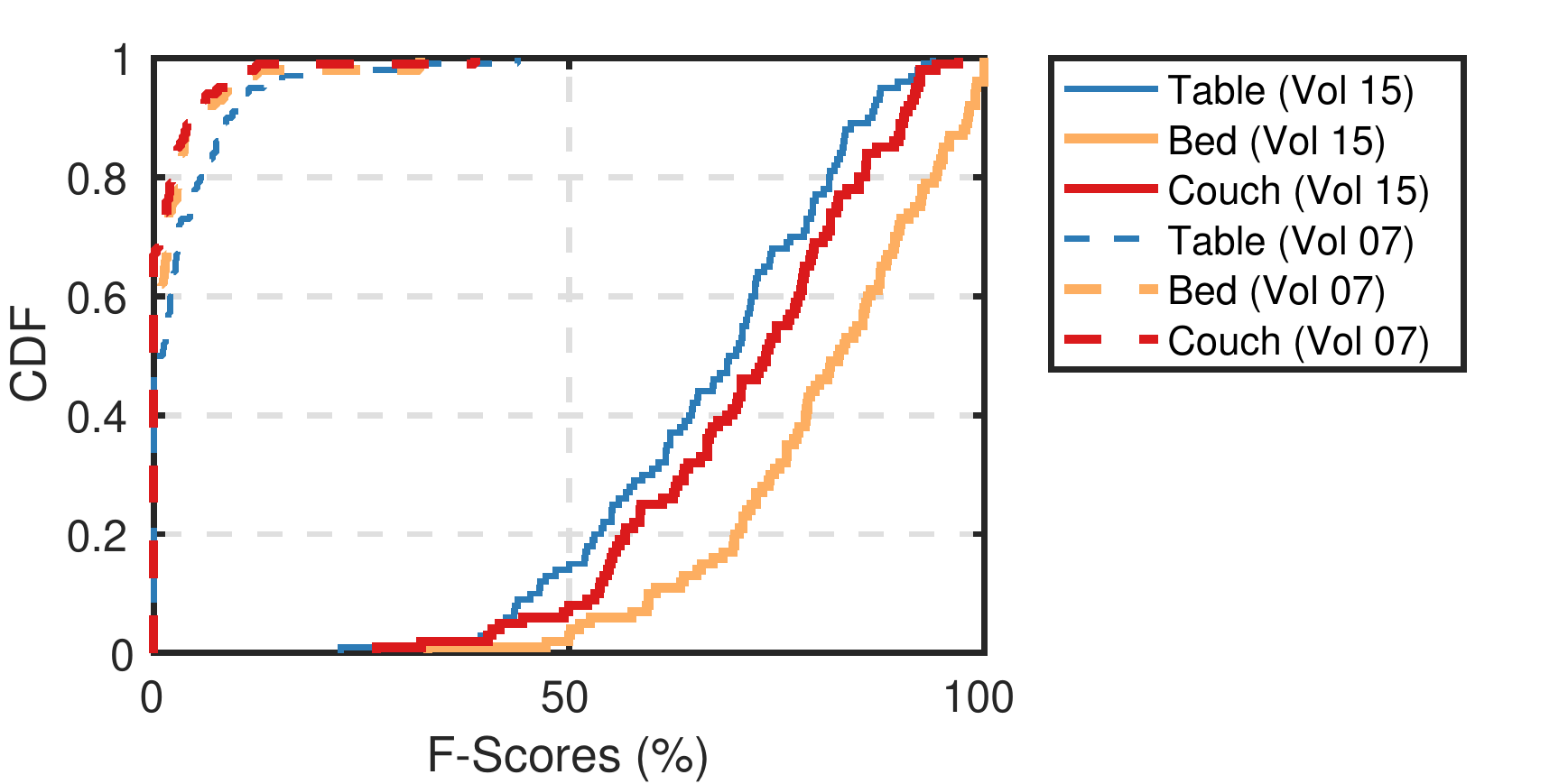}
			\caption{CDF of $CNN - LSTM$ classifier F-Scores for various test surfaces at max and min test volumes}
			\label{fig:cdf_surface_cnnLSTM_results}
		\end{subfigure}
	\end{subfigure}
	\caption{Impact of changes in test surface under the baseline training scenario. Testing is done using data collected when the phone plays the music at the highest and lowest volumes while resting on the table, bed or couch.}
	\label{fig:attack_surface_results}		
\end{figure*}

\subsubsection{Impact of changes in surface during testing (i.e., for each playback volume, the phone is placed on different surfaces)}
\label{sec:surface_results}

To study the impact of the phone placement surface on the effectiveness of the attack, we placed the phone on each of our 3 surfaces as music played at each of our music volumes. Figure \ref{fig:attack_surface_results} summarizes the results from these experiments for the highest and lowest volumes. Sub-figures \ref{fig:attack_surface_etkNN_results} and \ref{fig:cdf_surface_etkNN_results} respectively show the F-score returned by the ET+kNN and CNN-LSTM classifiers when only a single guess is made (we focus on the single most likely prediction since the 2nd and 3rd predictions do not produce any new notable patterns). Sub-figure \ref{fig:attack_surface_etkNN_results} shows that for both volumes, the table gave the highest F-score, followed by the bed and couch. Sub-figure \ref{fig:cdf_surface_etkNN_results} reveals a similar pattern for the lowest volume, but however reveals a different pattern for the highest volume (i.e., the bed performed best and the table performed worst). Overall, what these two figures tell us is that under training configuration \#1 (recall Table \ref{tab:training_configurations}, Page \pageref{tab:training_configurations}), none of the three phone placement surfaces consistently performs better than the others during the attack (i.e., during testing). Sub-figures \ref{fig:cdf_surface_etkNN_results} and \ref{fig:cdf_surface_cnnLSTM_results} show CDFs of the F-scores for both classifiers, and again further reinforce the observation that no single surface consistently performs best even when you look at the distribution of individual F-scores (i.e., at the highest volume, the red is in between the yellow and blue for the CNN-LSTM classifier, while the yellow is the one located in between for the ET+kNN classifier).

Overall, the patterns depicted here suggest that training configuration \#1 is able to produce a general-purpose attack that might appeal to an adversary who seeks to minimize surface-oriented biases for victims who might potentially use a wide range of phone placement surfaces. 

\begin{figure*}[h]
	\centering
	\begin{subfigure}[t]{0.95\textwidth}
		\begin{subfigure}[t]{.5\textwidth}
			\centering
			\includegraphics[width=1\linewidth]{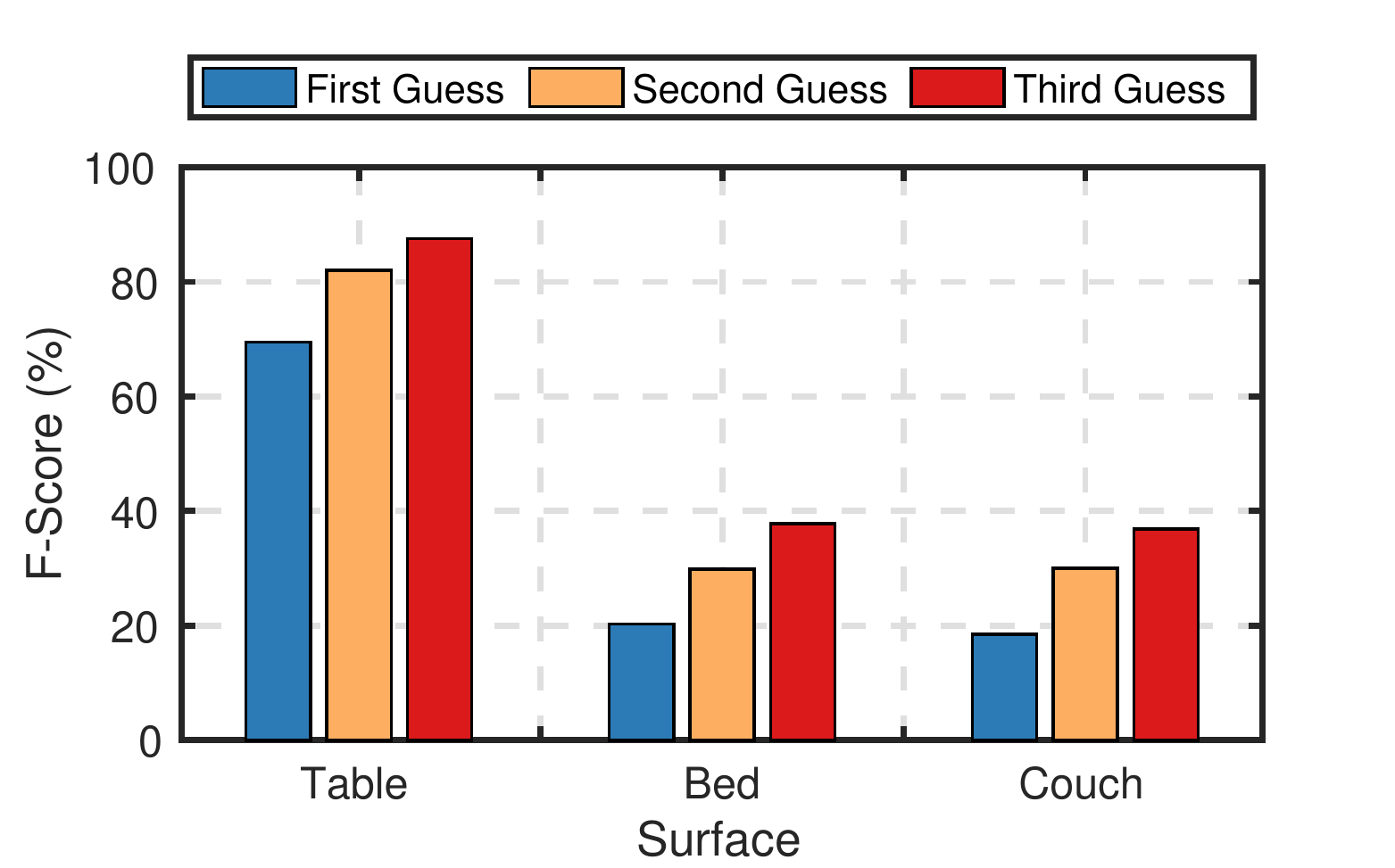}
			\caption{$ET + kNN$ classifier}
			\label{fig:table_training_et_knn_results}
		\end{subfigure}%
		~ 
		\begin{subfigure}[t]{0.5\textwidth}
			\centering
			\includegraphics[width=1\linewidth]{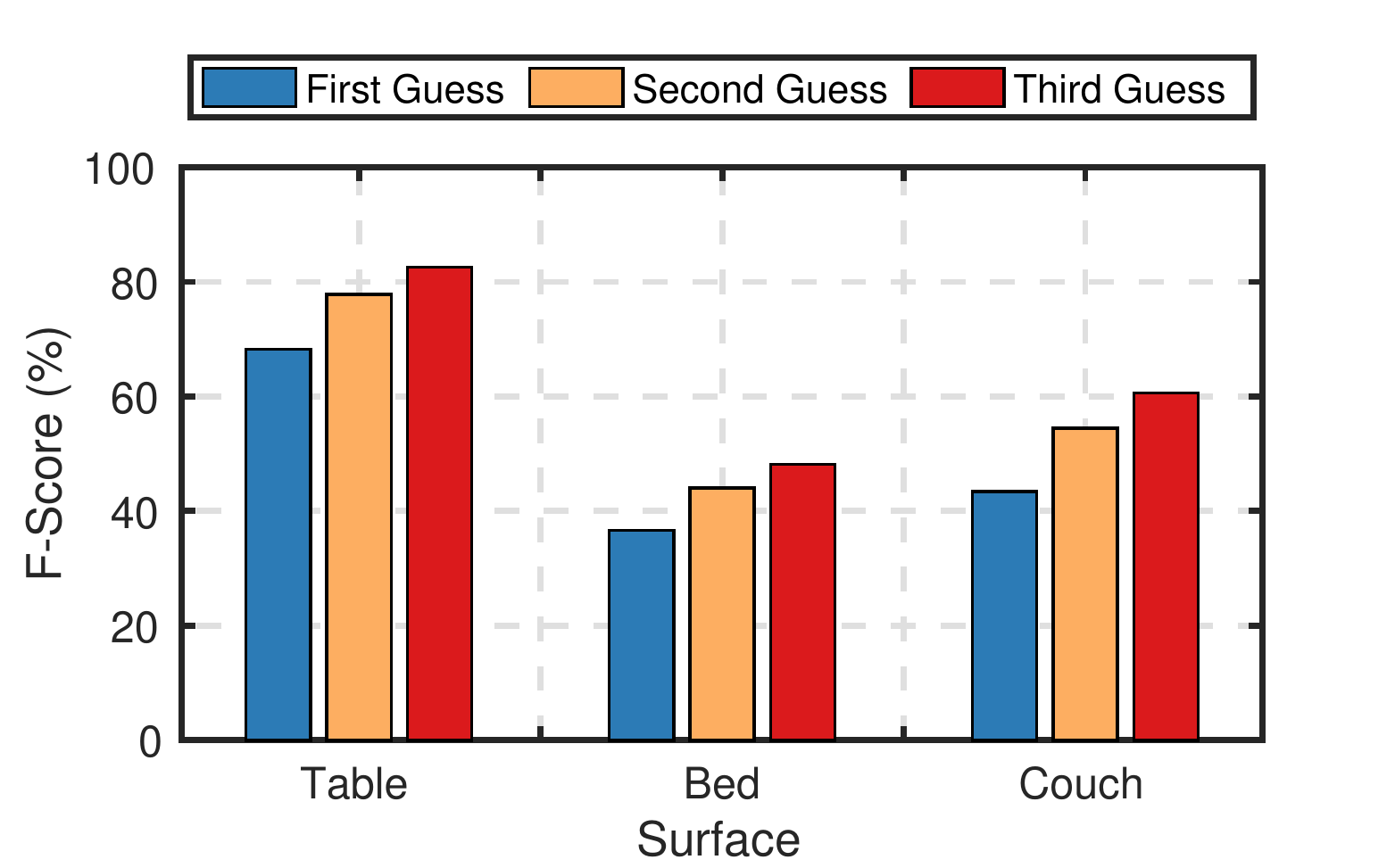}
			\caption{$CNN - LSTM$ classifier}
			\label{fig:table_training_cnn_lstm_results}
		\end{subfigure}
	\end{subfigure}
	
	\caption{Performance of surface-specific training scenario when training is done using data collected while the phone rests on the {\bf\underline{table}}. Testing is done using data collected when the phone plays music at maximum volume while resting on the table, bed and couch}
	\label{fig:table_training_results}	
\end{figure*}

\begin{figure*}[h]
	\centering
	\begin{subfigure}[t]{0.95\textwidth}
		\begin{subfigure}[t]{.5\textwidth}
			\centering
			\includegraphics[width=1\linewidth]{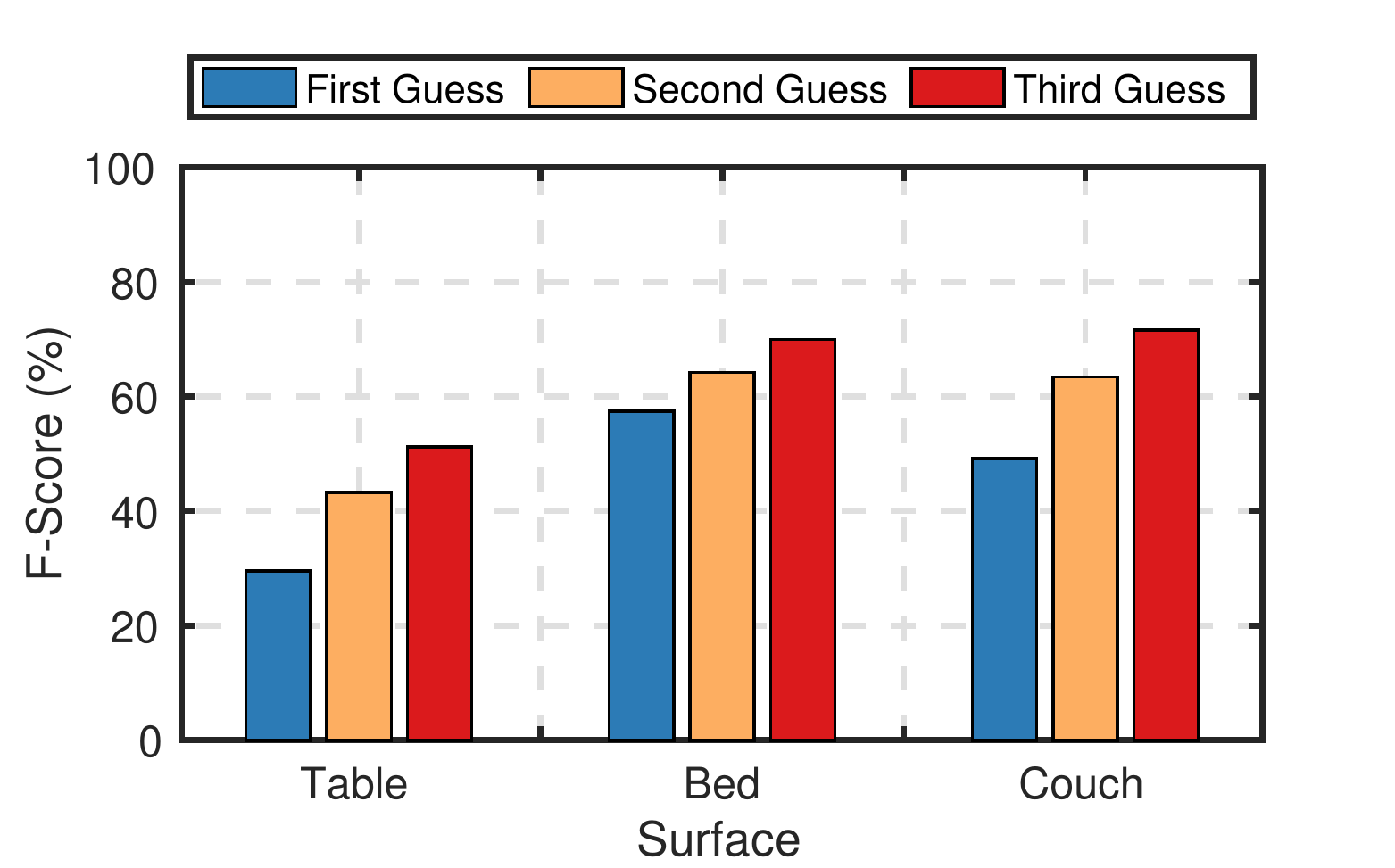}
			\caption{$ET + kNN$ classifier}
			\label{fig:bed_training_et_knn_results}
		\end{subfigure}%
		~ 
		\begin{subfigure}[t]{0.5\textwidth}
			\centering
			\includegraphics[width=1\linewidth]{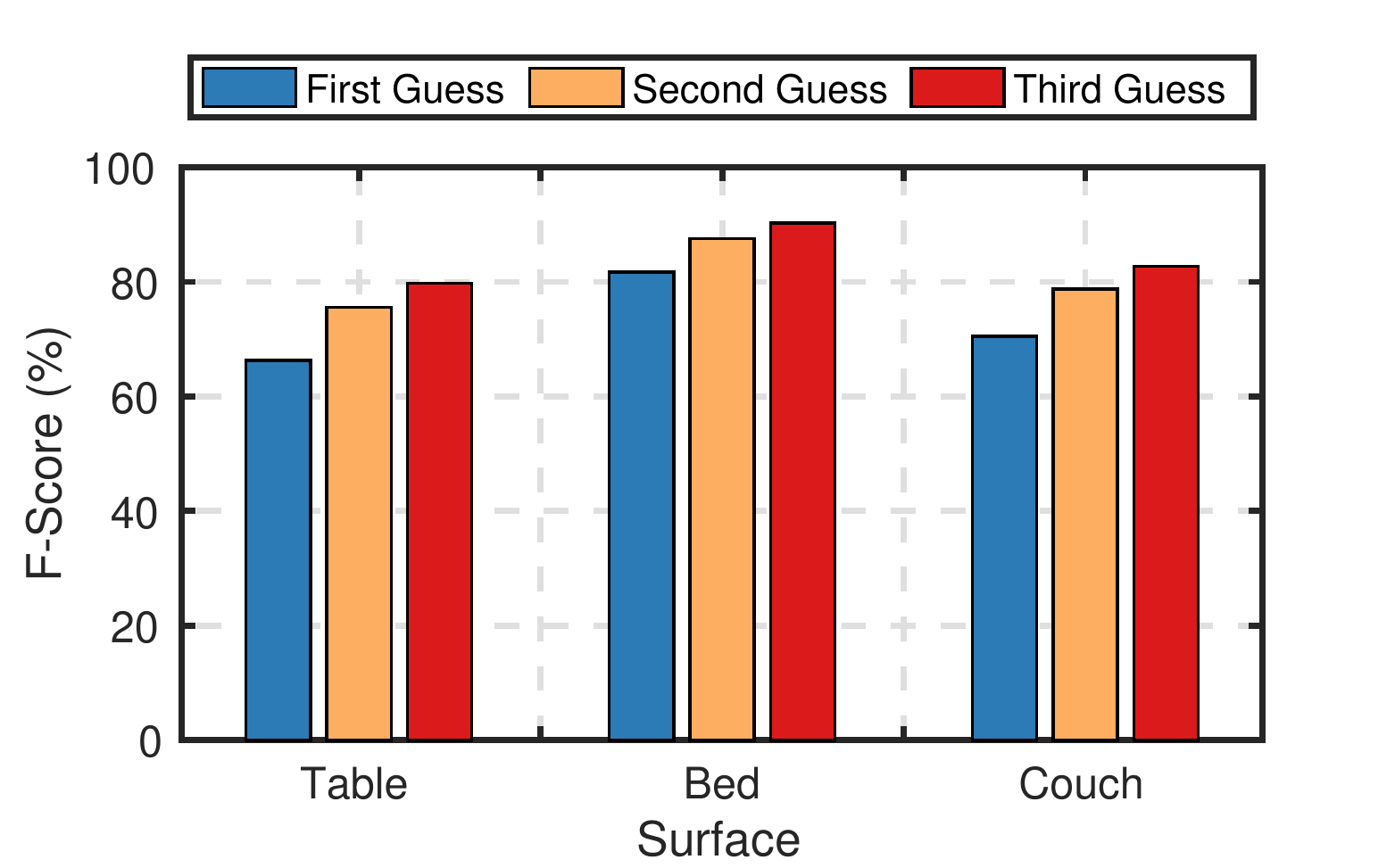}
			\caption{$CNN - LSTM$ classifier}
			\label{fig:bed_training_cnn_lstm_results}
		\end{subfigure}
	\end{subfigure}
	
	\caption{Performance of surface-specific training scenario when training is done using data collected while the phone rests on the {\bf\underline{bed}}. Testing is done using data collected when the phone plays music at maximum volume while resting on the table, bed and couch}
	\label{fig:bed_training_results}	
\end{figure*}

\begin{figure*}[h]
	\centering
	\begin{subfigure}[t]{0.95\textwidth}
		\begin{subfigure}[t]{.5\textwidth}
			\centering
			\includegraphics[width=1\linewidth]{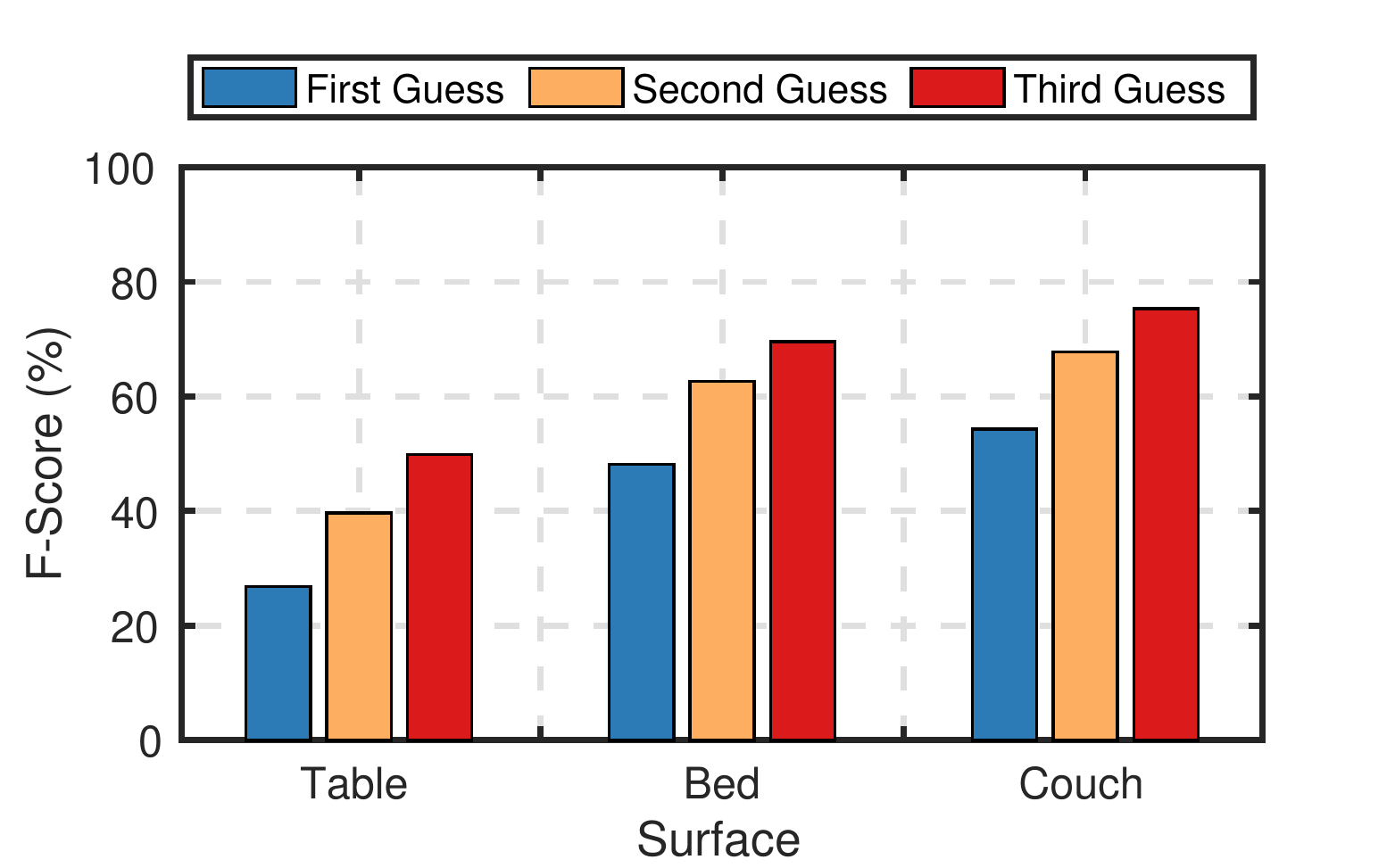}
			\caption{$ET + kNN$ classifier}
			\label{fig:couch_training_et_knn_results}
		\end{subfigure}%
		~ 
		\begin{subfigure}[t]{0.5\textwidth}
			\centering
			\includegraphics[width=1\linewidth]{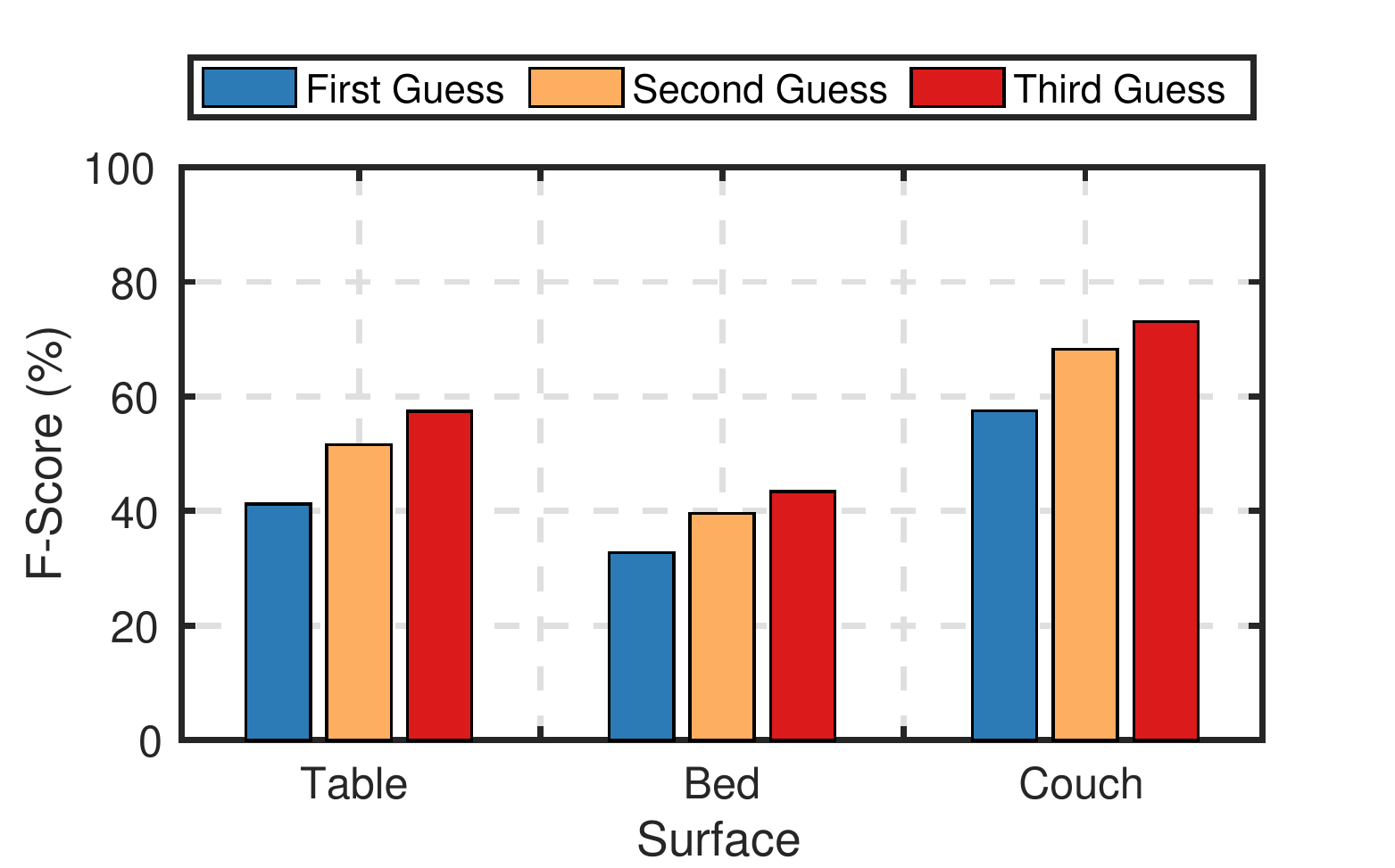}
			\caption{$CNN - LSTM$ classifier}
			\label{fig:couch_training_cnn_lstm_results}
		\end{subfigure}
	\end{subfigure}
	
	\caption{Performance of surface-specific training scenario when training is done using data collected while the phone rests on the {\bf\underline{couch}}. Testing is done using data collected when the phone plays music at maximum volume while resting on the table, bed and couch}
	\label{fig:couch_training_results}	
\end{figure*}

\subsection{Performance of Music Identification Attack Under Surface-Centric Training Scenario (i.e., Training Configurations II through IV, Table \ref{tab:training_configurations})}
\label{surface-specific-1}
In this section we showcase the performance of the attack
when training was done based on data collected from a single phone placement surface. We only present results here for the testing configurations when music was played at maximum volume (see testing configurations I, VI and XI, Table \ref{tab:testing_configurations}, Page \pageref{tab:testing_configurations}) since the test configurations for the lower volumes depicted an effect of reduced playback volume similar to what has already been discussed in Section \ref{sec:combined_results}. Figures \ref{fig:table_training_results}, \ref{fig:bed_training_results} and \ref{fig:couch_training_results} respectively summarize the results from the table-centric, bed-centric and couch-centric training configurations described above. Figure \ref{fig:table_training_results} shows that for both classifiers, the attack performed best when data collected while the phone rested on the table was used to perform testing on a model which had been built (or trained) with data collected from the same (table) surface.  When testing data was obtained from the two other surfaces, the F-scores obtained were almost half of those seen for the table-table comparison. In Figures \ref{fig:bed_training_results} and \ref{fig:couch_training_results}, this pattern is somewhat subdued but still apparent as the bed gives F-scores in Figure \ref{fig:bed_training_results} that are at least as high as those obtained with the other 2 surfaces, and the couch gives F-scores that are at least as high as those obtained with the other 2 surfaces in Figure \ref{fig:couch_training_results}. 

The notion that a match between surfaces in the training and testing sets gives the best performance is not so surprising as one can expect a classifier to learn better if the training and testing conditions are similar. Perhaps a more interesting pattern from these 3 figures is that the cross-surface train-test match-ups still perform much better than a random attack (recall that a random attacker would get about 1\% F-score for a single guess), and in some cases almost as well as the single surface cases. This trait suggests that an attacker who for some reason decides to build a training model based on a single surface of choice could still feasibly attack a wide range of victims who rest their phones on various surfaces. Because the attacker would in practice very likely not know the surface on which the victim places the phone, this ability of the attack to perform well in cross-surface train-test settings is a strong indicator of its practicality.

\section{Experimental Results --- Novelty Detector}
\label{novelty-12}
Recall (from Section \ref{nd}) the scenario in which a song to be classified does not exist in the training set (i.e., in practice the attacker can only find so many songs for training). Certain attackers might handle this scenario by including a novelty detection step that flags unknown songs which will not be forwarded to the music identification step. Here, we present results from this anomaly detection step. Revisit Section \ref{nd} for how we train the anomaly detection module. Like in the previous section, we present the results under both the baseline and surface-specific training approaches.

\subsection{Performance of Novelty Detection Attack Under Baseline Training Scenario (i.e., Training Configuration I, Table 1)}
\label{novel-base}	
Table  \ref{tab:novelty_detector_performance} summarizes the performance of our novelty detection attack when data from multiple volumes and surfaces is combined into a single training set. Testing is done for all 3 surfaces and 5 different volumes. Observe that the highest F-score of 78.26\% is obtained when testing is done using a phone that plays music at the highest volume while placed on a table surface. The lowest F-score (50.11\%) on the other hand is obtained when testing is done on the couch at the lowest music volume.

Overall, for Volume 11 and above, the novelty detection attack performs well above random guessing\footnote{This being a two-class problem (i.e., clasifier determines whether a song is in the corpus or not), random guessing has an F-score of about 50\%} for all test surfaces under this training configuration. As long as the victim plays music at a high volume therefore, the novelty attack under this configuration would be useful as a means to augment the plain music identification attack.

\begin{table*}[h]
	\centering
	\begin{tabular}{lccccc}
		\toprule
		\multirow{2}{*}{\textbf{Test Surface}} & \multicolumn{5}{c}{\textbf{Volume level used during testing}}                           \\
		& \textbf{Vol 15} & \textbf{Vol 13} & \textbf{Vol 11} & \textbf{Vol 09} & \textbf{Vol 07} \\
		\midrule
		Table                             & 78.26           & 71.18           & 74.79           & 66.58           & 56.62           \\
		Bed                               & 71.97           & 73.12           & 68.29           & 55.06           & 55.07           \\
		Couch                             & 70.61           & 68.18           & 63.79           & 53.21           & 50.11           \\ \bottomrule
	\end{tabular}
	\caption{Performance of the novelty detection attack under the baseline training scenario.}
	\label{tab:novelty_detector_performance}
\end{table*}

\subsection{Performance of Novelty Detection Attack Under Surface-Centric Training Scenario (i.e., Training Configurations II through IV, Table \ref{tab:training_configurations})}

\begin{table*}[h]
	\centering
	\begin{tabular}{cllllll}
		\toprule
		\multirow{2}{*}{\textbf{Training Surface}} & \multicolumn{1}{c}{\multirow{2}{*}{\textbf{Test Surface}}} & \multicolumn{5}{c}{\textbf{Test Volume}}                                                \\
		& \multicolumn{1}{c}{}                                       & \textbf{Vol 15} & \textbf{Vol 13} & \textbf{Vol 11} & \textbf{Vol 09} & \textbf{Vol 07} \\ \midrule
		\multirow{3}{*}{Table}               & Table                                                       &     \textbf{86.21}       &        \textbf{78.08}    &     \textbf{75.69}       &      \textbf{68.25}      &    \textbf{60.42}        \\
		\textbf{}                       & Bed                                                        &    66.05       &    71.61        &      71.42      &      62.61      &      56.36      \\
		\textbf{}                       & Couch                                                      &    65.27        &    68.33        &    69.25        &     53.57       &     53.01       \\ 
		\midrule
		\multirow{3}{*}{Bed}   & Table                                                       &   50.65        &     55.11       &   62.15       &     58.12      &    53.01        \\
		& Bed                                                        &    \textbf{80.65}        &    \textbf{72.32}        &      \textbf{73.94}      &     \textbf{70.12}       &    \textbf{61.60}       \\
		& Couch                                                      &     74.91       &    73.38        &     72.03       &      60.76      &   53.31         \\ \midrule
		\multirow{3}{*}{Couch}               & Table                                                       &     51.59       &    57.33        &     59.41        &     54.15       &    53.20        \\
		\textbf{}                       & Bed                                                        &    70.83        &    66.20        &     66.54       &    61.57        &    54.58        \\
		\textbf{}                       & Couch                                                      &    \textbf{75.33}        &     \textbf{72.97}      &     \textbf{68.99}       &     \textbf{67.71}       &     \textbf{56.77}       \\ 
		\bottomrule
	\end{tabular}
	\caption{F-scores obtained with the novelty detection attack using surface-specific training data from smartphones without phone covers.}
	\label{tab:surface_specific_training_novelty_detector_performance}
\end{table*}

Table \ref{tab:surface_specific_training_novelty_detector_performance} shows the performance of the novelty detection attack under the surface-specific training configuration.  Similar to the patterns observed previously, the best F-scores are (unsurprisingly) obtained when the surface used for training is the same as the surface used for testing (see bolded text in the table). Another unsurprising pattern similar to patterns seen earlier is that the highest test volumes result in the highest F-scores overall. Finally, compared to the baseline training configuration (Section \ref{novel-base}) whose best F-score was less than 80\%, this configuration has a couple of cases that hit F-scores in the 80s. These higher F-scores might make it more appealing to the attacker, however it is noteworthy that the attacker would have to correctly match the victim's phone placement surface in order to reap these gains. 

	\section{Experimental Results --- Phone Covers as  Defense Against the Attack}
\label{sec:defense_mechanism}
In this section, we showcase results on how the vibration damping effect due to a phone cover impacts both the music identification and novelty detection attacks. Unlike other potential defense mechanisms (e.g., changes to sensor permission model), the use of phone covers is particularly interesting since it 
is a mechanism that many users already deploy for protection of their phones from physical damage. For each of these two kinds of attack, we showcase the impact of phone covers on both the baseline and surface-centric versions of the attacks.

\begin{table*}[h]
	\centering
	\begin{tabular}{llcccccc} 
		\toprule
		\multirow{3}{*}{\textbf{\begin{tabular}[c]{@{}l@{}} Attack\\Model \end{tabular}}}         & \multirow{3}{*}{\textbf{\begin{tabular}[c]{@{}l@{}} Volume\\Level \end{tabular}}} & \multicolumn{6}{c}{\textbf{Surface on which phone is placed during \underline{testing}}}                                                                                                                        \\
		&                                                                           & \multicolumn{2}{c}{\textbf{Table}}                                                                                                   & \multicolumn{2}{c}{\textbf{Bed}}                                                                                   & \multicolumn{2}{c}{\textbf{Couch}}                                                                                  \\ 
		\cmidrule(r){3-4}\cmidrule(r){5-6}\cmidrule(r){7-8}
		&                                                                           & 
		\multicolumn{1}{c}{\begin{tabular}[c]{@{}c@{}}FScore \\ Without \\ Defense \end{tabular}} & \multicolumn{1}{c}{\begin{tabular}[c]{@{}c@{}}FScore \\ With \\ Defense \end{tabular}} & \multicolumn{1}{c}{\begin{tabular}[c]{@{}c@{}}FScore \\ Without \\ Defense \end{tabular}} & \multicolumn{1}{c}{\begin{tabular}[c]{@{}c@{}}FScore \\ With \\ Defense \end{tabular}} & \multicolumn{1}{c}{\begin{tabular}[c]{@{}c@{}}FScore \\ Without \\ Defense \end{tabular}} & \multicolumn{1}{c}{\begin{tabular}[c]{@{}c@{}}FScore \\ With \\ Defense \end{tabular}}  \\ 
		\midrule
		\multirow{5}{*}{$ET + kNN$}                                 & Vol 15                                                                           &              70.93                  &              7.57                     &          62.86                       &            5.59                      &            61.39                     &             4.97                      \\
		& Vol 13                                                                           &            47.55                    &                4.17                   &              63.50                   &          3.67                        &         59.62                         &                 4.22                   \\
		& Vol 11                                                                           &                48.46                 &           3.09                         &             47.85                     &            2.70                       &           42.89                       &               3.38                     \\
		& Vol 09                                                                           &          28.75                       &         2.81                          &          18.56                        &            2.41                      &            3.41                      &               1.45                     \\
		& Vol 07                                                                           &          10.57                       &           2.42                         &           5.40                       &               0.97                    &         2.61                         &              1.83                      \\ 
		\midrule
		\multirow{5}{*}{$CNN - LSTM$}                                  & Vol 15                                                                           &             69.55                   &         4.31                           &              80.32                   &              7.39                     &          72.82                       &             4.31                       \\
		& Vol 13                                                                           &            70.99                    &               3.61                     &           76.32                      &            2.69                       &           78.73                       &              2.58                      \\
		& Vol 11                                                                           &            60.16                     &           1.31                         &            64.85                      &           1.32                        &         59.77                         &               0.98                    \\
		& Vol 09                                                                           &           33.32                      &          0.78                          &          9.46                        &           0.71                        &       2.27                           &              1.24                      \\
		& Vol 07                                                                           &           2.81                      &            0.57                        &          1.92                        &             0.69                      &        1.40                          &               0.81                     \\
		\bottomrule
	\end{tabular}
	\caption{Impact of phone covers on the music identification attack {\it{carried out by the na\"ive attacker}}. The table showcases the baseline scenario (i.e., when data from all surfaces and volumes is combined into a single training set). For each phone placement surface during testing, the table compares side-by-side the F-scores obtained when the test dataset is collected from a phone with and without a cover.}
	\label{tab:pc_attack_results}
\end{table*}

\subsection{Impact of Phone Covers on the Baseline Music Identification Attack}
\label{sub-main}
Here we present our results on the impact of phone covers on the baseline scenario of the music identification attack (i.e., the case where data from all volumes and surfaces is combined into a single training set -- recall description of this scenario in Section \ref{sec:attack_expt_results}). We divide these results into two sub-scenarios, namely: (1) the na\"ive attacker case, and, (2) the sophisticated attacker case. In sub-scenario (1), the attacker does not anticipate that the victim will use a phone cover. The attacker thus builds the training set from data collected from phones which have no covers (i.e., the baseline scenario in this case uses training data from all volumes and surfaces using data generated by phones {\it{not}} housed in covers). In sub-scenario (2), the attacker infers that the victim uses a phone cover (Section \ref{cover-def} will provide more details on this inference). Leveraging this knowledge, the attacker then builds the baseline training set from data collected from phones which have covers (i.e., the baseline training scenario in this case involves data from all volumes and surfaces generated by a phones housed in covers). Results from the two sub-scenarios are presented below in subsections \ref{sub-1} and \ref{sub-2}. Note that for both the na\"ive and sophisticated attackers, our primary {\it{testing}} configuration uses data collected from a phone with a cover/casing (i.e., we simulate an attack victim who takes steps to defend against the attack). To give context to these results however, we also include, side-by-side, results from alternate testing configurations where data is collected from a phone without a cover (i.e., we simulate a victim who does not employ any defense against the attack). 
\subsubsection{The na\"ive attacker case}
\label{sub-1}

Table \ref{tab:pc_attack_results} shows highlights of our results for the na\"ive attacker case. The forth column of this table (labelled ``FScore With Defense'') shows the F-scores when the attack defense mechanism was deployed --- i.e., testing was done for both classifiers when a phone housed in a cover was placed on the table while playing music. As earlier explained, for benchmarking purposes, the third column (labelled ``FScore Without Defense'') shows the results when the defence was not deployed --- i.e., testing was done for both classifiers when a phone housed in a cover was placed on the table while playing music. The numbers in this column are exactly the same as the F-scores previously reported in Figure \ref{fig:volume_attack_performance} (see first guess only). 

Observe that for all volumes, the F-scores in Column \#4 are a full order of magnitude below the F-scores seen in Column \#3. 
This trend shows that the phone cover quite significantly reduced the impact of the attack. Columns \#5 and \#6 are respectively constructed similarly to Columns \#3 and \#4, except that the surface on which the phone is placed during testing is a bed. In the same vein Columns \#7 and \#8 are respectively constructed similarly to Columns \#3 and \#4, except that the surface on which the phone is placed during testing is a couch. Observe that all column-pairs reveal the same pattern of the phone cover significantly reducing the impact of the attack. Overall, the results in Table \ref{tab:pc_attack_results} suggest that given our notion of a na\"ive attacker, phone covers can do a great deal to mitigate the attack.

\subsubsection{The sophisticated attacker case}
\label{sub-2}
Table \ref{tab:pc_training_attack_results} shows highlights of our results for the sophisticated attacker case. All columns have exactly the same meaning as previously described for Table  \ref{tab:pc_attack_results}, except that the attacker in this case uses phones housed in covers to build the training set (recall training description provided in Section \ref{sub-main}). When the victim listens to music while resting the phone on the table (i.e., columns \#3 and 4), the use of a cover approximately halves the F-scores for the top two volumes relative to when a cover is not used. For the lowest three volumes, the impact of the phone cover is much more significant as most of the F-scores are an order of magnitude lower than the case without the defense. This pattern likely results from the fact that high music volumes are able to produce a vibration pattern for each song that is able to overcome the damping effect due to the phone covers. Overall, the results in Table \ref{tab:pc_training_attack_results} suggest that given our notion of a sophisticated attacker, phone covers are perhaps a suitable defense {\it{only if}} the victim plays music at low volumes.

\begin{table*}[h]
	\centering
	\begin{tabular}{llcccccc} 
		\toprule
		\multirow{3}{*}{\textbf{\begin{tabular}[c]{@{}l@{}} Attack\\Model \end{tabular}}}         & \multirow{3}{*}{\textbf{\begin{tabular}[c]{@{}l@{}} Volume\\Level \end{tabular}}} & \multicolumn{6}{c}{\textbf{Surface on which phone is placed during \underline{testing}}}                                                                                                                        \\
		&                                                                           & \multicolumn{2}{c}{\textbf{Table}}                                                                                                   & \multicolumn{2}{c}{\textbf{Bed}}                                                                                   & \multicolumn{2}{c}{\textbf{Couch}}                                                                                  \\ 
		\cmidrule(r){3-4}\cmidrule(r){5-6}\cmidrule(r){7-8}
		&                                                                           & \multicolumn{1}{c}{\begin{tabular}[c]{@{}c@{}}FScore \\ Without \\ Defense \end{tabular}} & \multicolumn{1}{c}{\begin{tabular}[c]{@{}c@{}}FScore \\ With \\ Defense \end{tabular}} & \multicolumn{1}{c}{\begin{tabular}[c]{@{}c@{}}FScore \\ Without \\ Defense \end{tabular}} & \multicolumn{1}{c}{\begin{tabular}[c]{@{}c@{}}FScore \\ With \\ Defense \end{tabular}} & \multicolumn{1}{c}{\begin{tabular}[c]{@{}c@{}}FScore \\ Without \\ Defense \end{tabular}} & \multicolumn{1}{c}{\begin{tabular}[c]{@{}c@{}}FScore \\ With \\ Defense \end{tabular}}  \\ 
		\midrule
		\multirow{5}{*}{$ET + kNN$}                                 & Vol 15                                                                           &              70.93                  &              37.25                    &           62.86                      &              30.29                   &         61.39                        &            15.71                       \\
		& Vol 13                                                                           &          47.55                      &              24.93                     &          63.50                       &          16.54                        &       59.62                           &                6.28                   \\
		& Vol 11                                                                           &              48.46                   &             3.09                      &          47.85                        &            8.97                      &     42.89                             &              4.43                     \\
		& Vol 09                                                                           &             28.75                    &         1.53                           &         18.56                         &                3.84                  &      3.41                            &             1.62                      \\
		& Vol 07                                                                           &          10.57                       &         2.07                          &         5.40                        &             1.24                      &         2.61                         &             1.03                       \\ 
		\midrule
		\multirow{5}{*}{$CNN - LSTM$}                                  & Vol 15                                                                           &           69.55                     &              24.27                     &           80.32                      &               26.94                   &          72.82                       &           19.34                         \\
		& Vol 13                                                                           &          70.99                      &              17.98                  &        76.32                         &            21.22                     &       78.73                           &          12.07                         \\
		& Vol 11                                                                           &              60.16                   &                7.62                   &        64.85                          &          4.72                        &        59.77                          &         2.63                          \\
		& Vol 09                                                                           &           33.32                      &             3.51                       &          9.46                        &           2.02                        &                          2.27        &              1.72                     \\
		& Vol 07                                                                           &         2.81                        &             1.94                       &         1.92                         &           1.76                        &                        1.40          &          1.37                          \\
		\bottomrule
	\end{tabular}
	\caption{Impact of phone covers on the music identification attack {\it{carried out by the sophisticated attacker}}. The table showcases the baseline scenario (i.e., when data from all surfaces and volumes is combined into a single training set). For each phone placement surface during testing, the table compares side-by-side the F-scores obtained when the test dataset is collected from a phone with and without a cover.}
	\label{tab:pc_training_attack_results}
\end{table*}

\begin{figure*}[h]
	\centering
	\begin{subfigure}[t]{0.95\textwidth}
		\begin{subfigure}[t]{.5\textwidth}
			\centering
			\includegraphics[width=1\linewidth]{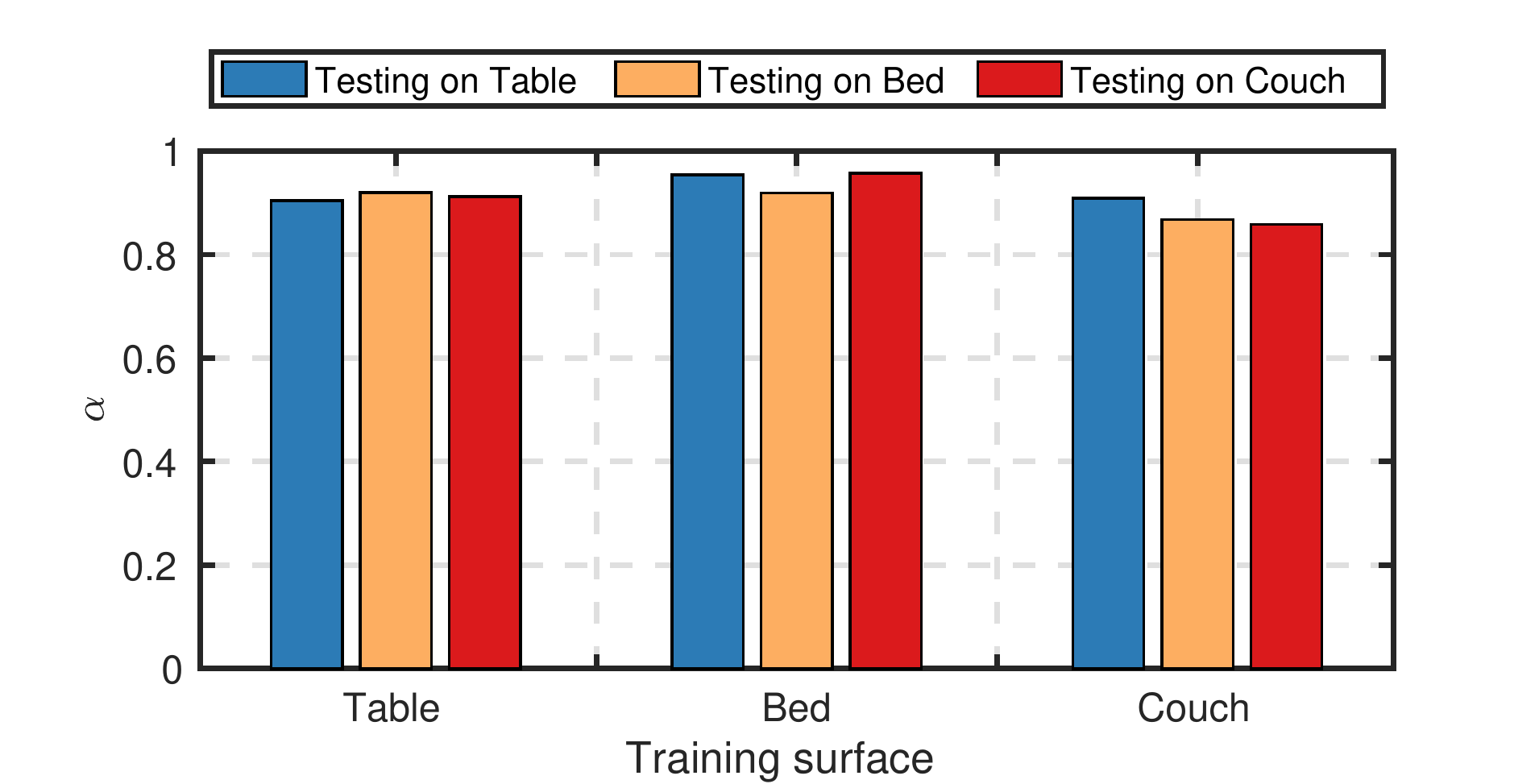}
			\caption{Na\"ive attacker}
			\label{fig:naive_attacker_surface_cnn_lstm_training_vol17}
		\end{subfigure}%
		~ 
		\begin{subfigure}[t]{0.5\textwidth}
			\centering
			\includegraphics[width=1\linewidth]{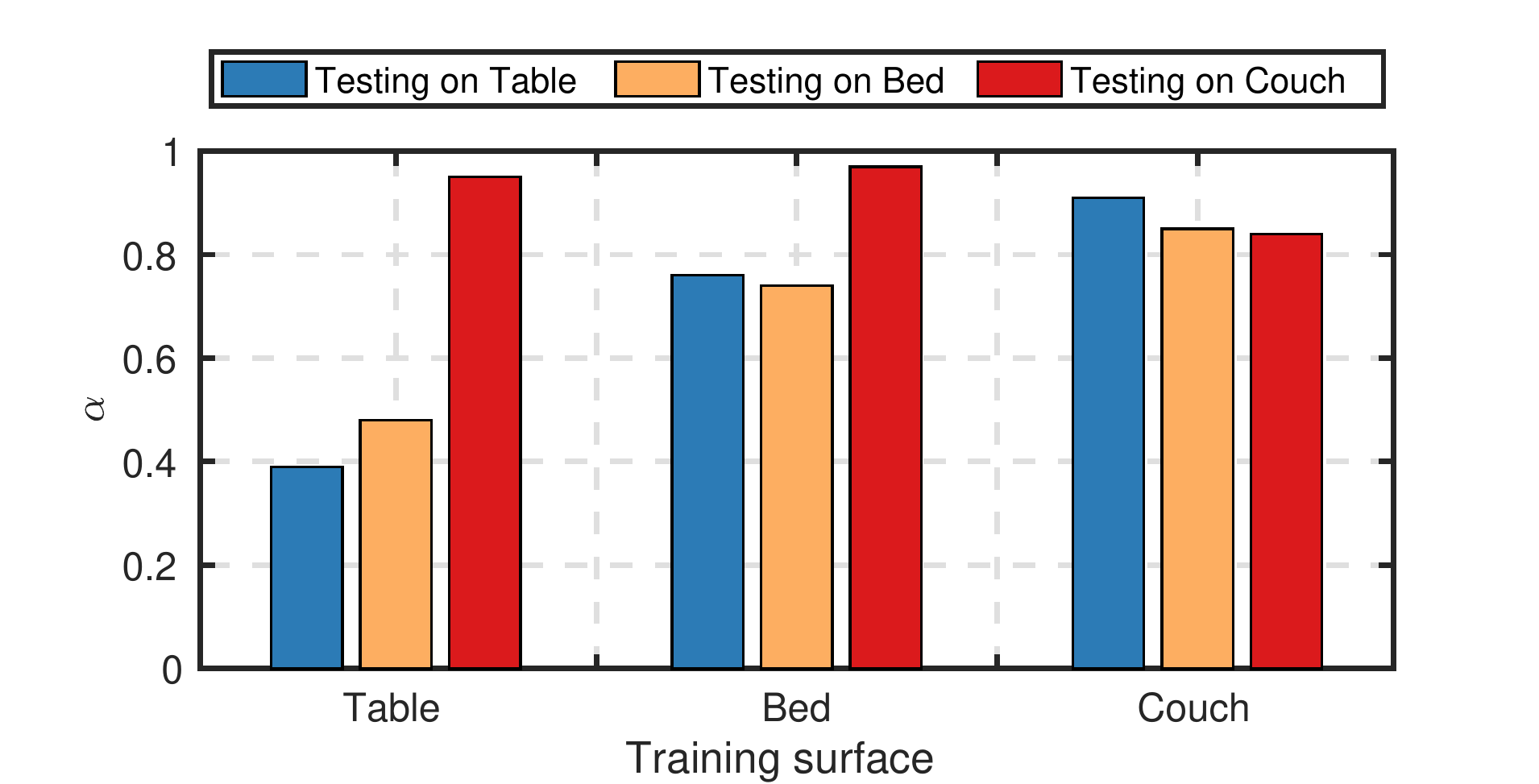}
			\caption{Sophisticated attacker}
			\label{fig:sophisticated_attacker_surface_cnn_lstm_training_vol17}
		\end{subfigure}
	\end{subfigure}
	
	\caption{Impact of phone covers on the music identification attack under the surface-specific training scenario. In both plots testing is done at the highest volume (Volume 17). A higher $\alpha$ corresponds to a more effective defense against the attack.}
	\label{fig:pc_surface_attack_results}	
\end{figure*}
	
\subsection{Impact of Phone Covers on the Surface-Centric Music Identification Attack}
\label{covers-scmi-attack}
In this section we evaluate the impact of phone covers on the surface-centric attack (i.e., attack designed such that training is tailored to individual surfaces as previously shown in in Table \ref{tab:training_configurations}, Page \pageref{tab:training_configurations}, Training Configurations II through IV). Like was done in the previous section, we again delineate between the na\"ive and sophisticated attacker scenarios (recall descriptions of these two scenarios in Section \ref{sub-main}). We however present the results in a different format (using bar charts instead of tables similar to Table \ref{tab:pc_training_attack_results}) because the surface-specific scenario has many different training-testing combinations that would each require a distinct table of results. The bar charts provide us a way to present these results in a compact form without losing much information. 

\subsubsection{The na\"ive attacker case}
Figure \ref{fig:naive_attacker_surface_cnn_lstm_training_vol17} summarizes the impact of phone covers on the surface-centric attack under the na\"ive attacker assumption. The plot is based on the highest testing volume since the other volumes provided no new significant insights. On the x-axis, three different surface-specific training options are represented. For each of these training configurations, testing is done in 3 different ways as captured by the color-coded legends. On the y-axis of these plots, the ratio $\alpha$ is computed from the formula, $\alpha$=$\frac{F_{no-defense} - F_{defense}}{F_{no-defense}}$, where $F_{defense}$ is the F-score attained by the attack when the victim employs the defense mechanism (i.e., victim uses a phone cover on the phone used to generate test data), and $F_{no-defense}$ is the F-score attained by the attack when the victim does not employ a defense mechanism (i.e., victim does not use a phone cover on the phone used to generate test data) for that particular pair of training and testing surfaces.

Theoretically $\alpha$ should vary between 0 and 1. When $\alpha$=0, this would mean that the defence does not serve its purpose --- i.e., deployment of the defence causes no reduction at all in the F-score seen before defense deployment (i.e., ${F_{defense}}\approx{F_{no-defense}}$). In general, {\it{the higher the value of $\alpha$, the more the defence mechanism reduces the impact of the attack}}. 

Observe that for all combinations of training and testing surface, $\alpha$ is close to 1. This indicates that given a na\"ive attacker, the defence largely mitigates the attack. For lower testing volumes (results omitted here), we observed a similar pattern. 

\subsubsection{The sophisticated attacker case}
Figure \ref{fig:sophisticated_attacker_surface_cnn_lstm_training_vol17} captures the impact of phone covers on the surface-centric attack under the sophisticated attacker assumption. Like in Figure \ref{fig:naive_attacker_surface_cnn_lstm_training_vol17}, we only report results corresponding to the highest testing volume and use the ratio $alpha$ to express the impact of the phone cover. The figure depicts mixed results --- i.e., some scenarios (e.g., the first two bars) gave values of $\alpha$ that are much lower than the corresponding scenarios in Figure \ref{fig:naive_attacker_surface_cnn_lstm_training_vol17} (implying that the sophisticated attacker was only marginally mitigated), while others (e.g., the last 3 bars) did not seem to differ much from the pattern seen in Figure \ref{fig:naive_attacker_surface_cnn_lstm_training_vol17} (indicating that the defense performed as well as it did for the na\"ive attacker). 

The overall message from the figure is that, depending on the surface used for training and the surface on which the victim places the phone during the attack, the sophisticated attacker could be much more successful than the na\"ive attacker. We conjecture that the couch surface fails to offer any advantage to the attacker because its soft material ``mutes'' much of the vibrations which are again further ``muted'' by the phone cover. The resultant vibration is thus mostly random noise, which has no consistent patterns that can be captured by the classifiers even in the case of a sophisticated attacker who uses the phone cover just like the victim does. For mostly similar reasons, the bed depicts a similar pattern, albeit slightly less pronounced than the couch.

\begin{table*}[h]
	\centering
	\begin{tabular}{llcccccc} 
		\toprule
		\multirow{3}{*}{\textbf{\begin{tabular}[c]{@{}l@{}} Attacker \end{tabular}}}         & \multirow{3}{*}{\textbf{\begin{tabular}[c]{@{}l@{}} Volume\\Level \end{tabular}}} & \multicolumn{6}{c}{\textbf{Surface on which phone is placed during \underline{testing}}}                                                                                                                        \\
		&                                                                           & \multicolumn{2}{c}{\textbf{Table}}                                                                                                   & \multicolumn{2}{c}{\textbf{Bed}}                                                                                   & \multicolumn{2}{c}{\textbf{Couch}}                                                                                  \\ 
		\cmidrule(r){3-4}\cmidrule(r){5-6}\cmidrule(r){7-8}
		&                                                                           & \multicolumn{1}{c}{\begin{tabular}[c]{@{}c@{}}FScore \\ Without \\ Defense \end{tabular}} & \multicolumn{1}{c}{\begin{tabular}[c]{@{}c@{}}FScore \\ With \\ Defense \end{tabular}} & \multicolumn{1}{c}{\begin{tabular}[c]{@{}c@{}}FScore \\ Without \\ Defense \end{tabular}} & \multicolumn{1}{c}{\begin{tabular}[c]{@{}c@{}}FScore \\ With \\ Defense \end{tabular}} & \multicolumn{1}{c}{\begin{tabular}[c]{@{}c@{}}FScore \\ Without \\ Defense \end{tabular}} & \multicolumn{1}{c}{\begin{tabular}[c]{@{}c@{}}FScore \\ With \\ Defense \end{tabular}}  \\ 
		\midrule
		\multirow{5}{*}{Na\"ive}                                 & Vol 15                                                                           &           78.26                     &              55.66                    &           71.97                     &           54.45                      &            70.61                     &             53.67                     \\
		& Vol 13                                                                           &               71.18                 &             54.02                      &             73.12                    &               51.86                   &         68.18                         &             53.67                     \\
		& Vol 11                                                                           &           74.79                      &              51.42                     &                68.29                  &           51.28                       &        63.79                          &                 51.18                  \\
		& Vol 09                                                                           &           66.58                      &            51.35                        &             55.06                     &          51.31                        &           53.21                       &              49.65                     \\
		& Vol 07                                                                           &            56.62                     &         49.42                          &             55.07                    &            49.57                       &               50.11                   &              49.99                      \\ 
		\midrule
		\multirow{5}{*}{Sophisticated}                                  & Vol 15                                                                           &          78.26                      &             57.58                      &           71.97                      &              56.47                    &         70.61                        &                52.66                    \\
		& Vol 13                                                                           &          71.18                      &           51.84                     &           73.12                      &           51.61                      &        68.18                          &               51.64                    \\
		& Vol 11                                                                           &           74.79                      &             50.34                      &             68.29                     &            51.17                      &           63.79                       &                52.31                   \\
		& Vol 09                                                                           &           66.58                      &           50.91                         &                 55.06                 &           51.62                        &             53.21                     &             51.46                      \\
		& Vol 07                                                                           &              56.62                  &           51.15                         &               55.07                   &          49.91                         &                  50.11                &               49.81                     \\
		\bottomrule
	\end{tabular}
	\caption{Impact of phone covers on the novelty detection attack. The table showcases the baseline scenario (i.e., when data from all surfaces and volumes is combined into a single training set). For each phone placement surface during testing, the table compares side-by-side the F-scores obtained when the test dataset is collected from a phone with and without a cover. The classifier used for all results in this table is the one-class SVM (recall details from Section \ref{nd})}
	\label{tab:pc_novelty_detector_performance}
\end{table*}	

\begin{figure*}[h]
	\centering
	\begin{subfigure}[t]{0.95\textwidth}
		\begin{subfigure}[t]{.5\textwidth}
			\centering
			\includegraphics[width=1\linewidth]{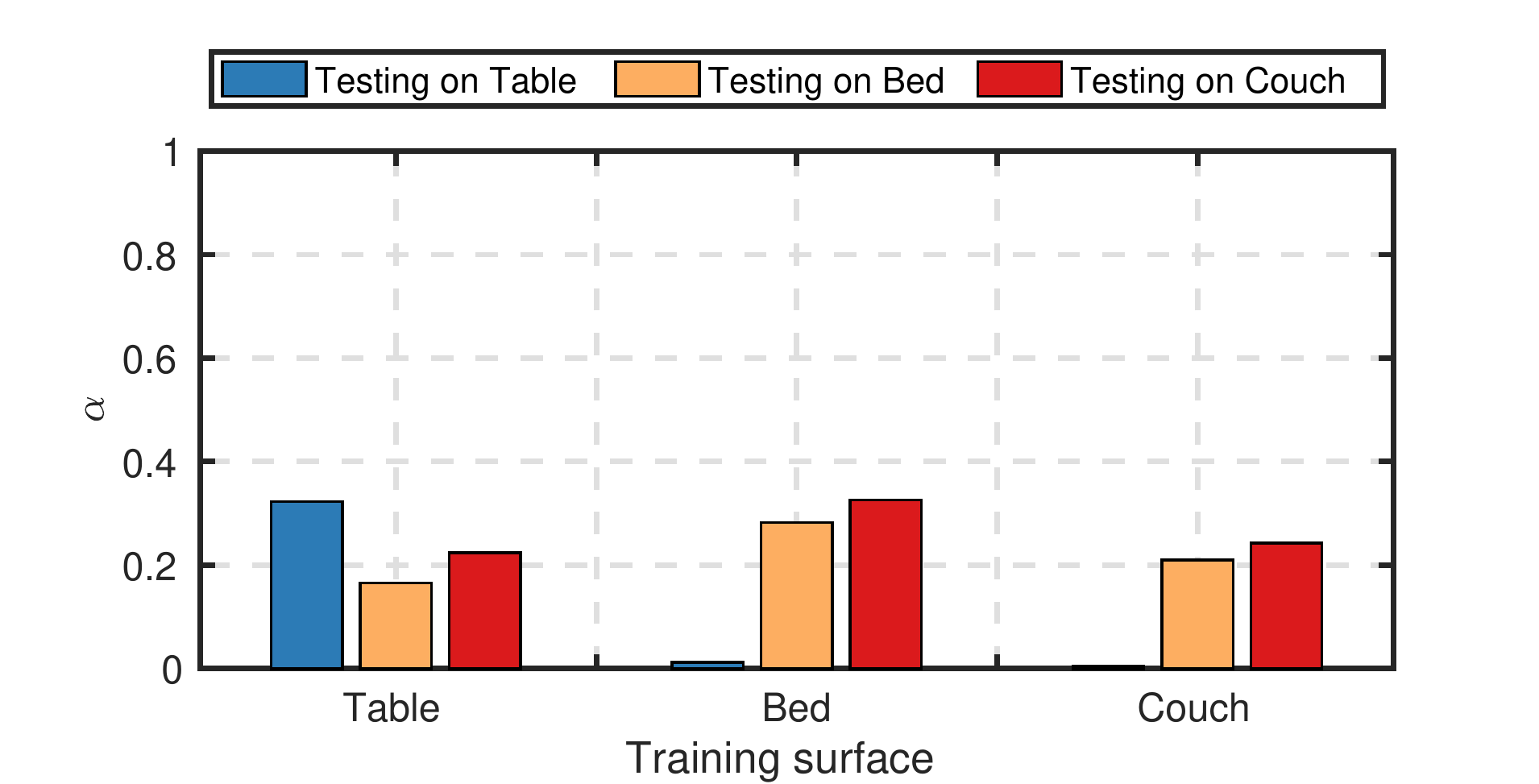}
			\caption{Na\"ive attacker}
			\label{fig:naive_attacker_surface_training_novelty_vol17}
		\end{subfigure}%
		~ 
		\begin{subfigure}[t]{0.5\textwidth}
			\centering
			\includegraphics[width=1\linewidth]{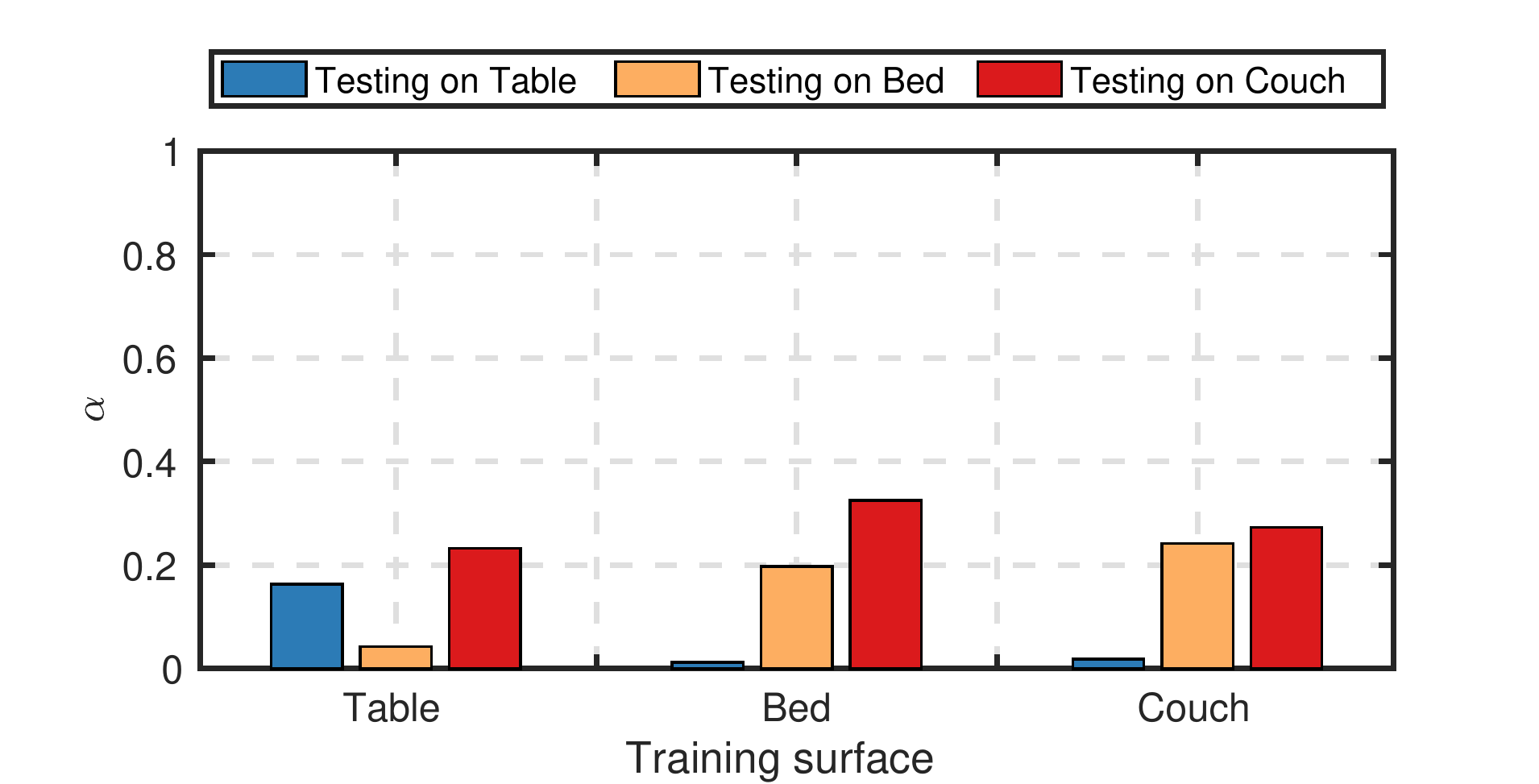}
			\caption{Sophisticated attacker}
			\label{fig:sophisticated_attacker_surface_training_novelty_vol17}
		\end{subfigure}
	\end{subfigure}
	
	\caption{Impact of phone covers on the novelty detection attack under the surface-specific training scenario. In both plots testing is done at the highest volume (Volume 17).}
	\label{fig:pc_surface_training_novelty_results}	
\end{figure*}

\subsection{Impact of Phone Covers on the Baseline Novelty Detection Attack}
\label{cover-baseline}
In this section, we showcase the impact of phone covers on the baseline form of the novelty detection attack. As done previously, we again separate the results into the novice and sophisticated attacker scenarios. 
\subsubsection{The na\"ive attacker case}
The upper half of Table \ref{tab:pc_novelty_detector_performance} summarizes the impact of phone covers on the novelty detection attack launched by the na\"ive attacker. Like was done in Section \ref{sub-main} (recall Table \ref{tab:pc_training_attack_results}), we test on 3 different surfaces and, for context, compare side-by-side the case when a defense is employed by the victim (i.e., the FScore With Defense column) with the case when no defense is employed (i.e., the FScore Without Defense column). All results in the table are generated based on the one-class SVM (recall details from Section \ref{nd}). The F-scores shown in the "FScore Without Defense" column are exactly those previously presented in Section \ref{novel-base}, Table \ref{tab:novelty_detector_performance}, where all training and analysis was done without involvement of phone covers (i.e., no defense involved).  As far as that column is concerned therefore, there is no distinction between the na\"ive and sophisticated attacker (which explains the similar numbers). This distinction is only made in the columns labelled "FScore With Defense" since it is the notion of defence associated with this column that makes one of our two attackers sophisticated, and the other na\"ive, as per our definitions. 

Observe that, when testing is done based on a phone placed on the table surface, the victim's use of a defence results in a reduced impact of the attack (reduction is between 5 and 20 points, depending on the volume). Because the novelty detection problem is a two-class problem (i.e., a song either belongs or does not belong to the corpus), F-Scores in the 50\% range indicate an almost total neutralization of the attack. The pattern seen in this table hence shows that the use of a phone cover during testing is able to largely overcome this attack.

\subsubsection{The sophisticated attacker case}
The lower half of Table \ref{tab:pc_novelty_detector_performance} summarizes the impact of phone covers on the novelty detection attack launched by the sophisticated attacker. Save for the fact that the ``FScore With Defense'' column assumes that the attacker incorporates phone covers into the training, every aspect of the experiment is similar to that previously described for the upper half of the table.  Like for the na\"ive attacker case, the table reveals that F-Scores drop to the 50's, meaning that the sophisticated attacker does not do much better than the na\"ive attacker.  This suggests that for the novelty detection problem, the use of a phone cover blurs the vibrations enough to even thwart the sophisticated attacker. 

\begin{table*}[h!]
	\centering
	\begin{tabular}{cllllll}
		\toprule
		\multirow{2}{*}{\textbf{Model}} & \multicolumn{1}{c}{\multirow{2}{*}{\textbf{Test Surface}}} & \multicolumn{5}{c}{\textbf{Test Volume}}                                                \\
		& \multicolumn{1}{c}{}                                       & \textbf{Vol 15} & \textbf{Vol 13} & \textbf{Vol 11} & \textbf{Vol 09} & \textbf{Vol 07} \\ \midrule
		\multirow{3}{*}{$ET + kNN$}               & Desk                                                       & 98.49           & 98.61           & 97.32           & 98.33           & 94.39           \\
		\textbf{}                       & Bed                                                        & 97.47           & 95.75           & 93.38           & 91.22           & 80.77           \\
		\textbf{}                       & Chair                                                      & 98.08           & 99.07           & 98.02           & 93.52           & 86.72           \\ 
		\midrule
		\multirow{3}{*}{$CNN - LSTM$}   & Desk                                                       & 95.17           & 96.88           & 99.07           & 97.39           & 90.19           \\
		& Bed                                                        & 92.34           & 97.29           & 97.13           & 82.42           & 53.79           \\
		& Chair                                                      & 97.20           & 92.14           & 85.39           & 97.71           & 84.49           \\ 
		\bottomrule
	\end{tabular}
	\caption{Phone cover detector performance over various test surfaces and test volumes. For each of the two classes (i.e., phone cover vs no phone cover), the data from all surfaces and volumes is combined into a single training set.}
	\label{tab:phone_cover_detector}
\end{table*}

\subsection{Impact of Phone Covers on the Surface-Centric Novelty Detection Attack}
In this section, we showcase the impact of phone covers on the surface-specific form of the novelty detection attack. We again separate our results into the novice and sophisticated attacker scenarios, and for our analysis use the ratio $\alpha$, which was previously introduced in Section \ref{covers-scmi-attack} under the surface-centric {\it{music identification}} attack.

\subsubsection{The naive attacker case}	
\label{naive-novelty}
Figure \ref{fig:naive_attacker_surface_training_novelty_vol17} shows the impact of phone covers on the surface-centric attack under the naive attacker assumption. Observe that all values of $\alpha$ are very low (0.35 or less). When testing is done on the table for models built using the other 2 surfaces (the last two stacks of bars), $\alpha$ gets even much lower (close to zero). While it is not so clear why the table stands out in this way, the plot overall points to the defence mechanism largely subduing the attack. 

\subsubsection{The sophisticated attacker case}	

Figure \ref{fig:sophisticated_attacker_surface_training_novelty_vol17} shows the impact of phone covers on the surface-centric attack under the sophisticated attacker assumption. The low values of $alpha$ are similar to those seen earlier in Figure \ref{fig:naive_attacker_surface_training_novelty_vol17}, pointing to the phone covers being able to defeat this attack as well.  
Overall, these results, along with those previously seen in Sections \ref{naive-novelty} and \ref{cover-baseline} indicate that the novelty detection attack is much less able to withstand the phone cover-based defense than the music identification attack (recall results in Sections \ref{covers-scmi-attack} and \ref{sub-main}).

\section{Experimental Results --- Detecting the Victim's Use of a Phone Cover}
\label{cover-def}
The sophisticated attacker described in Section \ref{sec:defense_mechanism} will likely seek to formally detect the victim's usage (or non-usage) of the phone cover in order to make an informed decision on how to train the classifiers that drive the music inference attack. In particular, given sensor data collected from the victim's phone, the sophisticated attacker will likely pursue the following 2 steps in chronological order: (1) Use the sensor data to infer whether the victim's phone has a casing, (2) Accordingly implement the music inference attack. 

Until this point, our results have focused on various forms of Step (2). Here, we present our findings on how well sensor readings might answer the question of whether the victim uses a phone cover or not (i.e., Step (1)). As previously described in Section \ref{dm} and the sections referred to therein, we combine data from all surfaces and volumes to train the two classifiers that we use to tackle this two-class problem (revisit Section \ref{dm} and sections referenced therein for training details). 

Table \ref{tab:phone_cover_detector} shows our results on how well these classifiers detect the presence of a phone cover on the target phone. The table shows test results for each surface and volume. Observe that at the two highest volumes, the F-score is greater than 90\% irrespective of the surface on which the phone is placed during testing. For the lower volumes, the F-scores are a mostly a mix of F-scores in the 90\'s and 80\'s. Overall these results indicate that the attacker will be able to detect the usage of a phone cover with very high accuracy. Given the motivation of a more robust music inference engine, we argue that the good performance of this module would motivate attackers to go through this step before undertaking the music inference attack itself. 

	\section{Discussion and Conclusion}
	In this paper we successfully demonstrate a sensor-based side-channel attack on user listening habits when the sound is played from the speakers of a mobile device. As part of this work we have explored novelty detection, the way in which different types of machine learning models effect the performance of the attack on a variety of surfaces and against different volume levels, and how users can protect themselves against such attacks. Before concluding this paper we first describe some insights we gained during the course of this research and ways in which future research could expand upon our work.

{\it Attack Impact: } We show that an attacker can use accelerometer data from a smartphone to classify songs that a user is listening to with an accuracy of over 80\%. As discussed in the introduction to our paper, this knowledge could be used to make educated guesses about a user's behavior, political preferences, or social traits \cite{Akshay2018, North2010, Rentfrow2003, Adrian10}. This type of information could be used for targeted advertising or disinformation campaigns to see what types of users are more likely to engage in the content (several of the classes could be the content itself).

We also found that although the ET + kNN classifier performed around the same as the DNN for the Table surface, the DNN outperformed the ET + kNN classifier for soft surfaces (by up to 20\%). This demonstrates that if an attacker aims to achieve robust performance on a wide variety of surfaces, particularly those with natural dampening properties, they will likely have to use a DNN.

{\it Attack Deployment: } As previously stated in our threat model in Section \ref{sec:threat_model}, we believe the attacker would likely use a hybrid deployment scenario. This would involve some processing on the phone before uploading the data to run the machine learning models used for executing the attack.

Potential steps executed on the phone include: (1) identifying if the phone is in a scenario in which the attack should be executed, (2) data quality analysis to throw away samples before further processing occurs, and (3) an optional compression step to reduce the footprint of the data being uploaded.

Once the sensor data is uploaded to the server (or node in a cluster), the attacker would then have significant flexibility in determining which model to run against the data. It's likely that any sophisticated attacker would have dozens of models for different versions of the attack and additional models used for determining which version of the attack to execute (e.g., for determining the surface or if the phone has a case).

If the attacker wishes to deploy the attack on a mobile device in its entirety then they would likely wish to reduce the footprint of the classification mechanism. For example with DNNs this is possible through pruning and various other neural network size reduction techniques which reduce the computational and storage overhead of the DNNs once placed on the device.

{\it Defensive Experiment Takeaways: } Although the defensive measure we put in place is highly effective, we did not analyze the effectiveness of different types of cases. In particular, the Otterbox case is one of the most durable and is more likely than a thin plastic case to protect the user against attacks on their listening habits. This suggests that if the attack (or a similar approach) were to be heavily utilized, it would be advantageous for the makers of lightweight cases to add padding for mitigating vibrations of the device.

Other solutions have been suggested in recent work, such as the modification of the phone software or hardware (e.g., \cite{anand2019spearphone}). However, the use of a phone case is a more user friendly solution and it is thus far quite robust against countermeasures on the part of the attacker. Our research found that the use of a phone case as a defensive measure remained viable even when the attacker trained on defensive data. This means that the Otterbox case manages to suppress enough vibrations that it makes many songs almost indistinguishable.

{\it Conclusion: } In this paper we demonstrate an attack that is capable of classifying music that a user is listening to via the speakers on their mobile device, an attack that could likely be extended to other forms of media with relative ease. We advance the state of the art in attacks on mobile device audio using motion sensors by achieving an accuracy of over 80\% using a corpus of 100 songs, a previously unexplored target of such attacks. Our research also contributes additional experiments that show defensive measures using vibration dampening phone cases can significantly reduce the impact of such an attack, even when the attacker uses defensive data during the training process. The usage of five volume levels and three surfaces makes our research a robust exploration of the kinds of environments in which this attack might be conducted. Our future research will focus on ways in which the attacker can further combat defensive measures and improve the effectiveness of their attack under non-ideal conditions such as low volume levels.

\section{Acknowledgment}

This research was supported by National Science Foundation Award Number: 1527795.

\balance

\bibliographystyle{IEEEtran}
\bibliography{IEEEabrv,songsbib}

\newpage

\onecolumn

\section{Appendices}

\subsection{List of Top 150 Selected Features}
\label{subsec:selected_features}

\begin{figure*}[h]
	\centering
	\begin{subfigure}[t]{0.95\textwidth}
		\begin{subfigure}[t]{.60\textwidth}
			\centering
			\includegraphics[width=1\linewidth]{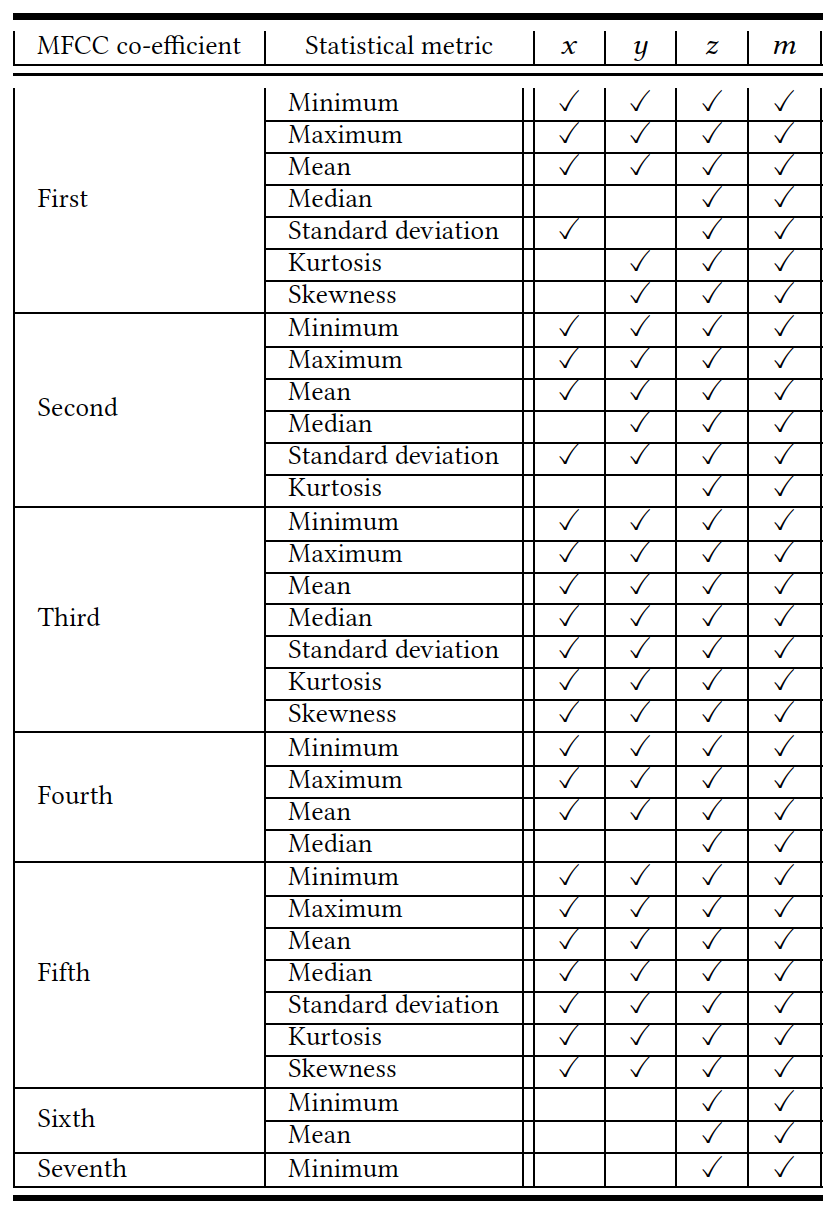}
			\caption{A list of statistical metrics computed on MFCC co-efficients. 
			}
			\label{fig:selected_mfcc_features}
		\end{subfigure}%
		~ 
		\begin{subfigure}[t]{0.45\textwidth}
			\centering
			\includegraphics[width=1\linewidth]{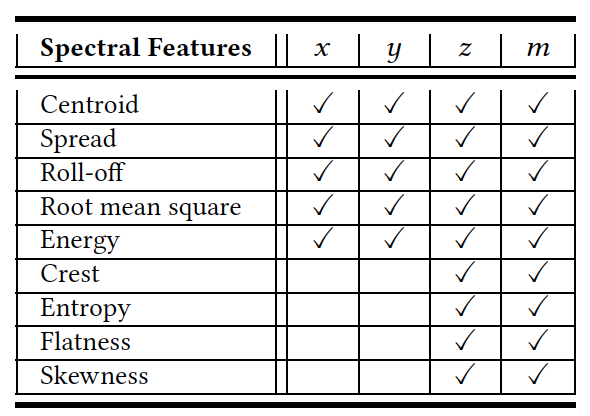}
			\caption{A list of spectral features. 
			}
			\label{fig:selected_spectral_features}
		\end{subfigure}
	\end{subfigure}
	
	\caption{A list of top 150 selected features used in our feature engineering-based framework.}
	\label{fig:top_150_features}	
\end{figure*}

\newpage

\subsection{List of Billboard Top 100 Songs used in our experiments}

\begin{table*}[h]
	\tiny
	\centering
	\begin{tabular}{|l|l|l||l|l|l|} \hline
		\toprule 
		\textbf{No.} & \textbf{Song Title}                             & \textbf{Artist}                               & \textbf{No.} & \textbf{Song Title}                & \textbf{Artist}                             \\  \hline \hline
		1.           & Nice For What                                   &  Drake                                & 51.          & IDGAF                              & Dua Lipa                           \\  \hline
		2.           & Psycho                                          & \begin{tabular}[t]{@{}l@{}}Post Malone Featuring Ty Dolla \\ \$ign\end{tabular} & 52.          & Sad!                               & XXXTENTACION                       \\  \hline
		3.           & I Like It                                       & Cardi B, Bad Bunny \& J Balvin       & 53.          & Woman, Amen                        & Dierks Bentley                     \\  \hline
		4.          & God's Plan                                      & Drake                                & 54.         & Call Out My Name                   & The Weeknd                         \\  \hline
		5.           & Girls Like You                                  & Maroon 5 Featuring Cardi B           & 55.          & Done For Me                        & Charlie Puth Featuring Kehlani     \\  \hline
		6.          & Lucid Dreams                                    & Juice WRLD                           & 56.          & Sit Next To Me                     & Foster The People                  \\  \hline
		7.                                 & Boo'd Up                                        & Ella Mai                                                  & 57.                               & Simple                                                  & Florida Georgia Line               \\ \hline
		8.                                 & The Middle                                      & Zedd, Maren Morris \& Grey                                & 58.                               & X                                                       & Nicky Jam x J Balvin               \\ \hline
		9.                                 & No Tears Left To Cry                            & Ariana Grande                                             & 59.                               & Ball For Me                                             & Post Malone Featuring Nicki Minaj  \\ \hline
		10.                                & Meant To Be                                     & Bebe Rexha \& Florida Georgia Line                        & 60.                               & Ghost Town                                              & Kanye West 
		\\ \hline
		11.                                & Yes Indeed                                      & Lil Baby \& Drake                                         & 61.                               & You Make It Easy                                        & Jason Aldean                       \\ \hline
		12.                                & This is America                                 & Childish Gambino                                          & 62.                               & Freeee (Ghost Town, Pt. 2)                              & Kids See Ghosts                    \\ \hline
		13.                                & Friends                                         & Marshmello \& Anne-Marie                                  & 63.                               & Mercy                                                   & Brett Young                        \\ \hline
		14.                                & Walk It Talk It                                 & Migos Featuring Drake                                     & 64.                               & Dura                                                    & Daddy Yankee                       \\ \hline
		15.                                & Mine                                            & Bazzi                                                     & 65.                               & KOD                                                     & J. Cole                            \\  \hline
		16.                                & In My Blood            & Shawn Mendes                                                                                               & 66.                               & Everything's Gonna Be Alright            & David Lee Murphy \& Kenny Chesney                                                             \\ \hline
		17.                                & Perfect                & Ed Sheeran                                                                                                 & 67.                               & Fire                                     & Kids See Ghosts                                                                               \\ \hline
		18.                                & Look Alive             & BlocBoy JB Featuring Drake                                                                                 & 68.                               & Esskeetit                                & Lil Pump                                                                                      \\ \hline
		19.                                & Never Be The Same      & Camila Cabello                                                                                             & 69.                               & Cudi Montage                             & Kids See Ghosts                                                                               \\ \hline
		20.                                & Better Now             & Post Malone                                                                                                & 70.                               & Praise The Lord (Da Shine)               & A\$AP Rocky Featuring Skepta                                                                  \\ \hline
		21.                                & Whatever It Takes      & Imagine Dragons                                                                                            & 71.                               & Fake Love                                & BTS                                                                                           \\ \hline
		22.                                & Back To You            & Selena Gomez                                                                                               & 72.                               & Japan                                    & Famous Dex                                                                                    \\ \hline
		23.                                & Be Careful             & Cardi B                                                                                                    & 73.                               & Kids See Ghosts                          & Kids See Ghosts                                                                               \\ \hline
		24.                                & Rockstar               & Post Malone Featuring 21 Savage                                                                            & 74.                               & Alone                                    & Halsey Featuring Big Sean \& S. Don                                                     \\ \hline
		25.                                & Delicate               & Taylor Swift                                                                                               & 75.                               & Powerglide                               & Rae Sremmurd \& Juicy J                                                                       \\ \hline
		26.                                & Heaven                 & Kane Brown                                                                                                 & 76.                               & I Lived It                               & Blake Shelton                                                                                 \\ \hline
		27.                                & Havan                  & Camila Cabello Featuring Y. Thug                                                                        & 77.                               & Overdose                                 & YoungBoy Never Broke Again                                                                    \\ \hline
		28.                                & I'm Upset              & Drake                                                                                                      & 78.                               & Lovely                                   & Billie Eilish \& Khalid                                                                       \\ \hline
		29.                                & Wait                   & Maroon 5                                                                                                   & 79.                               & Youngblood                               & 5 Seconds of Summer                                                                           \\ \hline
		30.                                & One Kiss               & Calvin Harris \& Dua Lipa                                                                                  & 80.                               & Dame Tu Cosita                           & \begin{tabular}[t]{@{}l@{}}Pitbull x El Chombo x Karol G \\ Featuring Curry Ranks\end{tabular} \\ \hline
		31.                                & All Mine               & Kanye West                                                                                                 & 81.                               & OTW                                      & Khalid, Ty Dolla \$ign \& 6LACK                                                               \\ \hline
		32.                                & Tequila                & Dan + Shay                                                                                                 & 82.                               & Zombie                                   & Bad Wolves                                                                                    \\ \hline
		33.                                & Plug Walk              & Rich The Kid                                                                                               & 83.                               & Me Niego                                 & Reik Featuring Ozuna \& Wisin                                                                 \\ \hline
		34.                                & Love Lies              & Khalid \& Normani                                                                                          & 84.                               & Lose It                                  & Kane Brown                                                                                    \\ \hline
		35.                                & New Rules              & Dua Lipa                                                                                                   & 85.                               & Violent Crimes                           & Kanye West                                                                                    \\ \hline
		36.                                & Te Bote                & \begin{tabular}[t]{@{}l@{}}Casper Magico, Nio Garcia, Darell, \\ Nicky Jam, Ozuna \& Bad Bunny\end{tabular} & 86.                               & I Was Jack (You Were Diane)              & Jake Owen                                                                                     \\ \hline
		37.                                & Freaky Friday          & Lil Dicky Featuring Chris Brown                                                                            & 87.                               & Sativa                                   & \begin{tabular}[t]{@{}l@{}}Jhene Aiko Featuring Swae Lee Or \\ Rae  Sremmurd\end{tabular}      \\ \hline
		38.                                & Taste                  & Tyga Featuring Offset                                                                                      & 88.                               & Rich Sex                                 & Nicki Minaj Featuring Lil Wayne                                                               \\ \hline
		39.                                & Reborn                 & Kids See Ghosts                                                                                            & 89.                               & Life Goes On                             & Lil Baby Featuring Gunna \& L. U. Vert                                                      \\ \hline
		40.                                & Yikes                  & Kanye West                                                                                                 & 90.                               & I Know You                               & Lil Skies Featuring Yung Pinch                                                                \\ \hline
		41.                                & All Girls Are The Same & Juice WRLD                                                                                                 & 91.                               & El Farsante                              & Ozuna \& Romeo Santos                                                                         \\ \hline
		42.                                & 4th Dimension          & \begin{tabular}[t]{@{}l@{}}Kids See Ghosts Featuring Louis \\ Prima\end{tabular}                                                                      & 92.                               & Wouldn't Leave                           & \begin{tabular}[t]{@{}l@{}}Kanye West Featuring \\ PARTYNEXTDOOR \end{tabular}                                                           \\ \hline
		43.                                & Chun-Li                & Nicki Minaj                                                                                                & 93.                               & Beautiful Crazy                          & Luke Combs                                                                                    \\ \hline
		44.                                & Pray For Me            & The Weeknd \& Kendrick Lamar                                                                               & 94.                               & Sin Pijama                               & Beck G + Natti Natasha                                                                        \\ \hline
		45.                                & One Number Away        & Luke Combs                                                                                                 & 95.                               & Babe                                     & Sugarland Featuring Taylor Swift                                                              \\ \hline
		46.                                & I Like Me Better       & Lauv                                                                                                       & 96.                               & Drip                                     & Cardi B Featuring Migos                                                                       \\ \hline
		47.                                & Feel The Love          & Kids See Ghosts                                                                                            & 97.                               & Big Bank                                 & \begin{tabular}[c]{@{}l@{}}YG Featuring 2 Chainz, Big Sean \& \\ Nicki   Minaj\end{tabular}     \\ \hline
		48.                                & Get Along              & Kenny Chesney                                                                                              & 98.                               & Singles You Up                           & Jordan Davis                                                                                  \\ \hline
		49.                                & Up Down                & \begin{tabular}[t]{@{}l@{}}Morgan Wallen Featuring Florida \\ Georgia Line\end{tabular}                     & 99.                               & Take Back Home Girl                      & Chris Lane Featuring Tori Kelly                                                               \\ \hline
		50.                                & Tati                   & 6ix9ine Featuring DJ Spinking                                                                              & 100.                              & Welcome To The Party                     & Diplo, French Montana \& Lil Pump                                                       \\ \hline
		
		\bottomrule   \hline                                                 
	\end{tabular}
	\caption{List of Billboard Top 100 Songs used in our experiments}
	\label{tab:song_list}
\end{table*}

\end{document}